\documentclass[12pt]{article}
\pdfoutput=1

\newlength{\abstractwidth}
\usepackage{epsf}
\usepackage{color}
\usepackage{graphicx}
\usepackage{tikz}
\usepackage{hyperref}

\usepackage{amsmath}
\usepackage{amssymb}
\usepackage{latexsym}

\renewcommand{\thefootnote}{\fnsymbol{footnote}}
\renewcommand{\thanks}[1]{\footnote{#1}}
\newcommand{\starttext}{
\setcounter{footnote}{0}
\renewcommand{\thefootnote}{\arabic{footnote}}}

\newcommand{\bea}{\begin{eqnarray}}
\newcommand{\eea}{\end{eqnarray}}
\newcommand{\be}{\begin{eqnarray}}
\newcommand{\ee}{\end{eqnarray}}

\def\cB{{\cal B}}
\def\cC{{\cal C}}

\def\cF{{\cal F}}
\def\cG{{\cal G}}
\def\cH{{\cal H}}
\def\cI{{\cal I}}

\def\cL{{\cal L}}
\def\cM{{\cal M}}

\def\cO{{\cal O}}

\def\cR{{\cal R}}
\def\cS{{\cal S}}

\def\cY{{\cal Y}}

\def\mA{\mathfrak{A}}
\def\mB{\mathfrak{B}}
\def\mC{\mathfrak{C}}
\def\mD{\mathfrak{D}}

\def\mI{\mathfrak{I}}
\def\mJ{\mathfrak{J}}

\def\mS{\mathfrak{S}}

\def\mb{\mathfrak{b}}

\def\mf{\mathfrak{f}}

\def\mt{\mathfrak{t}}

\def\ZZ{{\mathbb Z}}
\def\RR{{\mathbb R}}
\def\NN{{\mathbb N}}
\def\CC{{\mathbb C}}

\def\GG{{\mathbb G}}

\def\Re{{\rm Re \,}}
\def\Im{{\rm Im \,}}

\def\det{{\rm det \,}}

\def\half{{1\over 2}}
\def\p{\partial}

\def\f{\varphi}
\def\tet{\vartheta}
\def\ep{\varepsilon}
\def\om{\omega}

\def\Sep{{\Sigma _{ab}}}

\def\pbx{\p _{\bar x}}

\def\pbz{\p _{\bar z}}

\def\GA{\cG}
\def\kap{\kappa}
\def\oom{\overline{\om}}

\def\mod{{\cal F}^{(h)}  }

\def\no{\nonumber}
\def\sm{\smallskip}

\begin{document}
\starttext
\setcounter{footnote}{0}

\begin{flushright}
2017 December 17 \\
QMUL-PH-17-27 \\
DAMTP-2017-45
\end{flushright}

\vskip 0.3in

\begin{center}

{\Large \bf Higher genus modular graph functions,  } 

\vskip 0.1in

{\Large \bf string invariants, and their exact asymptotics}

\vskip 0.4in

{\large \bf Eric D'Hoker$^{(a)}$, Michael B. Green$^{(b)}$ and Boris Pioline$^{(c)}$}

\vskip 0.15in

{ \sl  (a) Mani L. Bhaumik Institute for Theoretical Physics}\\
{\sl  Department of Physics and Astronomy}\\
{\sl University of California, Los Angeles, CA 90095, USA}

\vskip 0.05in

{ \sl (b) Department of Applied Mathematics and Theoretical Physics }\\
{\sl Wilberforce Road, Cambridge CB3 0WA, UK}, and \\
{\sl Centre for Research in String Theory, School of Physics, }\\
{\sl Queen Mary University of London, Mile End Road, London, E1 4NS, England}

\vskip 0.05in

{\sl (c) Laboratoire de Physique Th\'eorique et Hautes Energies, CNRS UMR 7589,}\\
{\sl Universit\'e Pierre et Marie Curie, 4 place Jussieu, 75252 Paris, France}, and \\
{\sl Sorbonne Universit\'es, UPMC Universit\'e Paris 6, UMR 7589, F-75005 Paris, France}

\vskip 0.1in

{\tt \small dhoker@physics.ucla.edu, M.B.Green@damtp.cam.ac.uk,  pioline@lpthe.jussieu.fr}

\vskip 0.5in

\begin{abstract}
The concept and the construction of modular graph functions are generalized from genus-one to higher genus surfaces. The  integrand of the four-graviton superstring amplitude at genus-two  provides a generating function for a special class of such functions.  A general method is developed for analyzing the behavior of modular graph functions under non-separating degenerations in terms of a natural real parameter $t$. For arbitrary genus,  the Arakelov Green function and the Kawazumi-Zhang invariant degenerate to a Laurent polynomial in $t$ of degree $(1,1)$  in the limit $t\to\infty$.  For genus two, each coefficient of the low energy expansion of the string amplitude degenerates to a Laurent polynomial of degree $(w,w)$ in $t$, where $w+2$ is the degree of homogeneity in the kinematic invariants. These results are exact to all orders in $t$, up to exponentially suppressed corrections.  The non-separating degeneration of a general class of modular graph functions at arbitrary genus is sketched and similarly results in a Laurent polynomial in $t$ of bounded degree. The coefficients in the Laurent polynomial are generalized modular graph functions for a punctured Riemann surface of lower genus.

\end{abstract}
\end{center}

\newpage

\setcounter{tocdepth}{1} 
\tableofcontents

\newpage

\baselineskip=15pt
\setcounter{equation}{0}
\setcounter{footnote}{0}

\section{Introduction and overview}
\setcounter{equation}{0}
\label{sec1}

Modular graph functions for a genus-one Riemann surface map certain classes of graphs to real-analytic  $SL(2,\ZZ)$-invariant functions on the Poincar\'e upper half-plane $\cH$. They naturally arise in the low energy expansion of superstring amplitudes at genus one  \cite{Green:2008uj} and exhibit some remarkable properties. Modular graph functions generalize non-holomorphic Eisenstein series, which may be viewed as modular graph functions for one-loop graphs.  They  are intimately connected with elliptic polylogarithms and multiple zeta values \cite{DHoker:2015wxz}, and obey systems of differential equations  as well as  surprising algebraic relations  \cite{D'Hoker:2015foa,DHoker:2015sve,DHoker:2016mwo,DHoker:2016quv}. Notably, in the neighborhood of the cusp of the standard fundamental domain for $\cH/SL(2,\ZZ)$, a modular graph function of $\tau \in \cH$ behaves, up to exponentially suppressed terms, as a Laurent polynomial in $\Im \tau$ of bounded degree \cite{DHoker:2017zhq}, thereby generalizing the behavior of non-holomorphic Eisenstein series. 

\sm

In superstring perturbation theory, scattering amplitudes are approximated by power series in the string coupling constant with coefficients that are associated with Riemann surfaces of {\sl arbitrary} genus.  The tree-level contribution arising from genus-zero is well-known textbook material and will not be discussed further here. Although the genus-one amplitudes have also been known for many years \cite{Green:1981yb}, the systematic study of their low energy expansion is fairly recent \cite{Green:2008uj,Green:1999pv}, and was the main motivation for considering genus-one modular graph functions.
At genus two, only the simplest superstring amplitudes involving four massless particles have been calculated, for bosons in Type II and Heterotic string theories using the RNS formulation  in \cite{D'Hoker:2002gw,D'Hoker:2005jc,D'Hoker:2005ht}, and reproduced for bosons and extended to fermions in Type~II using the pure spinor formulation in \cite{Berkovits:2005df,Berkovits:2005ng}. These amplitudes reduce to integrals over the positions of the four vertex operators on the Riemann surface and over the three-dimensional moduli space of complex structures of genus-two Riemann surfaces. After integration over the vertex operator positions, the remaining integrands  depend on the remaining moduli  and are invariant under the genus-two modular group $Sp(4,\ZZ)$. Their  low energy expansion naturally provides modular invariants which will serve as prototypes for higher-genus modular graph functions.

\sm

The simplest non-trivial  genus-two modular graph function was  shown in \cite{D'Hoker:2013eea} to coincide with the Kawazumi-Zhang (KZ) invariant  \cite{Kawazumi,Zhang}. The latter is closely related to the Faltings invariant, which is of central importance in Arakelov geometry  \cite{zbMATH06139356}. The study of physical constraints on the genus-two string integrand, and the use of complex structure deformation theory, have revealed a wealth of unexpected properties satisfied by the KZ invariant, including an eigenvalue equation for the $Sp(4,\ZZ)$-invariant differential operators of degree two \cite{DHoker:2014oxd} and degree four  \cite{Pioline:2015qha}, and a theta-lift representation analogous to Borcherds' theta-lift for the Igusa cusp form \cite{Pioline:2015qha}. It was pointed out in \cite{D'Hoker:2013eea} that successively higher order terms in the low energy expansion of the genus-two superstring amplitudes provide an infinite number of novel generalizations of KZ-invariants. 

\sm

The first goal of this paper is to define and construct general modular graph functions on compact Riemann surfaces of arbitrary genus, motivated by the mathematical structure of genus-two superstring amplitudes. Key ingredients in our construction will be the canonical K\"ahler form and the Arakelov Green function (which is closely related to the scalar Green function commonly used in the string theory literature \cite{D'Hoker:1988ta}) on a Riemann surface $\Sigma_h$ of arbitrary genus $h$.
In terms of these objects, an infinite class of modular graph functions may be readily constructed by complete analogy with the case of genus-one. However, the invariants arising from genus-two string amplitudes do not fit into this class. Learning the lessons of string theory, we shall extend this class by generalizing the canonical K\"ahler form on which they are constructed to a family of forms of top degree on a product  of several copies of $\Sigma_h$. Special cases of this generalization were anticipated  by Kawazumi  in his work on the construction of invariant tensors on Teichm\"uller space \cite{Kawa1, Kawa2}. The modular graph functions thus constructed are functions on the moduli space of Riemann surfaces and, for low genus $h \leq 3$,  may equivalently be viewed as functions on the Siegel upper half space. We shall continue to use the terminology ``modular graph function" for arbitrary genus in order to emphasize their close relation  with their genus-one cousins \cite{DHoker:2015wxz}.

\sm

The second goal of this paper is to determine the behavior of higher-genus modular graph functions near the non-separating divisor in the Deligne-Mumford compactification of the moduli space  of compact Riemann surfaces. The motivation is, of course, to gain further insight into the structure of these higher invariants. In particular, experience in studying modular graph functions at genus one, and the KZ-invariant at genus two,  has taught us that the asymptotic behavior under a non-separating degeneration provides invaluable information towards uncovering and then proving the differential and algebraic relations these functions obey. The behavior under a separating degeneration  is also of obvious interest, but lies outside the scope of this paper.  

\sm

To achieve the second goal, we shall develop a general method for evaluating the behavior of modular graph functions on a Riemann surface $\Sigma_{h+1}$ of genus $h+1$ as the surface undergoes a non-separating degeneration to a compact surface $\Sigma_h$ to which  two punctures $p_a,p_b$ are added. We denote by $\Omega$ the period matrix of $\Sigma _{h+1}$ with respect to a canonical homology basis of cycles $\mA_I, \mB_I$ for $I=1,\cdots, h+1$ and choose the degenerating cycle to be $\mB_{h+1}$. The non-separating degeneration then corresponds to the limit where the entry $\sigma = \Omega _{h+1, h+1}$ of the period matrix tends to $i \infty$. The period matrix of $\Sigma _h$ with respect to the same homology basis is denoted by~$\tau$. 
The key features of our method are as follows.
\begin{description}
\item (a) The approach to the non-separating degenerating is parametrized by the positive real variable $t=\det (\Im \Omega)/\det( \Im \tau)$, such that the degeneration limit $\sigma\to i\infty$ corresponds to $t \to \infty$. Furthermore,  $t$ is invariant under the Fourier-Jacobi group $Sp(2h,\mathbb{Z})\ltimes (\mathbb{Z}^h + \tau \ZZ^h) \ltimes \mathbb{Z}$ which is the subgroup of the full modular group $ Sp(2h+2,\mathbb{Z})$ leaving the cusp at  $\sigma=i \infty$ invariant.
\item (b) A  real harmonic function $f$  on $\Sigma_h \setminus \{ p_a, p_b\}$ is used to provide a natural  definition of the discs in Fay's construction \cite{fay73} of a non-separating degeneration. The discs are centered at the punctures $p_a$ and $p_b$ and their boundaries are defined as level sets of the function~$f$ at the values $\pm 2 \pi t$. The holomorphic differentials on $\Sigma_{h+1}$ are expressed simply in terms of $f$ and the holomorphic differentials on $\Sigma_h$. The  function $f$  has been used earlier in string theory to endow the string worldsheet with a global light-cone time coordinate \cite{Giddings:1986rf,DHoker:1987hzc}.  
\item (c) Finally, we develop a variational method for calculating the power-behaved part of the $t$-derivative of any integral whose $t$-dependence arises solely from  the positions of  its boundary cycles near the two punctures of the degeneration. This method will turn out to be quite powerful in determining the possible power-like terms. Physically, the method is akin to the renormalization group methods in quantum field theory, with the parameter $t$ playing the role of the logarithm of a short-distance cutoff. 
\end{description}

\sm 

Using this method we shall prove four  theorems (numbered 1 through 4) which specify the behavior near the non-separating degeneration, as  $t \to \infty$, of  the following objects,
\begin{enumerate}
\itemsep 0.0in
\item the Arakelov Green function $\GA_h(x,y)$ for arbitrary genus;
\item the Kawazumi-Zhang invariant $\varphi_h$ for arbitrary genus;
\item the higher order string invariants $\cB_w(\Omega)$ generated by the Taylor expansion
of the genus-two four-graviton amplitude $\cB^{(2)} (s_{ij} |\Omega)$ defined in \eqref{B2g} and
studied in  \cite{D'Hoker:2013eea};
\item an infinite class of  modular graph functions generated by the Taylor expansion
of a putative ``higher genus amplitude" $\mod(s_{ij}|\Omega)$ defined in \eqref{h-loop1}  for arbitrary genus (but whose role in string theory  is as yet unclear).
\end{enumerate}
We prove that, in each of these cases, the expansion is given by a Laurent polynomial of bounded degree in $t$ plus  terms of order $\cO(e^{-2 \pi t})$.  The degree\footnote{We denote the degree of a Laurent polynomial by $(r,s)$ where the integers $r$ and $-s$ respectively stand for the exponents of the highest and lowest powers in $t$. } of the Laurent polynomial in $t$ is $(1,1)$  for Theorems~1 and~2, while  it is $(w,w)$  for Theorems~3 and~4, where $w$ is the weight of the modular graph function, defined as its degree of homogeneity in the  Green function. The Laurent polynomial is independent of $\Re\sigma$.  Its coefficients are modular graph functions of genus $h$, and generalizations thereof which depend both on the period matrix $\tau$ of $\Sigma_h$ and on the image $v$  of the  punctures under the Abel map. For $h=1$, such objects were referred to as {\sl single-valued elliptical multiple polylogarithms} in \cite{DHoker:2015wxz}.   

\sm

For arbitrary genus, Theorems 1 and 2 reproduce earlier mathematical  results by Wentworth  \cite{MR1105425} and de Jong \cite{zbMATH06355718} at  order $t^1$ and $t^0$ and extend those results by evaluating the contribution of order  $1/t$ and showing that the series vanishes for all higher powers of $1/t$. For genus two, Theorem 2 provides a direct derivation of the asymptotics of the genus-two KZ invariant  obtained in \cite{Pioline:2015qha}, without relying on the  theta lift representation.
 
 \sm

In a companion paper \cite{DGP}, we shall derive the explicit coefficients of the Laurent polynomial that arises in the  non-separating degeneration of the genus-two string invariant of weight $w=2$ contributing to the $D^8 \cR^4$  low-energy effective interaction. We shall also provide a detailed check that its complete degeneration limit, or tropical limit, agrees with the predictions of independent supergravity calculations.

\subsection{Organization}

The remainder of this paper is organized as follows. In section \ref{sec2} we review genus-one modular graph functions and then  proceed to discussing higher genus surfaces,  the canonical K\"ahler form, the Arakelov Green function, and the simplest infinite class of higher genus modular graph functions. Learning from the structure of the genus-two superstring amplitude, we generalize the construction as described in the preceding paragraph. In section \ref{sec3} we expound our method for calculating non-separating degenerations, introducing the degenerating parameter $t$ and worldsheet time function $f$. 
In section \ref{sec4} we obtain the non-separating degeneration of the Arakelov Green function and KZ-invariant at any genus, developing techniques along the way to evaluate the behavior of general modular graph functions. The non-separating degeneration of the low energy expansion of the genus-two string amplitude is presented in section \ref{sec5}, with the technical proof given in appendix A, while the degeneration of a natural class of higher-genus modular graph functions is given  in section \ref{sec6}. 

\subsection*{Acknowledgments}

ED thanks Bill Duke and D.H. Phong  for discussions on various related topics, Justin Kaidi for helpful comments on the draft, and  the Kavli Institute for Theoretical Physics at UCSB for their hospitality during the early stage of this work. MBG thanks Arnab Rudra for discussions on related topics \cite{AR}.
MBG and BP are grateful to the Mani L. Baumik Institute at UCLA for hospitality  during part of this project. All authors are grateful to the organizers of the workshop {\sl Automorphic forms, Mock modular forms and String theory} at the Banff International Research Station for providing a stimulating atmosphere in the last stage of this project. 

\sm

The research of ED  is supported in part by the National Science Foundation under research grants PHY-16-19926 and PHY-1125915, and by a Fellowship from the Simons Foundation.  MBG has been partially supported by STFC consolidated grant ST/L000385/1. The research of BP is supported in part  by French state funds managed by  ANR in the context 
of the LABEX ILP (ANR-11-IDEX-0004-02, ANR-10- LABX-63).

\newpage

\section{Modular graph functions at higher genus}
\setcounter{equation}{0}
\label{sec2}

In this section, we shall extend the construction of modular graph functions from genus-one to higher genus Riemann surfaces. To do so, we begin by summarizing the key results of  the genus-one case, and then review some basic elements of algebraic geometry on higher genus surfaces, including the canonical K\"ahler  form, and the Arakelov Green function.\footnote{ General references on Riemann surfaces and their function theory include the textbooks \cite{Siegel, FK,MR688651,Klingen}. More specialized accounts of the topics of direct interest to this paper may be found in  \cite{D'Hoker:1988ta, fay73}, while useful overviews of Arakelov geometry are in \cite{AlvarezGaume:1987vm, MR1065156}.}

\subsection{Summary of genus-one}

We parametrize a compact genus-one Riemann surface without boundary $\Sigma= \CC/(\ZZ+\ZZ \tau)$  with modulus  $\tau \in \cH$  by a complex coordinate $z=\alpha+\beta \tau$ where $\alpha, \beta \in \RR/\ZZ$. We choose canonical  homology cycles $\mA_1$ and $\mB_1$ respectively along the identifications $z\approx z+1$ and $z\approx z+ \tau$, and normalize the holomorphic Abelian differential  by $\om_1 = dz$. The volume form $\kappa$ of unit area, and the  Dirac $\delta$-function of unit weight are as follows,\footnote{Throughout we shall use the {\sl coordinate Dirac $\delta$-function}   normalized to $\int _\Sigma  {i \over 2} dz \wedge d\bar z \, \delta ^{(2)} (z)=1$. When no confusion is expected to arise, we will not exhibit the dependence on the modulus $\tau$ in differential forms such as $ \om_1= \om_1(z|\tau) $ and $\kappa = \kappa (z |\tau)$, but we will exhibit the dependence for scalar functions such as $g(z|\tau)$.}
\bea
\label{mu1}
\kappa  = { i \over 2 \tau_2} dz \wedge d \bar z = d\alpha \wedge d\beta
\hskip 1in \delta ^{(2)} (z) = { 1 \over \tau_2} \delta (\alpha ) \delta (\beta)
\eea
The scalar Green function $G(x,y|\tau)= G(y,x|\tau) $ for the torus of modulus $\tau$ is symmetric and  formally the inverse of the scalar Laplace operator. Assuming translation invariance on the torus, it reduces to a function $g$ of the difference $x-y$, namely  $G(x,y|\tau)= g(x-y|\tau)$, where $g$ is uniquely defined by,
\bea
\pbz \p_z  g(z|\tau) =  - \pi  \delta ^{(2)} (z) + {\pi  \over \tau_2} 
\hskip 1in 
\int _\Sigma \kappa (z) g(z|\tau)=0
\eea
The Green function $g$ may be expressed as a double sum,
\bea
\label{green1}
g(z|\tau) = \sum _{(m,n) \not= (0,0)} { \tau_2\over \pi |m + n \tau |^2} \, e^{2 \pi i (m\beta -n \alpha )}
\eea
or in terms of the Jacobi elliptic function $\tet_1$ and the Dedekind function $\eta$, 
\bea
\label{green2}
g(z|\tau) = - \ln \left | { \tet _1 (z |\tau) \over \eta (\tau) } \right |^2 + { 2 \pi \over \tau_2} \left ( \Im z \right )^2
\eea
From either one of these representations, it is manifest that $g(z|\tau)$ is doubly periodic in $z$ with periods $\ZZ+ \ZZ\tau$ and
modular,   in the sense that
\bea
\label{modg}
g\left ( { z \over c \tau + d} \, \Bigg | \, { a \tau + b \over c \tau + d} \right ) = g (z|\tau)
\eea
for $a,b,c,d \in \ZZ$ with $ad-bc=1$ parametrizing an arbitrary $SL(2,\ZZ)$ transformation.

\sm

The genus-one four-graviton scattering amplitude corresponds to the special case $N=4$ of the following  family of integrals \cite{Green:1981yb}, 
\bea
\label{1-loop}
\cB^{(1)} (s_{ij}|\tau) = \prod _{i=1}^N \int _\Sigma  \kappa (z_i) \, 
\exp \left \{ \sum _{1 \leq i < j \leq N} s_{ij} \, G(z_i,z_j|\tau) \right \}
\eea
Here, the superscript on $\cB^{(1)}$ stands for genus one, $s_{ij}$ are complex parameters, and $G(z_i,z_j|\tau)$ is a scalar Green function on the torus, both of which will be characterized more precisely  below. Using the short-distance behavior of the Green function $G(z_i,z_j|\tau) \approx - \ln |z_i-z_j|^2$ one  verifies that the integral representation used to define $\cB^{(1)} (s_{ij}|\tau) $ is absolutely convergent provided $\Re (s_{ij} ) < 1$ for all $i,j $. Therefore, $\cB^{(1)} (s_{ij}|\tau) $ admits a  Taylor series expansion in the variables $s_{ij}$ at 0 which has unit radius  of convergence in each variable $s_{ij}$. Throughout this paper,  we shall be interested in considering the Taylor expansion and therefore always assume the stronger condition $|s_{ij} |< 1$. 

\sm

The genus-one contribution to the physical four-graviton string amplitude is obtained by setting $N=4$ in (\ref{1-loop}), identifying  the parameters $s_{ij}$ with   the physical kinematic invariants constructed from the string scale $\alpha '$ and the momenta  $k_i$ of  massless gravitons by $s_{ij} = - \alpha ' (k_i+k_j)^2/4$, integrating $\cB^{(1)} (s_{ij}|\tau)$ in $\tau$ over the  moduli space of genus-one Riemann surfaces, and multiplying by an overall kinematic factor. This integral is absolutely convergent only when all $s_{ij}$ are purely imaginary, but may be analytically continued throughout the $s_{ij}$ complex plane in each variable, as was shown in detail in  \cite{DHoker:1994gnm}. In addition to the poles in $s_{ij}$ at positive integers which already occurred in $\cB^{(1)}(s_{ij}|\tau)$ for each value of $\tau$, the integration over $\tau$ and its analytic continuation in $s_{ij}$ produce further singularities in $s_{ij}$, including branch cuts which reach all the way down to $s_{ij}=0$. In the present paper, we shall only consider the behavior of the Taylor series of $\cB^{(1)}(s_{ij}|\tau)$ near $s_{ij}=0$, which is not sensitive to those extra singularities. These considerations have been discussed here for genus one but apply to higher genus string amplitudes as well, though the analytic continuations obtained in  \cite{DHoker:1994gnm} for genus one have not been carried out to higher genus.

\sm

Mathematically, one may relax the condition $N=4$, thus jettisoning the direct connection to physical string amplitudes, but retain the constraint of momentum conservation $\sum _{i=1}^N  k_i=0$, so that the variables $s_{ij}$  satisfy $\sum _{i=1}^N s_{ij}=0$ for all $j=1,\cdots, N$. When these conditions are met, the integrals in the definition of $\cB^{(1)} (s_{ij}|\tau)$ are invariant under shifting $G(z_i,z_j|\tau) \to G(z_i,z_j|\tau) + \gamma (z_i|\tau) + \gamma (z_j|\tau)$ for an arbitrary function $\gamma$. This symmetry is a reflection of the underlying conformal invariance of the correlation function which define $\cB^{(1)} (s_{ij}|\tau)$ in string theory. We may use such a transformation to make a particularly convenient choice for the Green function $G(z_i,z_j|\tau) = g(z_i-z_j|\tau)$. The integrals giving the function $\cB^{(1)} (s_{ij}|\tau)$ are well-defined for arbitrary values of $N$ and independent complex parameters $s_{ij}$, subject only to the conditions $|s_{ij}| <1$ for absolute convergence of the Taylor expansion in powers of $s_{ij}$.  Conformal invariance will be guaranteed as long as we use a Green function $G(z_i,z_j|\tau)$  whose integral on the torus vanishes, so that $G(z_i,z_j|\tau) = g(z_i-z_j|\tau)$. Under these assumptions, the function $\cB^{(1)} (s_{ij}|\tau)$ is a generating function for all modular graph functions \cite{DHoker:2015wxz,D'Hoker:2015foa}. More precisely, the coefficient of the  term $\prod _{i<j} s_{ij} ^{n_{ij}}$ in the Taylor expansion of $\cB^{(1)} (s_{ij}|\tau)$ will be given by a modular graph function associated with a graph with $n_{ij}$ edges between the points $z_i$ and $z_j$.

\subsection{Higher genus Riemann surfaces}

Throughout we shall consider  compact Riemann surfaces without boundary. The topology of such a surface  $\Sigma$ is completely specified by its genus $h$. The homology group $H_1(\Sigma , \ZZ)$ of one-cycles on a surface $\Sigma$ of genus $h$ is isomorphic to $\ZZ^{2h}$ and supports an integer-valued, anti-symmetric, non-degenerate intersection pairing which we denote by $\mJ(\cdot, \cdot)$.   A canonical basis of $\mA_I$ and $\mB_I$ one-cycles may be chosen in $H_1(\Sigma , \ZZ)$ for which $\mJ$ takes the form of the standard symplectic matrix, 
\begin{align}
\label{AB}
\mJ(\mA_I, \mA_J) & = \mJ(\mB_I, \mB_J)  =  0
\no \\
\mJ(\mA_I, \mB_J) & = - \mJ(\mB_J, \mA_I)  =  \delta _{IJ}
\end{align}
for $I,J =1, \cdots, h$. For $h=2$, this choice is illustrated in figure 1. A canonical basis of holomorphic Abelian differentials $\om_I$ for $I=1,\cdots, h$ for the Dolbeault cohomology group $H^{(1,0)} (\Sigma)$ may be normalized  on $\mA$-cycles, and we have, 
\bea
\label{omnorm}
\oint _{\mA_I} \om_J = \delta _{IJ} 
\hskip 1in 
\oint _{\mB_I} \om_J = \Omega _{IJ} 
\eea
The complex variables $\Omega _{IJ}$ denote the components of the period matrix $\Omega$ of the surface $\Sigma$.  By the Riemann relations, $\Omega$ is symmetric, and has positive definite imaginary part,
\bea
\Omega ^t = \Omega 
\hskip 1in 
Y = \Im \Omega >0
\eea
as a result of the following pairing relation,
\bea
\int_\Sigma \omega_I \wedge \overline{\omega_J} = -2i\, Y_{IJ}
\eea
The choice of canonical $\mA$ and $\mB$-cycles is not unique: a new choice of canonical  $\mA'$ and $ \mB'$-cycles is obtained by applying an arbitrary modular transformation $M \in Sp(2h,\ZZ)$,  such that $M$ satisfies $M^t \mJ M= \mJ$, and the cycles are transformed as follows,
\bea
\left ( \begin{matrix} \mB'_I \cr \mA_I' \cr \end{matrix} \right )= 
\sum _J \left ( \begin{matrix} A_{IJ} & B_{IJ} \cr C_{IJ} & D_{IJ} \cr \end{matrix} \right )  
\left ( \begin{matrix} \mB_J \cr \mA_J \cr \end{matrix} \right )
\hskip 1in M = \left ( \begin{matrix} A & B \cr C & D \cr \end{matrix} \right ) 
\eea
Under a modular transformation $M$, the matrix of holomorphic Abelian differentials $\omega$,  the  period matrix $\Omega$, and its imaginary part $Y$, transform as follows, 
\bea
\om ' & = & \om \, (C \Omega +D) ^{-1} 
\no \\
\Omega '  & = & (A \Omega +B) (C\Omega +D)^{-1}
\no \\  
Y' & = & \left ((C\Omega + D)^\dagger \right )^{-1} \, Y \, (C\Omega +D)^{-1}
\eea
where we have grouped the components $\omega _I$ into a row matrix  $\omega$, and denoted transformed quantities with a prime superscript, which we shall do throughout.  The moduli space of complex structures of compact Riemann surfaces of  genus $h$  will be denoted by $\cM_h$.

\sm

\begin{figure}
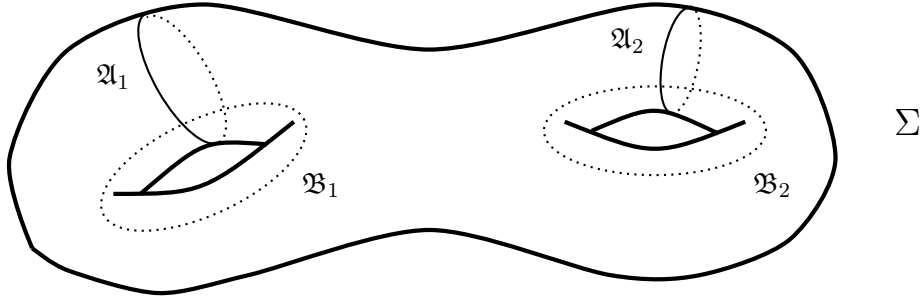

\begin{center}
\tikzpicture[scale=1.2]
\scope[xshift=-5cm,yshift=0cm]
\draw[ultra thick] plot [smooth] coordinates {(0.6,0.6) (0.35,1.6) (1,2.8) (2.5, 3.3)  (5,2.8) (7.5, 3.3) (9,2.8) (9.5,1.6) (9,0.7) (8,0.3) (7, 0.3) (5,0.8) (3, 0.3) (2,0.1) (1,0.35) (0.6,0.6) };
\draw[ultra thick] (1.5,1.2) .. controls (2.5, 1.2) .. (3.5,2);
\draw[ultra thick] (1.8,1.2) .. controls (2.5, 1.8) .. (3.2,1.75);
\draw[ultra thick] (6.5,2) .. controls (7.5, 1.6) .. (8.5,2);
\draw[ultra thick] (6.8,1.9) .. controls (7.5, 2.2) .. (8.2,1.87);
\draw[thick, dotted,  rotate=25] (2.9,0.3)  ellipse (35pt and 15pt);
\draw[thick, dotted,  rotate=00] (7.5,1.9)  ellipse (35pt and 14pt);
\draw[thick, dotted,  rotate=30] (3.2,0.2)  arc (-90:90:0.3 and 0.8);
\draw[thick,  rotate=30] (3.2,1.84)  arc (90:270:0.3 and 0.82);
\draw[thick, dotted,  rotate=-10] (7.2,3.4)  arc (-90:90:0.2 and 0.6);
\draw[thick,  rotate=-10] (7.2,4.58)  arc (90:270:0.2 and 0.59);
\draw (1.5,2.5) node{\small $\mA_1$};
\draw (7.2,2.9) node{\small $\mA_2$};
\draw (3.8,1.3) node{\small $\mB_1$};
\draw (8.8,1.3) node{\small $\mB_2$};
\draw (10.3,2) node{{\large $\Sigma$}};
\endscope
\endtikzpicture
\caption{Canonical homology cycles on a compact genus-two Riemann surface $\Sigma$.}
\end{center}
\label{fig:1}
\end{figure}

The rank $h$ Siegel upper half space $\cH_h$ may be defined as the space of all $h\times h$ complex-valued symmetric matrices whose imaginary part is positive definite. Alternatively, a more geometrical definition is as the coset space of the non-compact group $Sp(2h,\RR)$ by its maximal compact subgroup $SU(h) \times U(1)$,
\bea
\cH_h = Sp(2h,\RR)/ \left ( SU(h) \times U(1) \right )
\eea
The presence of the $U(1)$ factor  implies that $\cH_h$ is a K\"ahler manifold and its $Sp(2h,\RR)$-invariant K\"ahler metric is given as follows,
\bea
ds^2 = \sum _{I,J,K,L} (Y^{-1})^{IK} (Y^{-1})^{JL} d\Omega _{IJ} d\bar \Omega _{KL}
\eea
where $Y^{-1}$ is the inverse of the matrix $Y$. 

\sm

The moduli space $\cM_h$ for $h=1,2,3$  may be identified with $\cH_h/Sp(2h,\ZZ)$ provided we remove from $\cH_h$  for $h=2,3$ all elements which correspond to disconnected surfaces, and take into account the effect of automorphisms including the involution on the hyper-elliptic locus for $h=3$. For $h\geq 4$, the moduli space $\cM_h$ is a complex co-dimension $\half (h-2)(h-3)$ subspace of $\cH_h/Sp(2h,\ZZ)$ known as the Schottky locus.

\subsection{The canonical K\"ahler form}

Our goal is to generalize the generating function for genus-one modular graph functions given in (\ref{1-loop}) to higher genus. For the torus, the basic ingredients in (\ref{1-loop}) are the volume form $ \kappa (z)$ and the Green function $G(z,w|\tau)=g(z-w|\tau)$, which we shall generalize  to arbitrary genus in this subsection. In the next subsection, we shall introduce the simplest non-trivial class of higher genus modular graph functions, while in the subsequent subsections  we shall extend the construction  to yet more general classes.

\sm

On the torus, the volume form $\kappa $ and the Green function $g(z-w|\tau)$ were specified uniquely by their translation invariance, plus the requirements of unit area of $\kappa$ and vanishing integral of $g$ against $\kappa$. On a  higher genus compact Riemann surface $\Sigma$ there are no continuous symmetries, and thus no translation invariance. However, we may map $\Sigma$ holomorphically into its Jacobian variety $J(\Sigma) = \CC^h /(\ZZ^h + \Omega \ZZ^h)$ via the Abel map, 
\bea
\label{zeta}
\zeta : \Sigma \to J(\Sigma) \hskip 1in \zeta _I (z)= \int _{z_0} ^z \omega _I - \Delta _I(z_0) 
\eea
where the index $I$ runs over $I=1,\cdots, h$ and   $\Delta _I(z_0)$ is the Riemann vector,
\bea
\Delta _I (z_0) = \half - \half \Omega _{II} + \sum _{J \not= I} \oint _{A_J} \omega_J (z) \int ^z _{z_0} \omega_I\,
\eea
which is such that $\zeta_I(z)$ is independent  of $z_0$. The map $\zeta$ is well-defined since upon moving the point $z$ around $\mA$ and $\mB$-cycles shifts $\zeta$ by elements in the lattice $\ZZ^h + \Omega \ZZ^h$. We note that for genus one the Abel map is bijective, while for higher genus it can be used to construct a bijection between $h$ copies of $\Sigma$ and $J(\Sigma)$. 

\sm

The Jacobian variety $J(\Sigma)$ is a flat K\"ahler manifold, and we may use translation invariance on $J(\Sigma)$ and modular invariance  to specify its canonical K\"ahler form $K $ by,
\bea
K = {i \over 2} \sum _{I,J} (Y^{-1})^{IJ} d\zeta_I \wedge d\bar \zeta_J
\eea
The form $ \kap$ on the Riemann surface $\Sigma$ is defined as the pull-back under the Abel map  of 
the canonical K\"ahler form $K$ on $J(\Sigma)$, 
\bea
\label{muh}
\kap   = {1 \over h} \, \zeta_\star (K) = { i \over 2h} 
\sum _{I,J} (Y^{-1}) ^{IJ} \omega_I  \wedge \overline{ \omega_J}
\eea
up to an overall factor of $1/h$, which has been included  in order to normalize its integral to unit area, $\int _\Sigma \kap =1$. The  K\"ahler form $\kap$ is referred to as the canonical K\"ahler form and its associated metric as the Arakelov metric. As they are constructed solely out of conformal invariant Abelian differentials, the Arakelov metric and K\"ahler form are conformal invariant as well. Clearly, when considered for genus one, the two-form $\kap$ defined in (\ref{muh})  coincides with the two-form $\kappa$ of (\ref{mu1}), whence our use of the same letter to designate it.

\subsection{The Arakelov Green function}

On a Riemann surface $\Sigma$ with period matrix $\Omega$, the Arakelov Green function $\GA(x,y|\Omega)$ is a real-valued symmetric function  on $\Sigma \times \Sigma$ which provides an inverse to the scalar Laplace operator on $\Sigma$ equipped with the Arakelov metric. In terms of local complex coordinates $(z,\bar z)$, we have $ \kap = {i \over 2}  \kap _{z\bar z} \, dz \wedge d\bar z$ and, 
\bea
\label{AraG}
 \pbz  \, \p_z \, \GA(z,y|\Omega )  & = &  - \pi \, \delta^{(2)} (z,y) +  \pi \, \kap _{z \bar z}  (z)  
\no \\
 \int _\Sigma \kap (z) \, \GA(z,y |\Omega) & = & 0
\eea
where  $\delta ^{(2)}(z,y)$ is the coordinate Dirac $\delta$-function normalized by ${ i \over 2} \int _\Sigma  dz \wedge d \bar z \, \delta ^{(2)} (z,y)=1$. An explicit expression for $\GA$ may be obtained by relating it to another Green function $G$ which is often used in string theory \cite{D'Hoker:1988ta}, and defined by, 
\bea
\label{G}
G(x,y|\Omega) = - \ln |E(x,y|\Omega)|^2 + 2 \pi \, \Im \left ( \int _y ^x \omega _I \right ) (Y^{-1})^{IJ} \, \Im \left ( \int _y ^x \omega _J \right )
\eea
The prime form $E(x,y|\Omega )$ is a differential form of weight $(-\half,0)_x \otimes (-\half, 0)_y$ defined on the covering space of $\Sigma \times \Sigma$ by the following formula \cite{fay73},
\bea
\label{E}
E(x,y|\Omega ) = { \theta [\nu] \left ( \int ^x _y \om |\Omega \right ) \over \eta _\nu (x|\Omega) \, \eta _\nu(y|\Omega)} 
\eea
Here $\nu$ is an odd half-characteristic, $\eta _\nu$ is the  holomorphic differential of weight $(\half, 0)$ whose square is the holomorphic Abelian differential given by,
\bea
\eta _\nu (x|\Omega)^2 = \sum _I \omega _I(x) \, \p_I \tet [\nu] (0|\Omega)
\eea 
The Riemann $\tet$-function for $\Omega \in \cH_h$ and  $v \in J(\Sigma)$ with characteristics $\nu = \nu '' + \Omega \nu'$ for  $\nu', \nu '' \in \RR^h$  is given by,
\bea
\label{tet}
\tet [\nu] (v|\Omega) = \sum _{n \in \ZZ^h} \exp \left \{ i \pi (n + \nu') ^t \Omega (n+\nu') + 2 \pi i (n+\nu')^t (v+\nu'') \right \}
\eea 
Though the individual factors in  the prime form depend on $\nu$, the full $E(x,y|\Omega)$ is independent of the choice of $\nu$ as long as it is an odd half characteristic. In view of the fact that the prime form, in the coordinates $x,y$ used above, behaves as $E(x,y|\Omega) \approx x-y + \cO((x-y)^3)$, one readily establishes that $G$ has the following Laplacian, 
\bea
\label{DeltaG}
 \pbz  \, \p_z \, G(z,y|\Omega)  =  - \pi \, \delta^{(2)} (z,y) +  \pi \, h \,  \kap _{z \bar z}  (z|\Omega )  
\eea
In view of the extra factor of $h$ in the second term on the right-hand side of (\ref{DeltaG}) compared to (\ref{AraG}), we see that $\GA$ and $G$ agree  (up to an additive constant) only for genus one. For higher genus, $G(x,y|\Omega )$, just as $E(x,y|\Omega)$ is defined only on the universal cover of $\Sigma \times \Sigma$, but does not transform as a proper scalar on $\Sigma \times \Sigma$. Nonetheless, $G$ may be used to provide an explicit formula for $\GA$, as the two are related as follows \cite{D'Hoker:1988ta}, 
\bea
\label{GAG}
\GA(x,y|\Omega) = G(x,y|\Omega) - \gamma (x|\Omega) - \gamma (y|\Omega) + \gamma ^1(\Omega)
\eea
where $\gamma (x|\Omega)$ and $\gamma^1(\Omega)$ are obtained by substituting (\ref{GAG}) into (\ref{AraG}) and using (\ref{DeltaG}), 
\bea
\label{gamma}
\gamma (x|\Omega) =  \int _\Sigma \kap (z) G(x,z|\Omega) 
\hskip 1in 
\gamma^1(\Omega)  = \int _\Sigma \kap (z) \gamma (z|\Omega)
\eea
In view of the fact that $G(x,y|\Omega)$ is defined only on the covering space of $\Sigma \times \Sigma$, it is understood in the definition of $G(x,y|\Omega)$ that a particular choice has been made for the fundamental domain for $\Sigma$ on which the expression for $G(x,y|\Omega)$ holds, and that the integration domains in (\ref{gamma}) are carried out over that same fundamental domain.

\subsection{A simple class of higher genus  modular graph functions}

Having naturally generalized the genus-one volume form $\kappa$ and Green function $g(z-w|\tau)$ to the Arakelov volume form $\kap$ and the Arakelov Green function $\GA$ on a surface $\Sigma$ of arbitrary genus $h \geq 2$, we may now define a generating function for a simple class of modular  graph functions on a Riemann surface $\Sigma$ of arbitrary genus $h$ with period matrix $\Omega$, 
\bea
\label{h-loop}
\mod  (s_{ij}|\Omega) = \prod _{i=1}^N \int _\Sigma  \kap (z_i) \, 
\exp \left \{ \sum _{1 \leq i < j \leq N} s_{ij} \, \GA(z_i, z_j|\Omega) \right \}
\eea
Given the behavior under modular transformations of $\kap$ and $\GA$, it is clear that $\mod  (s_{ij}|\Omega) $ is invariant under modular transformations by $Sp(2h,\ZZ)$ of its argument $\Omega$, and thus intrinsically depends on the complex structure of the surface $\Sigma$. More general classes of higher genus modular graph functions will be introduced more formally in sections \ref{sec:2.8} and \ref{sec:2.9} below.

\sm

The analytic properties of $\mod  (s_{ij}|\Omega) $ in the variables $s_{ij}$ are identical to those of its genus-one counterpart, and its Taylor series at $s_{ij}=0$ is absolutely convergent when $|s_{ij} | <1$ for all $i,j=1,\cdots, N$.
As in the graphical expansion of the genus-one case, a modular graph function  vanishes when it is associated with a graph containing a vertex on which a single edge ends. 

\sm

However, there are key differences from the genus-one case. First, due to the absence of translation invariance on $\Sigma$ for $h \geq 2$, the Green function $\GA(z,w|\Omega)$ depends on both points $z,w$ separately and, unlike the genus-one Green function $g(z-w|\tau)$, does not admit a simple Fourier series expansion.  As a result, the graphical expansion of the generating function $\mod (s_{ij}|\Omega)$ in powers of $s_{ij}$ differs in structure from that of the expansion of the genus-one generating  function $\cB^{(1)} (s_{ij}|\tau)$.  For genus one any connected graph which becomes disconnected upon deleting a single edge vanishes,  and any connected genus-one graph which becomes disconnected by removing a single vertex factorizes into a product of subgraph components. Neither of these properties holds in general for genus greater than one. Furthermore, as a result of the lack of translation invariance, the modular graph functions generated by  $\mod  (s_{ij}|\Omega) $ for $h \geq 2$ have no representation in terms of  Kronecker-Eisenstein sums, or any known generalization thereof.

\sm

Finally, we stress that for  $h=2$ the function $\mod  (s_{ij}|\Omega) $ differs from the genus-two superstring  integrand  which was evaluated in \cite{D'Hoker:2002gw,D'Hoker:2005jc,D'Hoker:2005ht}, and which will be presented in the next subsection.

\subsection{The genus-two four-graviton amplitude}

At higher genus, the only superstring amplitudes which have been computed explicitly so far are four-graviton amplitudes and their fermionic counterparts at genus two.   The integrand of the four-graviton  amplitude, integrated over vertex operator insertion points but unintegrated over the remaining bosonic moduli, is given by
\bea
\label{B2g}
\cB^{(2)} (s_{ij} |\Omega) 
=  { 1 \over 16} \int _{\Sigma ^4} { \cY \wedge \bar \cY \over (\det Y)^2}
\exp \left \{ \sum _{1 \leq i < j \leq N} s_{ij} \, \GA(z_i, z_j|\Omega) \right \}
\eea
As in the case of genus-one, the physical genus-two amplitude is obtained after multiplication by the well-known kinematic factor \cite{Green:1981yb}, and integration over $\Omega$. The superscript on $\cB^{(2)}$ stands for the genus, and the holomorphic form $\cY$ on $\Sigma ^4$ is given as follows,
\bea
\label{BY}
3 \cY & =  &
(t-u) \Delta (z_1,z_2) \wedge \Delta(z_3,z_4) 
\no \\ &&
+  (s-t) \Delta (z_1,z_3) \wedge \Delta (z_4,z_2)
\no \\ &&
+ (u-s) \Delta (z_1,z_4) \wedge \Delta (z_2,z_3)
\eea
where $\Delta(x,y) $ is given by, 
\bea
\label{Delta}
\Delta (x,y) = \om_1(x) \wedge \om_2(y) - \om _2(x) \wedge \om_1(y) = \ep ^{IJ} \om_I(x) \wedge \om _J(y)
\eea
and where $\ep^{IJ}$ is defined by $\ep^{JI}=-\ep^{IJ}$ and  $\ep ^{12}=1$.
The kinematics of four massless gravitons allows us to parametrize the six variables $s_{ij}$ in terms of just three,
\bea
\label{stu}
s= s_{12}=s_{34} \hskip 0.7in 
t = s_{14} = s_{23} \hskip 0.7in
u=s_{13} = s_{24} 
\eea
which in turn satisfy the kinematic relation $s+t+u=0$.  

\sm 

For given $\Omega$, the integral over $\Sigma ^4$ is absolutely convergent whenever $\Re(s_{ij} )<1$ for all $i,j=1,2,3,4$. Furthermore,  $\cB ^{(2)}  (s_{ij} |\Omega)$ is invariant under $Sp(4,\ZZ)$ modular transformations, and is thus intrinsically defined on the Riemann surface $\Sigma$. As a result, $\cB ^{(2)}  (s_{ij} |\Omega)$  admits a Taylor series expansion in powers of $s_{ij}$ with unit radius of convergence or, equivalently, in powers of the Green function~$\GA$. The symmetry under permutations of the points $z_i$ of the integral guarantees that the expansion in powers of $s,t,u$ may be arranged in terms of symmetric polynomials of these variables,
\bea
0 & = & s+ t+u
\no \\
\sigma _2 & = & s^2 + t^2 +u^2
\no \\
\sigma _3 & = & s^3 + t^3+u^3=3 stu
\eea
The resulting expansion is as follows,
\bea
\label{Bexp}
\cB^{(2)}  (s_{ij} |\Omega) = \sum_{p,q=0}^\infty  \cB_{(p,q)}  (\Omega) \, { \sigma _2^p \, \sigma _3^q \over p! \, q!}
\eea
The coefficients $\cB_{(p,q)} (\Omega)$ are referred to as {\sl genus-two string invariants} since they are invariant under $Sp(4,\ZZ)$ and thus intrinsically depend only on the surface $\Sigma$ and not on the period matrix chosen to represent $\Sigma$.  They  are real-analytic functions of $\Omega$ away from the degeneration divisors.   The weight $w$ of the invariant is defined as the expansion order in powers of the Green functions $\GA$ in the corresponding graph. Equivalently, in view of the fact that the measure factor $\cY \wedge \bar \cY$ is bilinear in $s,t,u$, the weight is also equal to the expansion order in powers of $s,t,u$ minus two, and is related to the powers $p$ and $q$ by,
\bea
w = 2p + 3q -2
\eea
Manifestly, we have $\cB_{(0,0)}  (\Omega)=0$ so that $w\geq 0$.

\sm

The coefficients in the expansion of the genus-two string amplitude $\cB^{(2)}  (s_{ij} |\Omega) $  do not belong to the simple class of modular graph functions  generated by $\cF^{(2)}  (s_{ij} |\Omega)$   presented in (\ref{h-loop}) of  the preceding subsection. In the next subsection, we shall extend the simple class to a more general class to which $\cB^{(2)}  (s_{ij} |\Omega) $ does belong, and which may be defined for arbitrary genus. In preparation for this generalization, we shall express the integration measure over $\Sigma ^4$ in $\cB^{(2)}  (s_{ij} |\Omega)$ in an alternative and illuminating way. We begin by recasting the holomorphic form $\cY$ in a tensorial form,
\bea
\cY & = & \cY^{I_1\, I_2 \, I_3 \, I_4} 
\, \omega _{I_1} (z_1) \wedge   \omega _{I_2} (z_2) \wedge 
\omega _{I_3} (z_3) \wedge \omega _{I_4} (z_4)
\no \\
\bar \cY & = & \cY^{J_1 \, J_2 \, J_3 \, J_4} 
\, \overline{ \omega} _{J_1} (z_1) \wedge   \overline{ \omega} _{J_2} (z_2) 
\wedge \overline{ \omega} _{J_3} (z_3) \wedge \overline{ \omega} _{J_4} (z_4)
\eea
where the tensor $\cY^{I_1\, I_2\, I_3\, I_4}$ is common to both $\cY$ and $\bar \cY$ (we do not complex conjugate the variables $s, t, u$) and is given by,
\bea
3 \,\cY ^{I_1\, I_2\, I_3\, I_4} = 
(t-u) \, \ep^{I_1\,I_2} \, \ep^{I_3\,I_4}
+  (s-t) \, \ep^{I_1\,I_3} \, \ep^{I_4\,I_2}
+ (u-s) \, \ep^{I_1\,I_4} \, \ep^{I_2\,I_3}
\eea
With the help of this tensor,  the measure factor takes the following form,
\bea
\label{YY}
{\cY \wedge \bar \cY \over (\det Y)^2} = {  \cY ^{I_1\, I_2\, I_3\, I_4} \, \cY ^{J_1\, J_2\, J_3\, J_4} \over (\det Y)^2} \,\prod _{i=1}^4 \omega _{I_i} (z_i)\wedge \overline{\omega} _{J_i}(z_i)
\eea
Next, we use the following identity,
\bea
{ \ep ^{I\,I'} \ep ^{J\, J'} \over \det Y} = (Y^{-1})^{IJ} (Y^{-1})^{I'J'} - (Y^{-1})^{IJ'} (Y^{-1})^{I'J}  
\eea
to express the prefactor on the right side of (\ref{YY}) as a manifestly modular invariant quadri-linear combination in $Y^{-1}$ with coefficients bilinear in $s,t,u$. The measure factor  of (\ref{h-loop}) corresponds to the term amongst them of the form $(Y^{-1}) ^{I_1J_1} (Y^{-1}) ^{I_2J_2} (Y^{-1}) ^{I_3J_3} (Y^{-1}) ^{I_4J_4} $.

\subsection{Genus-two higher string invariants}

In this section, we shall review the lowest weight string invariants, as derived in  \cite{D'Hoker:2013eea}. 
The resulting expressions may be simplified whenever a vertex occurs on which no Green function ends, since then standard formulas for integrating paired holomorphic and anti-holomorphic one-forms may be used. The key formulas are as follows,\footnote{Henceforth, when no confusion is expected to arise, we shall  omit the explicit  dependence on $\Omega$ and the wedge product symbol both of which will  be understood throughout. We shall also use the Einstein convention by which a repeated upper and lower index is contracted.}  
\bea
\label{int1}
\int _\Sigma \om_I \, \oom_J  & = & -2i \, Y_{IJ}
\no \\
\int _{\Sigma _u} \Delta (x,u) \, \overline{\Delta (u,y)} & = & 
2 i \, (\det Y) \, \delta _J{} ^I \, \om _I (x) \, \oom^J(y)
\no \\
\int _{\Sigma _u} \int _{\Sigma _v}  \Delta (x,u) \, \overline{\Delta (u,v)} \, \Delta (v,y) & = &
- 4 \, (\det Y) \, \Delta (x,y)
\eea
where the notation  $\Sigma _u$ indicates that the integrand is to be integrated in the variable $u$. 
Here and throughout, we use the notation, 
\bea
\oom ^J = (Y^{-1})^{JI}  \oom_I
\eea
We note that for $x=y$ the right-hand side of the second line in (\ref{int1}) reduces to $8 \, (\det Y) \, \kap (x)$, where $\kap $ is the canonical volume form defined in (\ref{muh}), and considered here for $h=2$.

\begin{figure}
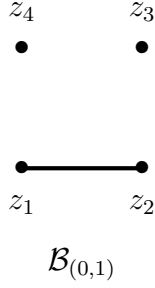

\begin{center}
\tikzpicture[scale=1.6]
\scope[xshift=-5cm,yshift=0cm]
\draw (-3,0) node{$\bullet$};
\draw (-3,1) node{$\bullet$};
\draw (-2,0) node{$\bullet$};
\draw (-2,1) node{$\bullet$};
\draw (-3,-0.3) node{$z_1$};
\draw (-2,-0.3) node{$z_2$};
\draw (-3,1.3) node{$z_4$};
\draw (-2,1.3) node{$z_3$};
\draw (-2.5, -0.8) node{$\cB_{(0,1)} $};
\draw[ultra thick] (-3,0) -- (-2,0);
\endscope
\endtikzpicture
\caption{The contributions to  $\cB_{(0,1)}$ arise from the above graph with coefficient $s_{12}$ plus the graphs obtained by all permuting  the points $(1,2) $ to $(i,j)$ multiplied by the coefficient $s_{ij}$.}
\end{center}
\label{fig:2}
\end{figure}

For the case of weight $w=0$, corresponding to $p=1, q=0$, there are no Green functions in the integrand, and all four points $z_i$ may be integrated out explicitly. For weight $w=1$, we have $p=0, q=1$, there is a single Green function, and two points may be integrated out, as illustrated by the graph in figure~2. Collecting these results, we have,
\bea
B_{(1,0)}  & = & 2
\no \\
B_{(0,1)}  & = & 64 \, \f
\eea
where $\f$ was shown in \cite{D'Hoker:2013eea} to be the Kawazumi-Zhang invariant \cite{Kawazumi,Zhang} for genus two,
\bea
\label{KZ}
\f  = - { 1 \over 4} \delta _{J_2}{}^{I_1}   \delta _{J_1}{}^{I_2} 
\int _{\Sigma ^2} \om _{I_1}  (z_1) \, \oom^{J_1}(z_1) \, \om _{I_2} (z_2) \, \oom^{J_2} (z_2) \, \GA(z_1,z_2) 
\eea
Note that $\f$ nicely illustrates the general structure provided in (\ref{YY}).

\sm

\begin{figure}[h]
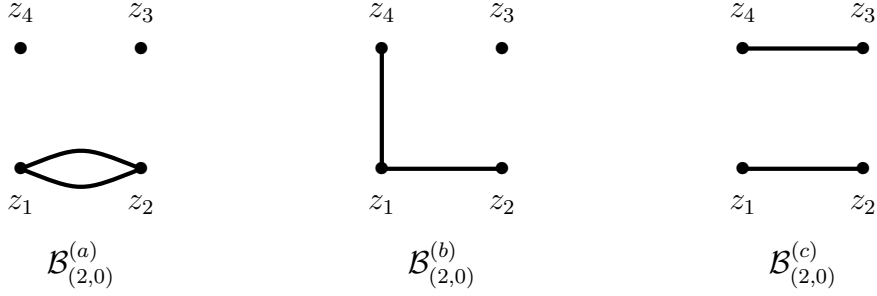

\begin{center}
\tikzpicture[scale=1.6]
\scope[xshift=-5cm,yshift=0cm]
\draw (0,0) node{$\bullet$};
\draw (0,1) node{$\bullet$};
\draw (1,0) node{$\bullet$};
\draw (1,1) node{$\bullet$};
\draw (0,-0.3) node{$z_1$};
\draw (1,-0.3) node{$z_2$};
\draw (0,1.3) node{$z_4$};
\draw (1,1.3) node{$z_3$};
\draw (0.5, -0.8) node{$\cB_{(2,0)} ^{(a)}$};
\draw[ultra thick] (0,0) .. controls (0.5, -0.2) .. (1,0);
\draw[ultra thick] (0,0) .. controls (0.5, 0.2) .. (1,0);
\draw (3,0) node{$\bullet$};
\draw (3,1) node{$\bullet$};
\draw (4,0) node{$\bullet$};
\draw (4,1) node{$\bullet$};
\draw (3,-0.3) node{$z_1$};
\draw (4,-0.3) node{$z_2$};
\draw (3,1.3) node{$z_4$};
\draw (4,1.3) node{$z_3$};
\draw (3.5, -0.8) node{$\cB_{(2,0)} ^{(b)}$};
\draw[ultra thick] (3,0) -- (4,0);
\draw[ultra thick] (3,0) --  (3,1);
\draw (6,0) node{$\bullet$};
\draw (6,1) node{$\bullet$};
\draw (7,0) node{$\bullet$};
\draw (7,1) node{$\bullet$};
\draw (6,-0.3) node{$z_1$};
\draw (7,-0.3) node{$z_2$};
\draw (6,1.3) node{$z_4$};
\draw (7,1.3) node{$z_3$};
\draw (6.5, -0.8) node{$\cB_{(2,0)} ^{(c)}$};
\draw[ultra thick] (6,0) -- (7,0);
\draw[ultra thick] (6,1) --  (7,1);
\endscope
\endtikzpicture
\caption{
The contributions to  $\cB_{(2,0)}^{(a)}$, $\cB_{(2,0)}^{(b)}$, $\cB_{(2,0)}^{(c)}$ arise respectively from the above graphs $(a), (b), (c)$ with coefficient $(s_{12})^2, s_{12}s_{14}$ and $s_{12}s_{34}$,  plus the graphs obtained by permuting  the points and multiplying by the corresponding coefficient bilinear in the $s_{ij}$.}
\end{center}
\label{fig:3}
\end{figure}

For higher values of the weight $w$, several contributions arise, some of which may be simplified by integrating out some of the points $z_i$. For $w=2$, we have $p=2, q=0$, so that $B_{(2,0)}(\Omega)$ may be evaluated by expanding $\cB^{(2)}  (s_{ij} |\Omega)$ to quartic order in $s,t,u$. Setting $t=-s, u=0$, and evaluating $\cY =  s \Delta (z_1,z_3) \wedge \Delta (z_4,z_2) $, we obtain,
\bea
\cB_{(2,0)} =
 {1 \over 64} \int _{\Sigma ^4} { | \Delta (z_1,z_3) \Delta (z_4,z_2)|^2 \over (\det Y)^2} \Big ( \GA(z_1,z_2)+\GA(z_3,z_4) - \GA(z_2,z_3) - \GA(z_1,z_4) \Big )^2
 \label{quart}
\eea
This contribution naturally decomposes into three modular functions corresponding to the distinct graphs exhibited in figure~3, 
\bea
\cB_{(2,0)}  = \cB_{(2,0)}^{(a)}   -2  \cB_{(2,0)}^{(b)}   + \cB_{(2,0)} ^{(c)}
\eea
where each contribution is given as follows, 
\bea
\cB_{(2,0)}^{(a)}  & = &  {1 \over 16} \int _{\Sigma ^4} { | \Delta (z_1,z_3) \Delta (z_4,z_2)|^2 \over (\det Y)^2} \, \GA(z_1,z_2)^2
\no \\
\cB_{(2,0)}^{(b)}  & = &  {1 \over 16} \int _{\Sigma ^4} { | \Delta (z_1,z_3) \Delta (z_4,z_2)|^2 \over (\det Y)^2} \, \GA (z_1,z_2) \GA(z_1,z_4) 
\no \\
\cB_{(2,0)}^{(c)}  & = &  {1 \over 16} \int _{\Sigma ^4} { | \Delta (z_1,z_3) \Delta (z_4,z_2)|^2 \over (\det Y)^2} \, \GA (z_1,z_2) \, \GA (z_3,z_4)
\label{arak}
\eea
We note that  conformal invariance allows us to replace $\cG(z_i,z_j)$ in  (\ref{quart})  by the scalar Green function $G(z_i,z_j)$. Our use of the Arakelov Green function $\GA$ guarantees that each individual contribution in (\ref{arak}) is  well-defined and conformally invariant.

\sm

The integrals over the points $z_3$ and $z_4$  in $\cB_{(2,0)}^{(a)}$ may be carried out  using (\ref{int1}), and the remaining integrals over the points $z_1$ and $z_2$ are governed by the canonical K\"ahler form,
\bea
\cB_{(2,0)}^{(a)}  =  4 \int _{\Sigma ^2} \kap(z_1) \, \kap (z_2) \, \GA(z_1,z_2)^2
\eea
Similarly, the integral over the point $z_3$   in $\cB_{(2,0)}^{(b)}$ may be carried out  using (\ref{int1}), and the remaining integral over the point $z_1$ is governed by the canonical K\"ahler form,
\bea
\cB_{(2,0)}^{(b)}   =    - \half \delta _{J_4}{}^{I_2}  \, \delta _{J_2}{}^{I_4} 
 \int _{\Sigma^3} \kap(z_1)  \, 
\om_{I_2} (z_2) \oom ^{J_2}(z_2) 
 \om_{I_4} (z_4) \oom ^{J_4}(z_4) \, \GA(z_1,z_2) \, \GA(z_1,z_4)\quad
\eea
A term proportional to $\kap(z_2) \kap (z_4)$ arises in the integrand as well, but cancels upon integration against $\GA(z_1,z_2)\GA(z_1,z_4)$ in view of the normalization of the Arakelov Green function in the second line of (\ref{AraG}).
Finally, in $\cB_{(2,0)}^{(c)}$ no points can be integrated out explicitly, but terms proportional to $\kap $ arise in the integrand for all four points, and again cancel upon integration against $\GA(z_1,z_2)\GA(z_3,z_4)$ in view of  (\ref{AraG}). As a result, we find, 
\bea
\cB_{(2,0)}^{(c)}   =   { 1 \over 16} 
\delta _{J_3}{}^{I_1}  \, \delta _{J_1}{}^{I_3}  \,  \delta _{J_4}{}^{I_2}  \delta _{J_2}{}^{I_4}  \, 
\int _{\Sigma^4} \prod _{i=1}^4 \om_{I_i} (z_i) \oom^{J_i}(z_i)
\GA(z_1,z_2) \, \GA(z_3,z_4) \quad
\eea
The pattern which emerges from the structure of $\cB_{(0,1)}, \cB_{(2,0)}^{(a)} , \cB_{(2,0)}^{(b)} $ and $\cB_{(2,0)}^{(c)} $ and their generalizations will be described in the next subsection.

\subsection{General classes of higher genus modular graph functions}
\label{sec:2.8}

The general construction of a genus-$h$ modular graph function associated with a graph with $N$ vertices labelled by $i=1,\cdots, N$ proceeds as follows.
\begin{enumerate}
\itemsep=  0 in
\item Each vertex $i$ is decorated by a pair of indices $(I_i, J_i)$   (where $I_i, J_i =1,\cdots, h$) and is associated with a factor of the volume form $\om_{I_i}(z_i) \, \oom ^{J_i}(z_i)$ in the integrand;
\item Between two distinct vertices $i \not = j$ we have $n_{ij} \geq 0$ edges producing a factor of  $\GA(z_i, z_j)^{n_{ij}}$ in the integrand;
\item No edge can begin and end on the same vertex;
\item The indices $I_i$ and $J_i$ are contracted to a modular invariant with the help of a linear combination of tensors  $\prod _{i=1}^N \delta_{J_{\sigma(i)}}{} ^{I_i}$ with coefficients depending only on the permutation $\sigma \in \mS_N$. Indices $I$ are contracted with factors $\om_I$ while indices $J$ are contracted with factors $\oom^J$ in accord with modular invariance.
\end{enumerate}
More explicitly, the modular graph functions take the form, 
\bea
\cC [n_{ij}; c(\sigma)] = c^{I_1 \cdots I_N}_{ J_1 \cdots J_N}  \int _{\Sigma ^N} \prod _{i=1}^N \om_{I_i} (z_i) \, \oom^{J_i} (z_i) \prod _{1 \leq i < j \leq N} \GA (z_i,z_j)^{n_{ij}}
\label{cagain}
\eea
and $c^{I_1 \cdots I_N}_{ J_1 \cdots J_N}$ is an invariant tensor built out of a linear combination of products of Kronecker $\delta$-symbols, which may be expressed as follows,
\bea
c^{I_1 \cdots I_N}_{ J_1 \cdots J_N} = \sum _{\sigma \in \mS_N} c(\sigma) \prod _{i=1}^N \delta  _{J_{\sigma (i)}}{}^{I_i}
\eea
where $c (\sigma)$ is an arbitrary scalar function of the permutation $\sigma \in \mS_N$.
The above construction gives a generalization of the construction of modular tensors on Teichm\"uller space initiated by Kawazumi in \cite{Kawa1, Kawa2}.  Since the singularities at $z_i=z_j$ are logarithmic, the integrals in  (\ref{cagain})  are absolutely convergent.

\subsection{Further generalizations}
\label{sec:2.9}

From our experience with  string theory, we know that it is appropriate at higher genus $h$, and higher multiplicity $N$, to also allow for differentials of the scalar Green function~$\GA$.  Instead of having a holomorphic one-form at a vertex $i$, it is permissable to have the differential $dz_i \p_{z_i} \GA(z_i, z_j)$, or the differential $dz_i \p_{z_i} \wedge dz_j \p_{z_j} \GA(z_i, z_j)$ at vertex $i$.
One may further encounter products of two Szeg\"o kernels (for their definition, see for example \cite{D'Hoker:1988ta}) at a point $z_i$, such as $S(z_i,z_j) S(z_i,z_k)$. It is well-known however that such products may be converted into a sum of the derivative of a scalar Green function and a contribution proportional to $\om_I(z_i)$ using Fay's trisecant identity (see for example \cite{fay73,D'Hoker:2002gw}). In such cases however, absolute convergence is no longer guaranteed, and non-trivial analytic continuations in $s_{ij}$ may be required to define the corresponding integrals over the points $z_i$ \cite{DHoker:1994gnm}.


\section{Parametrizing non-separating degenerations}
\setcounter{equation}{0}
\label{sec3}

A torus undergoes a single type of degeneration as its modulus $\tau$ tends to a cusp (usually chosen to be $\tau \to i \infty$) and degenerates to a sphere with two punctures.  A compact Riemann surface $\Sigma_{h+1}$ of higher genus $h+1$ with $h \geq 1$ admits two different basic types of degeneration: separating and non-separating.  Each type of degeneration  is specified by the vanishing of a complex parameter and corresponds to a divisor inside the Deligne-Mumford compactification of the moduli space of genus $h$ surfaces.

\sm

The separating divisor is reached by letting a cycle with trivial homology degenerate so that the surface becomes disconnected  into  a pair of surfaces whose genera add up to $h+1$. The effects of separating degenerations on modular graph forms will not be analyzed here.

\sm

\begin{figure}[h]
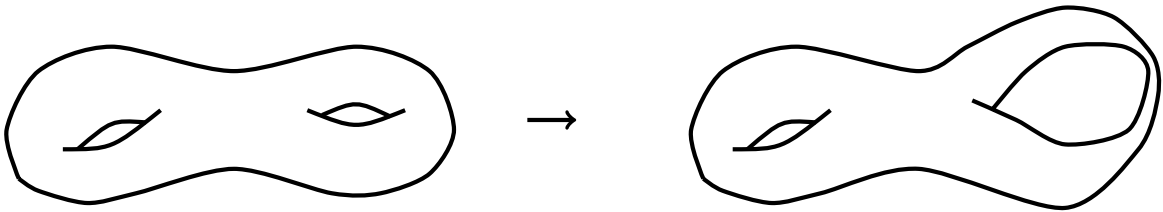

\begin{center}
\tikzpicture[scale=0.65]
\scope[xshift=-5cm,yshift=0cm]
\draw[ultra thick] plot [smooth] coordinates {(0.6,0.6) (0.35,1.6) (1,2.8) (2.5, 3.3)  (5,2.8) (7.5, 3.3) (9,2.8) (9.5,1.6) (9,0.7) (8,0.3) (7, 0.3) (5,0.8) (3, 0.3) (2,0.1) (1,0.35) (0.6,0.6) };
\draw[ultra thick] (1.5,1.2) .. controls (2.5, 1.2) .. (3.5,2);
\draw[ultra thick] (1.8,1.2) .. controls (2.5, 1.8) .. (3.2,1.75);
\draw[ultra thick] (6.5,2) .. controls (7.5, 1.6) .. (8.5,2);
\draw[ultra thick] (6.8,1.9) .. controls (7.5, 2.2) .. (8.2,1.87);
\draw [->, ultra thick] (11,1.8) -- (12,1.8);
\draw[ultra thick] plot [smooth] coordinates {(14.6,0.6) (14.35,1.6) (15,2.8) (16.5, 3.3)  (19,2.8) (20, 3.3) (21,3.8) (22, 4.1) (23,3.9) (23.8,3.1) (23.9,2.3) (23.5, 1.2) (22, 0) (19,0.8) (17, 0.3) (16,0.1) (15,0.35) (14.6,0.6) };
\draw[ultra thick] plot [smooth] coordinates {(20.5,2) (21.2, 2.8) (22, 3.3) (23.2, 3.3)   (23.7,2.8) (23.3, 1.6) (22,1.3) (21, 1.8)  (20.1, 2.2)};
\draw[ultra thick] (15.2,1.2) .. controls (16.2, 1.2) .. (17.2,2);
\draw[ultra thick] (15.5,1.2) .. controls (16.2, 1.8) .. (16.9,1.75);
\endscope
\endtikzpicture
\end{center}
\caption{The non-separating degeneration of a genus-two surface.}
\label{fig:a}
\end{figure}

The non-separating divisor is reached by letting a non-trivial homology cycle degenerate so that the surface sees one of the handles stretch to a thin cylinder. The  Riemann surface resulting from a basic non-separating degeneration has  genus $h$ and inherits two punctures.  The unique degeneration of the torus is actually of this type. The process is illustrated for the case of $h=1$ in figure \ref{fig:a}.
Different divisors, of the same or of different types may intersect, resulting in more drastic degenerations taking place in  higher co-dimension.  

\sm

In this section, we begin by reviewing the effect of a  non-separating degeneration on modular forms and modular graph functions at genus one. We then review how the non-separating degeneration of holomorphic Siegel modular forms at higher genus gives rise to an expansion in terms of holomorphic Jacobi modular forms. Finally, we set up the parametrization of the non-separating degeneration of non-holomorphic modular functions at higher genus in preparation for a full analysis in the next section.

\subsection{Genus-one modular forms and modular graph functions}

The degeneration as $\tau \to i \infty$ of  holomorphic forms transforming under $SL(2,\ZZ)$ or under one of its arithmetic subgroups is a classic result. Holomorphic Eisenstein series $\GG_w$ of weight $w \in \NN$ are defined  for $w \geq 3$ by the following absolutely convergent infinite sums,
\bea
\GG _w (\tau) = {(w-1)! \over 2 (2\pi i)^w} \sum _{(m,n) \in \ZZ^2} ' { 1 \over (m+ n \tau)^w}
\label{gwdef}
\eea
and admit an exact expansion  near the cusp, given by,
\bea
\GG_w(\tau) = - { B_w \over 2w} + \sum _{k=1}^\infty \sigma _{w-1}(k) q^k 
\eea
The prime superscript on the sum in (\ref{gwdef}) signifies the omission of the $m=n=0$ term; $B_w$ is the Bernoulli number; $q=e^{2 \pi i \tau}$; and $\sigma _s (n) = \sum _{d |n} d^s$ is the divisor sum. Thus, the behavior of $\GG_w(\tau)$ near the cusp is given by a constant plus exponentially suppressed terms. 

\sm

Non-holomorphic Eisenstein series $E_w$ are defined by the Kronecker-Eisenstein sums,
\bea
\label{Eisen}
E _w (\tau) =  \sum _{(m,n) \in \ZZ^2} ' \left ( { \tau_2 \over \pi |m+ n \tau|^2} \right )^w
\eea
Their asymptotic behavior near the cusp $\tau \to i \infty$ is also a classic result given by,
\bea
E_w (\tau) = - {B_{2w} \over (2w)!} (-4 \pi \tau_2)^w 
+{ 4 \, (2w-3)! \, \zeta (2w-1)  \over (w-2)! (w-1)! \, (4 \pi \tau_2)^{w-1}} + \cO(|q|)
\eea
Modular graph functions at genus one generalize non-holomorphic Eisenstein series to higher Kronecker-Eisenstein sums and have by now been studied fairly extensively  \cite{Green:2008uj,D'Hoker:2015foa,DHoker:2015wxz, Basu:2015ayg, Basu:2016kli, Brown:2017qwo,Brown2,Kleinschmidt:2017ege}. For example, a simple infinite family is given by modular graph functions depending on $a_1, \cdots, a_\ell \in \NN$ with $\ell \geq 2$ and may be defined as follows,
\bea
C_{a_a,\cdots, a_\ell}(\tau) = \sum_{{(m_r, n_r) \in \ZZ^2 \atop r=1, \cdots, \ell}} '
\delta \left ( \sum _{r=1}^\ell m_r \right ) \delta \left ( \sum _{r=1}^\ell n_r \right ) 
\prod _{r=1}^\ell \left ( { \tau_2 \over \pi |m_r+\tau n_r|^2} \right )^{a_r}
\eea
where  the Kronecker $\delta$-symbols restrict the sums of $m_r,n_r$ to vanish. For $\ell=2$, we recover the non-holomorphic Eisenstein series of (\ref{Eisen}) of weight $w=a_1+a_2$. For $\ell =3$, we have $w=a_1+a_2+a_3$ and the degeneration near the cusp is of the form,
\bea
C_{a_1, a_2, a_3} (\tau) = c_w (-4 \pi \tau_2)^w + { c_{2-w} \over (4 \pi \tau_2)^{w-2} } + \sum _{k=1}^{w-1} c_{w-2k-1} { \zeta (2k+1) \over (4 \pi \tau_2)^{2k+2-w}}
+ \cO(|q|)
\eea
where the coefficient $c_{2-w}$ is a sum over bilinears in odd $\zeta$-values with rational coefficients, while all other coefficients $c_w$ and $c_{w-2k-1}$ are rational numbers \cite{DHoker:2017zhq}.  The property possessed by all genus-one modular functions  is that their asymptotic behavior near the cusp $\tau \to i \infty$ is governed by a Laurent polynomial in $\tau_2$ of finite degree, plus exponentially suppressed terms. The degree of the polynomial is at most $(w,w-1)$, where  the weight $w$ is the total number of Green function factors in the modular graph function.

\subsection{Parametrization near a non-separating divisor}

We consider the non-separating degeneration of a compact Riemann surface  $\Sigma _{h+1}$ of genus $h+1$ without boundary. By an $Sp(2h+2,\ZZ)$ modular transformation, we may take the degenerating cycle to be a non-zero integer multiple of the canonical cycle $\mB_{h+1}$. Denoting the period matrix for this surface by $\Omega$, it is natural to block-decompose it and single out the dependence on the cycles $\mA_{h+1}$ and $\mB_{h+1}$, as follows,
\bea
\label{Om}
\Omega = \left ( \begin{matrix} \tau & v \cr v^t & \sigma \cr \end{matrix} \right )
\eea
where $\tau$ is an $h \times h$ symmetric matrix, $v$ is a column matrix of height $h$, and $\sigma$ is a  number, all of which are complex. The non-separating divisor corresponding to the cycle $\mB_{h+1}$ degenerating with the other cycles remaining fixed is characterized by $\sigma \to i \infty$. In this limit, the surface $\Sigma _{h+1}$ degenerates to a surface of genus $h$ with two punctures $p_a, p_b$, whose underlying compact surface will be denoted by $\Sigma _h$. The entry $\tau$ tends to the period matrix of $\Sigma _h$, while $v$ tends to the Abel map of the punctures, 
\bea
v_I = \int ^{p_b} _{p_a} \om _I = \zeta_I(p_b)-\zeta_I(p_a) \hskip 1in I=1,\cdots, h
\eea
where the holomorphic Abelian differentials $\om_I$ for $I=1,\cdots, h$ are those corresponding to the compact surface  $\Sigma_h$, and the Abel map $\zeta$ was defined in (\ref{zeta}).

\sm

The subgroup of $Sp(2h+2,\ZZ)$ which preserves the asymptotic limit $\sigma \to i \infty$ is a semi-direct product of the modular group $Sp(2h,\ZZ)$ of the compact surface $\Sigma _h$ with the group of translations of $v$ by the lattice $\ZZ^h + \tau \ZZ^h$ as well as integer shifts in $\sigma$ \cite{MR781735}, 
\bea 
Sp(2h+2,\ZZ) \to Sp(2h,\ZZ) \ltimes (\ZZ^h + \tau \ZZ^h) \ltimes \ZZ
\eea
Recall that the elements $M$ of $Sp(2h+2,\ZZ)$ may be parametrized as follows,
\bea
\label{em}
M = \left ( \begin{matrix} A & B \cr C & D \cr \end{matrix} \right )
\hskip 1in 
M^t \mJ M=\mJ
\eea
where $A,B,C,D$ are $(h+1) \times (h+1)$ matrices with integer entries, and $\mJ$ is the symplectic matrix of rank $h+1$. The subgroup $Sp(2h,\ZZ)$ corresponds to the following elements, 
\bea
\label{emm}
A= \left ( \begin{matrix} a & 0 \cr 0 & 1 \cr \end{matrix} \right )
\hskip 0.3in
B= \left ( \begin{matrix} b & 0 \cr 0 & 0 \cr \end{matrix} \right )
\hskip 0.3in
C= \left ( \begin{matrix} c & 0 \cr 0 & 0 \cr \end{matrix} \right )
\hskip 0.3in
D= \left ( \begin{matrix} d & 0 \cr 0 & 1 \cr \end{matrix} \right )
\quad
\eea
 while $a,b,c,d$ are $h \times h$ sub-matrices respectively of $A,B,C,D$. The action of the semi-simple part $Sp(2h,\ZZ)$ on the entries of the period matrix is given by,
\bea
\label{taup}
\tau ' & = & (a \tau +b)(c \tau +d)^{-1} 
\no \\
v' & = & \big ( (c \tau +d)^t \big )^{-1} v
\no \\
\sigma ' & = & \sigma - v^t (c \tau +d)^{-1} c v
\eea
while the action of the $\ZZ^{h+1} + \tau \ZZ^h$ factor is by translations of $v$ and $\sigma$ at fixed $\tau$.

\sm

A standard parametrization of the Riemann surface $\Sigma _{h+1}$ near its non-separating divisor is obtained by constructing $\Sigma _{h+1}$ from a compact surface $\Sigma _h$.  The canonical holomorphic Abelian differentials on $\Sigma_h$ will be denoted by $\om_I$ with $I=1,\cdots, h$ and the associated period matrix of $\Sigma _h$ will be denoted $\tau$. Next, we remove two points $p_a,p_b$ and denote the line integral between these punctures by $v _I = \int ^{p_b} _{p_a} \om _I$.  Local holomorphic coordinates $z_a$ and $z_b$, respectively near $p_a$ and $p_b$, are chosen such that $z_a=0$ at $p_a$ and $z_b=0$ at $p_b$. 

\sm

The standard construction \cite{fay73} of $\Sigma _{h+1}$ proceeds by introducing a  complex parameter $\mt$ and drawing coordinate circles $\mC_a'$ and $\mC_b''$ centered respectively at $p_a$ and $p_b$ and defined by $|z_a|= |\mt| $ and $|z_b|=|\mt|$ as well as coordinate circles $\mC_a''$ and $\mC_b'$ centered respectively at $p_a$ and $p_b$ and defined by $|z_a|= |z_b|=1 $ (see figure 5). The discs enclosed by  $\mC_a'$ and $\mC_b''$ are removed and the annulus  enclosed between $\mC_a'$ and $\mC_a''$ is identified with the annulus enclosed between $\mC_b'$ and $\mC_b''$ by identifying points with local coordinates $z_a, z_b$ according to the following rule,
\bea
z_a \, z_b = \mt
\eea
The cycles $\mC_a'$ and $\mC_b'$ are identified with one another, as are $\mC_a''$ and $\mC_b''$,
and the resulting cylinder spanned between $\mC_a'\approx \mC_b'$ and $\mC_a''\approx \mC_b ''$ is usually referred to as the funnel or plumbing fixture of the construction.  The locations of the punctures $p_a,p_b$ and the parameter~$\mt$, together with the moduli of the surface $\Sigma _h$ encoded in the independent entries of the period matrix $\tau$, provide all the moduli of the surface $\Sigma _{h+1}$.

\begin{figure}[h]
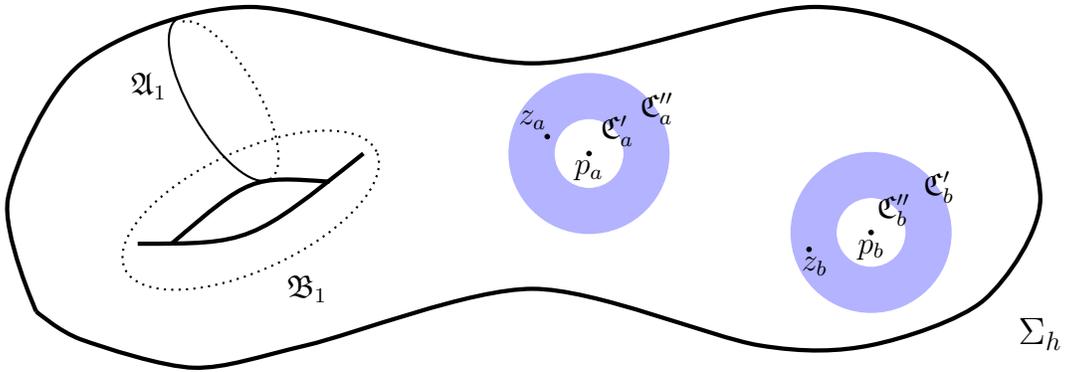

\begin{center}
\tikzpicture[scale=1.5]
\scope[xshift=-5cm,yshift=0cm]
\draw[ultra thick] plot [smooth] coordinates {(0.6,0.6) (0.35,1.6) (1,2.8) (2.5, 3.3)  (5,2.8) (7.5, 3.3) (9,2.8) (9.5,1.6) (9,0.7) (8,0.3) (7, 0.3) (5,0.8) (3, 0.3) (2,0.1) (1,0.35) (0.6,0.6) };
\draw[ultra thick] (1.5,1.2) .. controls (2.5, 1.2) .. (3.5,2);
\draw[ultra thick] (1.8,1.2) .. controls (2.5, 1.8) .. (3.2,1.75);
\draw[thick, dotted,  rotate=25] (2.9,0.3)  ellipse (35pt and 15pt);
\draw[thick, dotted,  rotate=30] (3.2,0.2)  arc (-90:90:0.3 and 0.8);
\draw[thick,  rotate=30] (3.2,1.84)  arc (90:270:0.3 and 0.82);
\draw (1.6,2.6) node{$\mA_1$};
\draw (3,0.8) node{$\mB_1$};
\draw (9.5,0.4) node{{\large $\Sigma_h$}};
\draw [very thick, fill, color=blue!30] (5.5,2) circle (0.7);
\draw [very thick, fill, color=blue!30] (5.5,2) circle (0.5);
\draw[ black, fill, white] (5.5,2) circle (0.3);
\draw[ black, fill] (5.5,2) circle (0.02);
\draw (5.5, 1.85) node{$p_a$};
\draw[ black, fill] (5.13,2.15) circle (0.02);
\draw (5, 2.3) node{$z_a$};
\draw (5.75, 2.2) node{$ \mC_a'$};
\draw (6.1, 2.4) node{$\mC_a''$};
\draw [very thick, fill, color=blue!30] (8,1.3) circle (0.7);
\draw [very thick, fill, color=blue!30] (8,1.3) circle (0.5);
\draw[ black, fill, white] (8,1.3) circle (0.3);
\draw[ black, fill] (8,1.3) circle (0.02);
\draw (8, 1.15) node{$p_b$};
\draw[ black, fill] (7.45,1.15) circle (0.02);
\draw (7.5, 1) node{$z_b$};
\draw (8.6, 1.7) node{$\mC_b'$};
\draw (8.2, 1.5) node{$\mC_b''$};
\endscope
\endtikzpicture
\caption{Fay's construction of a surface near a non-separating divisor in terms of a surface $\Sigma_h$, punctures $p_a,p_b$, and the annuli bounded by $\mC'_a, \mC''_a$ and  by $\mC''_b, \mC'_b$ identified for $h=1$.}
\end{center}
\label{fig:4}
\end{figure}

\subsection{Degeneration of Siegel modular forms to Jacobi forms}

A Siegel modular form $\Phi$ of degree $h+1$ and weight $k$ is a holomorphic  function on   $\cH_{h+1}$  which transforms  under $Sp(2h+2,\ZZ)$, as parametrized by (\ref{em}), by,
\bea
\Phi (\Omega ') = \big ( \det (C \Omega+D)  \big )^k \Phi (\Omega)
\eea
Decomposing a general element $\Omega \in \cH_{h+1}$ (which for large enough $h$ is not necessarily the period matrix of a Riemann surface) with respect to $\cH_h$ as in (\ref{Om}), and letting $\sigma \to i \infty$, 
we may expand $\Phi (\Omega)$ in positive integer powers of $e^{2 \pi i \sigma}$,
\bea
\Phi (\Omega) = \sum _{n=0}^\infty \phi_n (v |\tau) \, e^{ 2 \pi i n \sigma}
\eea
Implementing the modular transformation properties of $\Phi (\Omega) $ under the group $Sp(2h,\ZZ)$ which leaves the limit $\sigma \to i \infty$ invariant, as parametrized by (\ref{emm}), we deduce the  transformation properties of $\phi _n (v|\tau)$,
\bea
\phi _n ( v' |\tau') = \big ( \det (c \tau +d) \big )^k \, \exp \Big \{ 2 \pi i n \, v^t \, (c \tau+d)^{-1} c\, v \Big \} 
\, \phi _n (v |\tau)
\eea
where $v'$ and $\tau'$ were given in terms of the transformation parameters in (\ref{taup}). The transformation law for $\phi _n (v |\tau)$ corresponds to the degree-$h$ generalization of Jacobi forms. In particular, the limit of a degree-two Siegel modular form produces a standard degree-one Jacobi form of weight $k$ and index $n$ \cite{MR781735}.

\subsection{Parametrization by a worldsheet time function $f$}

Fay's construction of a genus $h+1$ surface near a non-separating divisor in terms of a genus $h$ Riemann surface with two punctures is general \cite{fay73}. Its practical implementation is rendered considerably more  convenient if one has a natural intrinsic construction of the cycles $\mC_a', \mC_a'', \mC_b'$ and $\mC_b''$. In this subsection, we shall introduce such a construction in terms of a real-valued harmonic function $f$ on the surface $\Sigma _h \setminus \{ p_a, p_b \}$ and parametrize the degeneration by a real parameter $t$ which is invariant under the Fourier-Jacobi group $Sp(2h,\ZZ)\ltimes (\ZZ^h + \Omega \ZZ^h) \ltimes \ZZ$ which is the subgroup of the modular group $Sp(2h+2,\ZZ)$ under which the non-separating degeneration cycle $\mB_{h+1}$ is invariant.

\sm
 
We consider a compact Riemann surface $\Sigma _h$ with a basis of canonical homology cycles $\mA_I, \mB_I$ for $I=1,\cdots, h$ as in (\ref{AB}), dual holomorphic Abelian differentials $\om_I$  and period matrix $\tau$ with elements  $\tau_{IJ}$ normalized as in (\ref{omnorm}),
 \bea
 \oint _{\mA_I} \om _J = \delta _{IJ} \hskip 1in \oint _{\mB_I} \om _J = \tau_{IJ}
 \hskip 1in 
 I,J = 1 , \cdots, h
 \eea
We delete two points $p_a, p_b$ from $\Sigma _h$  and define the corresponding Abelian integrals as follows,
\bea
\label{vu}
v_I = \int ^{p_b} _{p_a} \om _I 
\hskip 1in 
v = u_1 + \tau u_2 
\hskip 1in u_1, u_2 \in \RR^h
\eea
Fay's construction  so far provides $\tau$ and $v$ which give $\half h(h+3)$ of the $\half (h+1)(h+2)$ entries of the period matrix $\Omega $ of $\Sigma_{h+1}$. The single missing entry corresponds to the parameter $\sigma$ of the degeneration. We will now see that a suitable parameterization which reduces to $\sigma$ close to the non-separating degeneration limit is the key to analyzing  the exact asymptotics.

\subsubsection{The function $f$}

The key ingredient of our construction is the existence of a meromorphic Abelian differential on $\Sigma _h$ with simple poles at the punctures $p_a,p_b$ and holomorphic everywhere else, whose periods on the surface $\Sigma _h \setminus \{ p_a, p_b\}$ are all purely real. Such a differential is standard in mathematics \cite{FK} and was used long ago to formulate string perturbation theory and string field theory in a Lorentz-covariant manner on worldsheet light-cone diagrams by introducing a global proper time on the worldsheet \cite{Giddings:1986rf,DHoker:1987hzc}. The Abelian differential $\om_t$ is defined by, 
\bea
\label{omt}
\om_t = { 1 \over 2 \pi i} \, \om_{p_b,p_a} - \sum_{I=1}^h u_{2I} \, \om _I 
\eea
The Abelian differential $\om_{p_b,p_a}$ has simple poles at $p_b,p_a$ with respective  residues  $ +1$ and $-1$,  and has standard normalization on $\mA$-cycles so that,  
\bea
\oint _{\mA_I} \om _{p_b,p_a}=0 
\hskip 0.6in 
{ 1 \over 2 \pi i} \oint _{\mB_I} \om _{p_b,p_a}=  v_I 
\eea
where $v$ and $u_2$ were introduced in (\ref{vu}) with $v_I$ and $u_{2I}$ denoting their components. One verifies that the integrals of $\om_t$ around canonical homology cycles are given by, 
\bea
\label{norm}
\oint _{\mA_I} \om_t = - u_{2I} 
\hskip 0.6in 
\oint _{\mB_I} \om_t =  u_{1I} 
\hskip 0.6in 
\oint _{\mC_b} \om_t = -\oint _{\mC_a} \om_t =  1
\eea
where $\mC_b$ and $\mC_a$ are simple cycles enclosing $p_b$ and $p_a$ respectively. One also verifies the following Riemann relations, 
\bea
 \int _{\Sigma _h} \om_{p_b, p_a} \wedge \oom _I = 4 \pi \, \Im v_I
\hskip 1in
\int _{\Sigma _h} \om_t \wedge \oom _I=0
\eea
Since all the periods of $\om_t$ on $\Sigma _h \setminus \{ p_a, p_b\}$ are real, the imaginary part of the corresponding Abelian integral, normalized as follows,
\bea
\label{fzf}
f(z) - f(w) = 4 \pi \, \Im \int^z _w \om _t
\eea
defines a real single-valued harmonic function $f(z)$ on $\Sigma _h \setminus \{ p_a, p_b\}$, up to an additive constant. The normalization  factor  $4\pi$ has been included for later convenience. A helpful alternative formula for $f$ is in terms of the Arakelov Green function $\cG_h(z,w|\tau)$ on the surface $\Sigma_h$. We shall  fix the overall additive constant in $f$ by setting,  
\bea
\label{fGreen}
f(z) = \cG_h(z,p_b|\tau) - \cG_h(z,p_a |\tau) 
\eea
The difference $f(z)-f(w)$ is readily seen to agree with (\ref{fzf}).

\subsubsection{The funnel construction using $f$}

We shall now complete the construction of the surface $\Sigma _{h+1}$ near its non-separating divisor by supplementing  Fay's construction with the use of the function $f(z)$.  Key properties of this function are as follows: $f$ is a real harmonic function on $\Sigma _h \setminus \{ p_a, p_b\}$ which tends to $+ \infty$ as $z \to p_b$ and tends to $- \infty$ as $z \to p_a$. Thus, for sufficiently large $t_a'' < t_a < t_a'$ and sufficiently large $t_b' < t_b < t_b''$ we may define curves $\mC_a', \mC_a, \mC_a''$ and  $\mC_b', \mC_b, \mC_b''$ such that,\footnote{Throughout this paper, a curve will refer to a 1-dimensional real sub-manifold.}
\bea
\label{levels}
f(\mC_a')=-2\pi t_a'  & \hskip 1in f(\mC_b') = 2 \pi t_b'
\no \\
f(\mC_a)=-2\pi t_a  & \hskip 1in f(\mC_b) = 2 \pi t_b
\no \\
f(\mC_a'')=-2\pi t_a''  & \hskip 1in f(\mC_b'') = 2 \pi t_b''
\eea
Here, the notation $f(\mC)$ means that all the points on the curve $\mC$ are assigned the same value under the map $f$, so that $\mC_a'', \mC_a, \mC_a'$ and  $\mC_b', \mC_b, \mC_b''$ arise as (ordered) level sets for the  function~$f$, as depicted in figure 6 for $h=1$. The construction can be  used only for values of $t_a''$ and $t_b'$ large enough so that each level set for $f<-2\pi t_a''$, and each level set for $f> 2 \pi t_b'$ is a  connected curve. Since we are interested in the neighborhood of the non-separating divisor, we may assume that $t$ is large enough for this condition to hold. By inspection of figure 6, it is clear that for intermediate values of these parameters, the level sets become  disconnected when $h \geq 1$. 

\begin{figure}
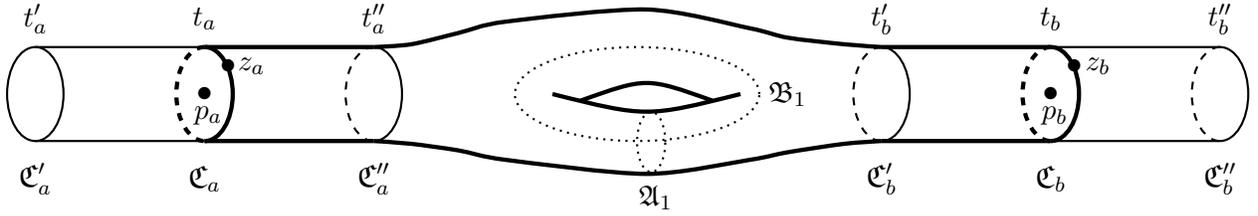

\begin{center}
\tikzpicture[scale=1.25]
\scope[xshift=-5cm,yshift=0cm]
\draw [thick] (0.3,-0.5) arc (0:360:0.3 and 0.5);
\draw [ultra thick] (1.8,-1) arc (-90:90:0.3 and 0.5);
\draw [ultra thick, dashed] (1.8,0) arc (90:270:0.3 and 0.5);
\draw [thick] (3.6,-1) arc (-90:90:0.3 and 0.5);
\draw [thick, dashed] (3.6,0) arc (90:270:0.3 and 0.5);
\draw [thick] (0,0) -- (1.8,0);
\draw [thick] (0,-1) -- (1.8,-1);
\draw [ultra thick] (1.8,0) -- (3.6,0);
\draw [ultra thick] (1.8,-1) -- (3.6,-1);
\draw[ultra thick] (5.5,-0.5) .. controls (6.5, -0.75) .. (7.5,-0.5);
\draw[ultra thick] (5.8,-0.57) .. controls (6.5, -0.32) .. (7.2,-0.57);
\draw [thick] (9,-1) arc (-90:90:0.3 and 0.5);
\draw [thick, dashed] (9,0) arc (90:270:0.3 and 0.5);
\draw [ultra thick] (10.8,-1) arc (-90:90:0.3 and 0.5);
\draw [ultra thick, dashed] (10.8,0) arc (90:270:0.3 and 0.5);
\draw [thick] (12.6,-1) arc (-90:90:0.3 and 0.5);
\draw [thick, dashed] (12.6,0) arc (90:270:0.3 and 0.5);
\draw [ultra thick] (9,0) -- (10.8,0);
\draw [ultra thick] (9,-1) -- (10.8,-1);
\draw [thick] (10.8,0) -- (12.6,0);
\draw [thick] (10.8,-1) -- (12.6,-1);
\draw[ultra thick] plot [smooth] coordinates {(3.6,0) (4,0.03) (4.5, 0.15) (5, 0.27) (6.5,0.4)  (7.6, 0.2) (8, 0.10) (8.6,0.03) (9,0)};
\draw[ultra thick] plot [smooth] coordinates {(3.6,-1) (4,-1.03) (4.5, -1.10)  (5, -1.2) (6.5,-1.35) (7.6, -1.2) (8, -1.12) (8.6,-1.03) (9,-1)};
\draw [thick, dotted] (7.7,-0.5) arc (0:360:1.3 and 0.5);
\draw [thick, dotted] (6.7,-1.01) arc (0:360:0.15 and 0.32);
\draw (0,-1.4) node{$\mC_a'$};
\draw (1.8,-1.4) node{$\mC_a$};
\draw (3.6,-1.4) node{$\mC_a''$};
\draw (0,0.3) node{\small $t_a'$};
\draw (1.8,0.3) node{\small $t_a$};
\draw (3.6,0.3) node{\small $t_a''$};
\draw (9,-1.4) node{$\mC_b'$};
\draw (10.8,-1.4) node{$\mC_b$};
\draw (12.6,-1.4) node{$\mC_b''$};
\draw (9,0.3) node{\small $t_b'$};
\draw (10.8,0.3) node{\small $t_b$};
\draw (12.6,0.3) node{\small $t_b''$};
\draw (1.8,-0.5) node{$\bullet$};
\draw (10.8,-0.5) node{$\bullet$};
\draw (1.84,-0.73) node{{\small $p_a$}};
\draw (10.84,-0.73) node{\small $p_b$};
\draw (2.05,-0.2) node{$\bullet$};
\draw (11.05,-0.2) node{$\bullet$};
\draw (2.3,-0.2) node{{\small $z_a$}};
\draw (11.3,-0.2) node{\small $z_b$};
\draw (8,-0.5) node{{\small $\mB_1$}};
\draw (6.6,-1.6) node{{\small $\mA_1$}};
\endscope
\endtikzpicture
\caption{The funnel construction near the non-separating divisor of the surface $\Sigma_{h+1}$ is illustrated for $h=1$.  The surface  $\Sep $ is obtained from $\Sigma _h$ by removing the discs with boundaries $\mC_a$ and $\mC_b$ centered at the punctures $p_a,p_b$ respectively. The surface $\Sigma _{h+1}$ is obtained from $\Sep$ by pairwise  identifying the cycles  $\mC_a \approx \mC_b$, (as well as $\mC_a' \approx \mC_b'$ and $\mC_a'' \approx \mC_b''$) which are homologous to $\mA_{h+1}$. The  cycle $\mB_{h+1}$ is obtained by connecting $z_a$ to $z_b$ via a curve on $\Sep$ and identifying the points $z_a \approx z_b$. The punctures $p_a, p_b$ lie on the underlying compact surface $\Sigma _h$, but do not belong to either $\Sep$ or $\Sigma_{h+1}$.}
\end{center}
\label{fig:5}
\end{figure}

The missing entry $\sigma$ of the period matrix $\Omega$ of $\Sigma_{h+1}$ in (\ref{Om}) (recall that we already have $\tau$ and $v$) is obtained from a line integral over $\mB_{h+1}$. The Abelian differential,
\bea
\om_{h+1} = { 1 \over 2 \pi i } \om _{p_b, p_a}
\eea
which is meromorphic on $\Sigma _h$, extends to the missing holomorphic Abelian differential on $\Sigma _{h+1}$. Its period on the cycle $\mA_{h+1}$ is guaranteed to be 1 by the normalization of its residues at $p_a$ and $p_b$ on the underlying surface $\Sigma _h$ by the normalizations provided in (\ref{norm}). Therefore, its line integral between the points $z_a$ and $z_b$ will provide its $\mB_{h+1}$-period,
\bea
 \oint _{\mB_{h+1}} \om_{h+1}  =  \int _{z_a} ^{z_b} \om_{h+1} = \sigma
\eea
Decomposing $\sigma = \sigma _1 + i \sigma _2$ with $\sigma _1, \sigma _2 \in \RR$, we find the  imaginary part $\sigma _2$ as follows,
\bea
\sigma _2 = { 1 \over 4 \pi} \big ( f(z_b) - f(z_a) \big ) +  u_2^t \, (\Im \tau ) \, u_2
\label{fdiff}
\eea
Using a not-so-well-known formula for the determinant of an $(h+1) \times (h+1)$ symmetric matrix
in terms of its block decomposition into $h$-dimensional and  1-dimensional parts, 
\bea
\det (\Im \Omega) = \det \left ( \begin{matrix} \Im \tau & \Im v \cr \Im v^t & \Im \sigma  \cr \end{matrix} \right )
= \det (\Im \tau) \Big ( \sigma _2 - \Im v^t \, (\Im \tau)^{-1} \, \Im v \Big )
\eea
along with the relation  $\Im v = (\Im \tau) u_2$ and some slight rearrangements, we may rewrite (\ref{fdiff}) in the following  form, 
\bea
\label{fft}
f(z_b) - f(z_a) = 4 \pi t
\eea
where $t$ is the following real positive combination of elements of the period matrix, 
\bea
\label{t}
t = { \det ( \Im \Omega)  \over \det (\Im \tau)} =  \sigma _2 - \Im v^t \, (\Im \tau)^{-1} \, \Im v
\eea
The parameter $t$ is invariant under the action of the modular subgroup $Sp(2h,\ZZ)$ on $\tau$ and $\Omega$ according to (\ref{taup}) and tends to $\infty$ upon letting $\sigma \to i \infty$ while keeping all other moduli fixed (or bounded). Thus, $t$ may be used to parametrize the neighborhood of the non-separating node, and will play a role in the asymptotics which is very analogous to the role $\tau_2$ played near the cusp for genus one. 

\sm

The construction of the surfaces $\Sigma _{ab}$ and $\Sigma _{h+1}$ from the compact surface $\Sigma _h$ proceeds as follows. We define the cycles $\mC_a, \mC_a', \mC_a''$ and $\mC_b, \mC_b', \mC_b''$ as the level sets of the function $f$ given in (\ref{levels}). For given degeneration modulus $t$, pairs of curves obey  the following rule, 
\bea
f(\mC_b) - f(\mC_a) = f(\mC_b') - f(\mC_a') = f(\mC_b'') - f(\mC_a'') = 4 \pi t
\eea 
or equivalently  $t_b - t_a = t_b' - t_a' = t_b '' - t_a ''= 2t$. Without loss of generality, we may choose $t_a=-t$ and $t_b = +t$, so that,
\bea
f(z) = - 2 \pi t + 4 \pi \, \Im \int _{z_a }^z \om _t
\eea
We shall define $\Sep$ as the genus $h$ Riemann surface with boundary cycles $\mC_a$ and $\mC_b$ specified by the level set of $f$ with values $\pm 2 \pi t$. 

\sm

The surface $\Sigma _{h+1}$ is obtained from $\Sep$ by identifying the cycles $\mC_a$ and $\mC_b$. This identification involves twisting the cycle $\mC_a$ relative to $\mC_b$ by the modulus $\sigma _1$. In this paper we shall be interested only in the contributions to the non-separating degeneration of modular graphs functions and string invariants that are power-behaved in $t$. All dependence on $\sigma _1$ is contained completely in contributions that are exponentially suppressed in $t$, so that the shift by $\sigma _1$, which is generally required to construct $\Sigma _{h+1}$,  may be entirely ignored for our evaluations in the non-separating degeneration where all contributions exponential in $t$ are omitted. The resulting surface is depicted in figure 6 for the case $h=1$.

\subsubsection{Constructing $f$ and $\Sep$ from $\Sigma _{h+1}$}

In the preceding part of this subsection, we have constructed the function $f$ and the surface $\Sep$ starting from the lower-genus surface $\Sigma _h$. It is also possible to do the converse and construct $f$ and $\Sep$ from $\Sigma _{h+1}$ in the neighborhood of the non-separating degeneration.  To do so, we consider a compact Riemann surface $\Sigma _{h+1}$ with a basis of canonical homology cycles $\mA_I, \mB_I$ for $I=1,\cdots, h+1$ as in (\ref{AB}), dual holomorphic Abelian differentials $\om_I$  and period matrix $\Omega$ with elements  $\Omega_{IJ}$ normalized as in (\ref{omnorm}),
 \bea
 \oint _{\mA_I} \om _J = \delta _{IJ} \hskip 1in \oint _{\mB_I} \om _J = \Omega_{IJ}
 \hskip 1in 
 I,J = 1 , \cdots, h+1
 \eea
 We shall now consider a family of such Riemann surfaces in the neighborhood of the non-separating degeneration obtained by degenerating the cycle $\mB_{h+1}$, while keeping all other cycles finite, and block-decompose the period matrix accordingly,
 \bea
 \label{Om1}
 \Omega = \left ( \begin{matrix} \tau & v \cr v^t & \sigma \cr \end{matrix} \right )
 \eea
 In a slight abuse of notation, we use the same symbols here as for the converse construction from $\Sigma_h$ to $\Sigma _{h+1}$ because, near the non-separating degeneration, the two will coincide up to terms which are exponentially suppressed in power of $t$. We consider the following holomorphic Abelian differential $\om _t$ on $\Sigma_{h+1}$, 
 \bea
\om_t = \om_{h+1} - \sum _{I=1}^h  (u_2)_I \, \om _I
\hskip 1in 
v = u_1 + \tau u_2
\eea 
where $\tau$ and $v$ are given by (\ref{Om}), and $u_1, u_2 \in \RR^h$. The Abelian differential $\om_t$ has been constructed so as to have real periods along the cycles $\mA_I$ for $I=1,\cdots, h+1$ and along $\mB_I$ for $I=1,\cdots, h$. Since $u_2$ is real, this is manifest for the $\mA$-cycles, while for the $\mB$-cycles it follows from the expressions for  $v$ which are equivalent under the Riemann relations, 
\bea
v_I = \oint _{\mB_I} \om_{h+1} = \oint _{\mB_{h+1} } \om_I 
\eea
The period of $\om_t$ around the remaining cycle $\mB_{h+1}$  is given by, 
\bea
\oint _{\mB_{h+1}} \om _t = \sigma - \sum _{I=1}^h (u_2)_I v_I
\eea
and its imaginary part is precisely equal to $t$ in view of (\ref{t}). Therefore, we may define a real harmonic function $f$, whose period around the cycle $\mB_{h+1}$ is $4 \pi t$ and whose periods around all other cycles vanish, by the following expression,
\bea
f(z) -f(w) = 4 \pi \, \Im \int ^z _w \om _t
\eea
To complete the construction, we now consider a cycle $\mC$ homologous to $\mA_{h+1}$, and cut the surface $\Sigma _{h+1}$ along $\mC$. We denote the resulting boundary cycles by $\mC_a$ and $\mC_b$. On this  cut surface with two boundary components the function $f$ is now single-valued. Since $f$ is harmonic, it will attain its minimum and maximum values on the boundary cycles. It suffices to continuously adjust the choice of the cycle $\mC$ so that the function $f$ on the resulting boundary cycles $\mC_a, \mC_b$ of $\Sep$ takes the values $\pm 2 \pi t$. We have now completely reproduced the surface $\Sep$, equipped with the function $f$, starting from a compact surface $\Sigma _{h+1}$.

\begin{figure}[h]
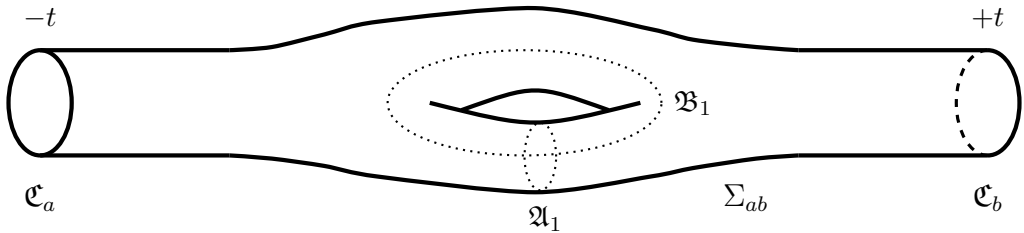

\begin{center}
\tikzpicture[scale=1.4]
\scope[xshift=-5cm,yshift=0cm]
\draw [ultra thick] (1.8,-1) arc (-90:90:0.3 and 0.5);
\draw [ultra thick] (1.8,0) arc (90:270:0.3 and 0.5);
\draw [ultra thick] (1.8,0) -- (3.6,0);
\draw [ultra thick] (1.8,-1) -- (3.6,-1);
\draw[ultra thick] (5.5,-0.5) .. controls (6.5, -0.75) .. (7.5,-0.5);
\draw[ultra thick] (5.8,-0.57) .. controls (6.5, -0.32) .. (7.2,-0.57);
\draw [ultra thick] (10.8,-1) arc (-90:90:0.3 and 0.5);
\draw [very thick, dashed] (10.8,0) arc (90:270:0.3 and 0.5);
\draw [ultra thick] (9,0) -- (10.8,0);
\draw [ultra thick] (9,-1) -- (10.8,-1);
\draw[ultra thick] plot [smooth] coordinates {(3.6,0) (4,0.03) (4.5, 0.15) (5, 0.27) (6.5,0.4)  (7.6, 0.2) (8, 0.10) (8.6,0.03) (9,0)};
\draw[ultra thick] plot [smooth] coordinates {(3.6,-1) (4,-1.03) (4.5, -1.10)  (5, -1.2) (6.5,-1.35) (7.6, -1.2) (8, -1.12) (8.6,-1.03) (9,-1)};
\draw [thick, dotted] (7.7,-0.5) arc (0:360:1.3 and 0.5);
\draw [thick, dotted] (6.7,-1.01) arc (0:360:0.15 and 0.32);
\draw (1.8,-1.4) node{$\mC_a$};
\draw (1.8,0.3) node{\small $-t$};
\draw (10.8,-1.4) node{$\mC_b$};
\draw (10.8,0.3) node{\small $+ t$};
\draw (8,-0.5) node{{\small $\mB_1$}};
\draw (6.6,-1.6) node{{\small $\mA_1$}};
\draw (8.5,-1.4) node{{$\Sep$}};
\endscope
\endtikzpicture
\caption{The surface $\Sep$ is obtained from $\Sigma_{h+1}$ in the vicinity of the non-separating divisor by cutting $\Sigma _{h+1}$ along a cycle homologous to $\mA_{h+1}$ and adjusting the position of the cycle so that $\mC_a$ and $\mC_b$  are level sets for $f= \pm 2 \pi t$,  shown here for the case $h=1$. }
\end{center}
\label{fig:6}
\end{figure}

\subsection{Relating integrals on the surfaces $\Sigma _{h+1}, \Sep$, $\Sigma _h$}

We shall be interested in evaluating the non-separating degeneration limit  $t\to\infty$ (keeping the complex structure moduli $\tau, v$ bounded during the degeneration) of a class of absolutely convergent multiple integrals over $\Sigma _{h+1}$.   Each integral is schematically of the following form, 
\bea
\int _{\Sigma _{h+1} } \om _I \wedge \oom_J \, \psi 
\eea
where $\om_I$ are the holomorphic Abelian differentials on $\Sigma _{h+1}$ with $I, J =1,\cdots, h+1$,  and $\psi$ is a smooth function on $\Sigma _{h+1}$. Our purpose is to extract the power-behaved dependence on $t$ exactly, evaluate the coefficients in this expansion in terms of convergent integrals on $\Sigma _h$, and neglect exponentially suppressed contributions $\cO(e^{-2 \pi t})$ as $ t \to \infty$. 

\sm

Since $\Sigma _{h+1}$ is obtained by identifying the boundary cycles $\mC_a$ and $\mC_b$ of the surface $\Sep$, we have the following exact relation between their integrals,
\bea
\label{omom}
\int _{\Sigma _{h+1} } \om _I \wedge \oom_J \, \psi = 
\int _{\Sep } \om _I \wedge \oom_J \, \psi 
\eea
We now wish to evaluate the asymptotics as $t \to \infty$. In this limit, the surface $\Sigma _{h+1}$ tends to the surface $\Sigma _h \setminus \{ p_a, p_b\}$, and therefore we need to make some assumptions on the behavior of $\psi$. Since $\psi$ is a polynomial combination of scalar Green functions on $\Sigma_{h+1}$ and integrals thereof, one can show that $\psi$ will be uniformly bounded by a finite power of $\log|z-p_a |$  or $\log|z-p_b |$ near $z=p_a,p_b$. We shall also assume that $\psi$ has no explicit $t$-dependence.

\sm

The fate of the integrals (\ref{omom}) will depend on $I$ and $J$. For $I,J=1,\cdots, h$ the differentials $\om_I, \om _J$ tend to holomorphic differential on $\Sigma _h$ up to exponential corrections $\cO(e^{-2 \pi t})$ which we neglect. As a result, we may recast the corresponding  integrals of (\ref{omom}) as follows, 
\bea
\label{DD}
\int _{\Sep } \om _I \wedge \oom_J \, \psi  = 
\int _{\Sigma _h } \om _I \wedge \oom_J \, \psi  
- \int _{\mD_a  } \om _I \wedge \oom_J \, \psi - \int _{\mD_b  } \om _I \wedge \oom_J \, \psi 
\eea
where $\mD_a$ and $\mD_b$ are the discs in $\Sigma _h$ whose boundaries are given by,
\bea
\p \mD_a = \mC_a \hskip 1in \p \mD_b = \mC_b
\eea
Since we are assuming that the behavior of $\psi$ near the punctures $p_a$ and $p_b$ is bounded by a finite power of a logarithm, the integrals over $\mD_a$ and $\mD_b$ are absolutely convergent, and we will show that they  are $\cO(e^{-2 \pi t})$ and thus negligible.

\subsubsection{The sizes of the discs $\mD_a$ and $\mD_b$}
\label{sec:3.5.1}

To show that the integrals over $\mD_a$ and $\mD_b$ in (\ref{DD}) are exponentially suppressed, we estimate the sizes of the discs $\mD_a$ and $\mD_b$, which may be derived from the definition of the boundary curves, $f(\mC_b)=-f(\mC_a)=2\pi t$, and the expression (\ref{fGreen}) for $f$ in terms of the Arakelov Green function $\GA_h$. The short-distance behavior of the Arakelov Green function is given by,
\bea
\label{gnear1}
\GA_h(z,w) = - \ln |z-w|^2 - \lambda  + \cO(z-w)
\eea 
for a $z,w$-independent $\lambda$. For arbitrary points $z_a \in \mC_a$ and $z_b \in \mC_b$, we have,
\bea
f(z_a) & = & -2 \pi t = \GA_h(z_a,p_b) - \GA_h(z_a,p_a)
\no \\
f(z_b) & = & +2 \pi t = \GA_h(z_b,p_b) - \GA_h(z_b,p_a)
\eea
As $t \to \infty$, these relations imply  $z_a \to p_a$ and $z_b \to p_b$, and we can apply the approximation of (\ref{gnear1}) to derive the following equations for $z_a,z_b$,
\bea
- \ln |z_a-p_a|^2 - \lambda - \GA_h(z_a,p_b) + \cO(z_a-p_a) = - 2 \pi t
\no \\
- \ln |z_b-p_b|^2 - \lambda - \GA_h(z_b,p_a) + \cO(z_b-p_b) = + 2 \pi t
\eea
Since $\GA_h(z_a,p_b)$ is smooth as $z_a \to p_a$ and $\GA_h(z_b,p_a)$ is smooth as $z_b \to p_b$, we may evaluate those Green functions at $z_a=p_a$ and $z_b=p_b$ up to terms of order $\cO(z_a-p_a)$ and $\cO(z_b-p_b)$ respectively. Thus, the curves $\mC_a$ and $\mC_b$ are approximately  coordinate-circles whose radii are exponentially small in $t$. We may parametrize the curves as follows,  
\bea
\label{CaCb}
z_a \in \mC_a & \hskip 0.4in &  |z_a-p_a|  = R 
\hskip 0.6in 
- 2 \ln R =  \, 2 \pi t + \lambda + \GA_h(p_a,p_b) 
\no \\
z_b \in \mC_b & \hskip 0.4in &  |z_b-p_b|  = R 
\eea
up to corrections which are exponentially suppressed  in $t$.  Because of this estimate, the correction terms $\cO(z_a-p_a) = \cO(z_b-p_b) = \cO(R)$ are exponentially suppressed in $t$ as well.

\subsubsection{Relating integrals on $\Sigma _{h+1}$ to integrals on $\Sigma _h$}
\label{sec:3.5.2}

We can now complete the argument for the exponential suppression of the integrals over $\mD_a$ and $\mD_b$ in (\ref{DD}). Since the integrand $\om _I \wedge \oom_J \, \psi$ is smooth, and the radius of the discs is exponentially suppressed, the integrals are  exponentially suppressed. The same holds true when one, but not both, of the indices $I$ or $J$ equals $h+1$. In that case the integrand has  simple  poles near the punctures, and hence is integrable, so that the argument still holds. In summary, we have whenever $(I,J) \not= (h+1, h+1)$,
\bea
\label{DDe}
\int _{\Sep } \om _I \wedge \oom_J \, \psi  = 
\int _{\Sigma _h } \om _I \wedge \oom_J \, \psi  
+ \cO(e^{-2\pi t})
\eea
We shall use this relation freely throughout the remainder of the paper. 

\sm

The remaining case is when $I=J=h+1$. Equivalently, we may consider the integral,
\bea
\label{omt}
\int _{\Sep } \om _t \wedge \oom_t \, \psi  = { 1 \over 4 \pi^2} \int _{\Sep } dz \wedge d\bar z\,  |\p_z f(z)|^2 \, \psi (z)
\eea   
where we have used formula (\ref{fzf}) to express $\om_t$ in terms of $f$. In this case we cannot use formula (\ref{DD}) because  the integrals of $\om _t \wedge \oom_t \, \psi$ over $\mD_a$ and $\mD_b$ will not  converge in general. The reason is that $\om_t$ has poles at $p_a$ and $p_b$, so that even though we are integrating over exponentially small discs $\mD_a$ and $\mD_b$, the integrals will grow with $t$ when $\psi$ is smooth at the punctures, or with powers of $t$ when $\psi$ grows as a power of a logarithm near the punctures. We wish to evaluate not merely the leading term in powers of $t$, but actually all power-behavior in $t$ exactly, up to exponentially suppressed contributions. To this end, we present a powerful method to do so in the next subsection.

\begin{figure}
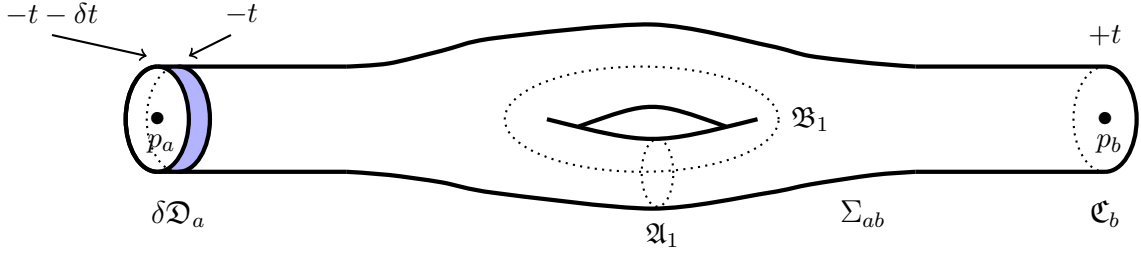

\begin{center}
\tikzpicture[scale=1.4]
\scope[xshift=-5cm,yshift=0cm]
\draw [ultra thick, fill, color=blue!30] (2,-1) arc (-90:270:0.3 and 0.5);
\draw [ultra thick, fill, color=white] (1.8,-1) arc (-90:270:0.3 and 0.5);
\draw [ultra thick] (2,-1) arc (-90:90:0.3 and 0.5);
\draw [ultra thick] (1.8,-1) arc (-90:90:0.3 and 0.5);
\draw [ultra thick] (1.8,0) arc (90:270:0.3 and 0.5);
\draw [thick, dotted] (2,0) arc (90:270:0.3 and 0.5);
\draw [ultra thick] (1.8,0) arc (90:270:0.3 and 0.5);

\draw [ultra thick] (1.8,0) -- (3.6,0);
\draw [ultra thick] (1.8,-1) -- (3.6,-1);
\draw[ultra thick] (5.5,-0.5) .. controls (6.5, -0.75) .. (7.5,-0.5);
\draw[ultra thick] (5.8,-0.57) .. controls (6.5, -0.32) .. (7.2,-0.57);
\draw [ultra thick] (10.8,-1) arc (-90:90:0.3 and 0.5);
\draw [thick, dotted] (10.8,0) arc (90:270:0.3 and 0.5);
\draw [ultra thick] (9,0) -- (10.8,0);
\draw [ultra thick] (9,-1) -- (10.8,-1);
\draw[ultra thick] plot [smooth] coordinates {(3.6,0) (4,0.03) (4.5, 0.15) (5, 0.27) (6.5,0.4)  (7.6, 0.2) (8, 0.10) (8.6,0.03) (9,0)};
\draw[ultra thick] plot [smooth] coordinates {(3.6,-1) (4,-1.03) (4.5, -1.10)  (5, -1.2) (6.5,-1.35) (7.6, -1.2) (8, -1.12) (8.6,-1.03) (9,-1)};
\draw [thick, dotted] (7.7,-0.5) arc (0:360:1.3 and 0.5);
\draw [thick, dotted] (6.7,-1.01) arc (0:360:0.15 and 0.32);
\draw (2,-1.4) node{$\delta \mD_a$};
\draw (0.8,0.5) node{\small $-t-\delta t$};
\draw[->, thick] (0.8,0.3) -- (1.7, 0.1);
\draw (2.6,0.5) node{\small $-t$};
\draw[->, thick] (2.5,0.3) -- (2.1, 0.1);
\draw (10.8,-1.4) node{$\mC_b$};
\draw (10.8,0.3) node{\small $+t$};
\draw (1.8,-0.5) node{$\bullet$};
\draw (10.8,-0.5) node{$\bullet$};
\draw (1.84,-0.73) node{{\small $p_a$}};
\draw (10.84,-0.73) node{\small $p_b$};
\draw (8,-0.5) node{{\small $\mB_1$}};
\draw (6.6,-1.6) node{{\small $\mA_1$}};
\draw (8.5,-1.4) node{{$\Sep$}};
\endscope
\endtikzpicture
\caption{The variational method evaluates the contribution from varying the boundary cycles of  $\Sep$ through a variation of $t$, here represented for the variation of the cycle $\mC_a$.}
\end{center}
\label{fig:7}
\end{figure}

\subsection{The variational method}

To extract the exact $t$-dependence of the integral over $\om_t \wedge \oom_t \psi$ in (\ref{DDe}), up to $t$-independent terms, and up to exponentially suppressed corrections, we use the fact that the integrand is independent of $t$, and that the only $t$-dependence of the integral arises from its integration domain $\Sep$. As a result, the variation of the integral, as $t$ is varied, localizes to an integral over a domain which is the union of the variations of the discs $\mD_a$ and $\mD_b$,
\bea
\delta \int _{\Sep } \om _t \wedge \oom_t \, \psi  = \int _{\delta \mD_a } \om _t \wedge \oom_t \, \psi
+ \int _{\delta \mD_b } \om _t \wedge \oom_t \, \psi
\eea
Given the parametrization of the boundary curves $\mC_a$ and $\mC_b$ in (\ref{CaCb}), we parametrize the infinitesimal variations of the discs as follows,
\bea
z_a \in \delta \mD_a & \hskip0.5in & z_a = p_a + R\, e^{ -s_a + i \theta _a}  
\hskip 1in 
0 \leq \theta _a, \theta _b \leq 2\pi
\no \\
z_b \in \delta \mD_b &  &  z_b = p_b + R \, e^{ -s_b + i \theta _b}   
\hskip 1in 
\, \, 0 \leq s_a, s_b <  \pi \delta t
\eea
where the ranges for $s_a$ and $s_b$ are obtained by varying $t$ in the expression for $R$  in (\ref{CaCb}). Since the integrations are now localized on exponentially small annuli, we may use the corresponding approximation to obtain the expressions for $\p_z f$ and $\om_t$ near the  punctures,
\bea
\label{derf}
 \p_z f (z) = { + 1 \over z-p_a}  + \cO(1)
 \hskip 1in 
 \p_z f (z) = { - 1 \over z-p_b}  + \cO(1)
\eea
The order $\cO(1)$ terms will produce exponentially suppressed contributions. Changing variables from $z_a$ to $s_a, \theta_a$ and from $z_b $ to $s_b, \theta _b$, the remaining integral is given as follows,
\bea
\delta \int _{\Sep } \om _t \wedge \oom_t \, \psi  & = &
{ - 2i \over 4 \pi^2} \int _0 ^{2 \pi} d \theta _a \int _0 ^{ \pi \delta t} d s_a \, \psi (p_a+ R \, e^{-s_a+i \theta _a}  )
\no \\ &&
+ { - 2i \over 4 \pi^2} \int _0 ^{2 \pi} d \theta _b \int _0 ^{ \pi \delta t} d s_b \, \psi (p_b+R \, e^{-s_b+i \theta _b}  ) + \cO(e^{-2 \pi t})
\eea
Taking the infinitesimal limit $\delta t \to 0$ allows us to compute the derivative of the integral which, after some minor simplifications, is given by, 
\bea
{ \p \over \p t}  \int _{\Sep } \om _t \wedge \oom_t \, \psi   = 
{ 1  \over 2 \pi i} \int _0 ^{2 \pi} d \theta  \Big ( \psi (p_a^\theta) +  \psi (p_b^\theta) \Big ) 
+ \cO(e^{-2 \pi t})
\eea
where we shall use the following abbreviations, with $R$ defined in (\ref{CaCb}),  throughout,
\bea
\label{papb}
p_a^\theta & = & p_a + R \, e^{i \theta} 
\no \\
p_b ^\theta & = & p_b + R \, e^{i \theta}
\eea
As announced earlier, this formula is valid for functions $\psi$ which may have a singularity at the punctures, as long as $\psi$ is regular elsewhere on $\Sigma _h$. 

\sm

To provide some examples, we apply the formula to the function $\psi=1$ and we find, 
\bea
{ \p \over \p t}  \int _{\Sep } \om _t \wedge \oom_t    = -2i
\eea
which may indeed be deduced from the Riemann bilinear relations $\int \om_t \wedge \oom_t = -2it$. Applying the formula to the function $\psi (z) = f(z)^n$ for an arbitrary integer $n \geq 0$, we use the fact that the integral evaluates the function on the curves $\mC_a$ and $\mC_b$ on which $f$ takes the respective constant values $- 2 \pi t$ and $2 \pi t$. As a result, the evaluation of the integral simply becomes,  
\bea
\label{fn}
{ \p \over \p t}  \int _{\Sep } \om _t \wedge \oom_t \, f^n   = -i \Big ( 1 + (-)^n \Big ) (2 \pi t)^n
\eea
In particular, the $t$-derivative of the integral vanishes for odd $n$.
We shall encounter this formula again in the subsequent section. Note that the function $\psi$ used in (\ref{fn}) does have non-trivial logarithmic singularities at the punctures.

\sm

We close this subsection by pointing out that the variational method introduced here is akin to the renormalization group methods of perturbative quantum field theory. Viewed from the vantage point of the surface $\Sigma_h$, the parameter $t$ plays the role of a short-distance cut-off on the integrals over Green functions which represent contributions to perturbative correlation functions in a scalar quantum field theory.
The field theory has derivative couplings, reflected in the derivatives of $f$ occurring in the differentials $\om_t$, as well as non-derivative couplings, reflected in the measure factors which do not involve $|\om_t|^2$. The variation in $t$ is the analogue of the variation in the cutoff of the renormalization group equations. More precisely, $t$ should be identified with $\ln \Lambda$ where $\Lambda$ is the usual quantum field theory cut-off with dimensions of mass,  while exponentially suppressed terms correspond to irrelevant contributions suppressed by negative powers of $\Lambda$.


\section{Degenerating the Green function and KZ-invariant}
\setcounter{equation}{0}
\label{sec4}

In this section, we shall use the parametrization of the surface $\Sigma _{h+1}$ near one of its non-separating nodes, developed in the preceding section, to evaluate the degeneration of the string Green function $G(x,y)$, the Arakelov Green function $\GA(x,y)$, and the Kawazumi-Zhang invariant $\varphi$. The non-separating degeneration must be considered with care as the limits involved are non-uniform across the degenerating surface. We consider the asymptotics where all the power contributions in $t$ are retained exactly, and all exponential contributions in $t$ are omitted.  Along the way, we shall gain more experience working with the  function $f$ and some of the calculational short-cuts it permits. Throughout, we shall denote the string Green functions by $G_{h+1}$ and $G_h$, the Arakelov Green functions by $\GA_{h+1}$ and $\GA_h$, and the Arakelov K\"aher forms by $\kappa _{h+1}$ and $\kappa _h$ on the surfaces $\Sigma _{h+1}$ and $\Sigma _h$, respectively.

\subsection{Degenerating Abelian differentials}

We shall analyze the non-separating degeneration of a compact surface $\Sigma _{h+1}$  when the cycle $\mB_{h+1}$ tends to infinity, keeping all other cycles fixed or bounded. Specifically, we shall send the $Sp(2h,\ZZ)$-invariant parameter $t$, defined in (\ref{t}), to infinity and study the result as a function of $t$, the moduli contained in the period matrix $\tau$ of  $\Sigma _h$, and the Abelian integral $v_I = \int _{p_a}^{p_b} \om _I$ for $I=1,\dots , h$. The Abelian differentials $\om_I$ on $\Sigma_{h+1}$ with $I=1,\cdots, h$ tend to the $h$ holomorphic Abelian differentials  on $\Sigma _h$ up to corrections which vanish exponentially in $t$ \cite{fay73}.  The remaining holomorphic Abelian differential $\om _{h+1}$ on $\Sigma_{h+1}$ equals $\om_{p_b, p_a}/(2 \pi i)$ up to exponentially suppressed corrections, or equivalently is represented by $\om_t$, 
\bea
\label{tom}
\om_t (z) = { i \over 2 \pi} dz \, \p_z f(z)
\eea
The function $f$ was expressed in terms of the Green function $\GA_h(x,y|\tau)$ in (\ref{fGreen}).

\subsection{Degenerating the string Green function $G_{h+1}$}

We begin by evaluation the degeneration of the Green function $G_{h+1}$ of (\ref{G}). Later on, we will use the result to evaluate the degeneration of the Arakelov Green function $\GA_{h+1}$ with the help of the relations (\ref{GAG}) and (\ref{gamma}), and of the Kawazumi-Zhang invariant. 

\sm

The first ingredient we need is the degeneration of the prime form on $\Sigma _{h+1}$ which simply tends to the prime form on $\Sigma _h$. This may be shown by choosing the odd spin structure $\nu$ in (\ref{E}) such that it restricts to an odd spin structure on $\Sigma _h$ and has vanishing characteristics on the degenerating cycle $\mB_{h+1}$, namely $\nu _{h+1}'=\nu _{h+1}''=0$. The holomorphic spinor $\eta _\nu(x)$ on $\Sigma _{h+1}$ then simply restricts to the holomorphic spinor on $\Sigma _h$, and the $\tet$-function similarly restricts to the $\tet$-function of one lower rank.

\sm

The second ingredient we need is the degeneration of the inverse of $Y$ in terms of $t$. Using the notation 
$\tau= \tau_1 + i \tau_2$  with $\tau_1, \tau_2$ real-valued $h \times h$ matrices, $v=v_1 + i v_2$  with $v_1, v_2$ real-valued $h \times 1$ matrices, and $\sigma = \sigma _1 + i \sigma _2$ where $\sigma_1, \sigma _2 \in \RR$, we have,
\bea
\label{invY}
\Omega = \left ( \begin{matrix} \tau & v \cr v^t & \sigma  \cr \end{matrix} \right )
\hskip 0.6in 
Y = \left ( \begin{matrix} \tau_2 & v_2 \cr v_2^t & \sigma _2 \cr \end{matrix} \right )
\hskip 0.6in 
Y^{-1}  = \left ( \begin{matrix} \tau_2^{-1}  & 0 \cr 0 & 0 \cr \end{matrix} \right )
+ { 1 \over t}  \left ( \begin{matrix} u_2 u_2^t  & -u_2 \cr -u_2^t & 1 \cr \end{matrix} \right )
\eea
where we use the subscript 2 to indicate the imaginary part, and set  $v_2= \tau_2 u_2$.

\sm

The third ingredient is the rearrangement, with the help of (\ref{invY}),  of the quadratic form involving Abelian integrals in the second  term on the right-hand side of (\ref{G}). Writing this term out more explicitly, we have,
\bea
2 \pi \, \sum _{I,J=1}^{h+1} \Im \! \int _y ^x \omega _I \, (Y^{-1})^{IJ} \, \Im \! \int _y ^x \omega _J & = &
2 \pi \, \sum _{I,J=1}^h \Im \! \int _y ^x \omega _I \, (Y^{-1})^{IJ} \, \Im \! \int _y ^x \omega _J
\no \\ &&
+ { 2 \pi \over t} \left [  \int _x ^y \left ( \om_{h+1} - \sum_{I=1}^h u_{2I} \om_I \right ) \right ]^2
\eea
The term on the right-hand side of the first line is precisely what is needed to produce the Green function $G_h(x,y|\tau)$.  The one-form under the integral on the second line is precisely $\om_t$, and the imaginary part of its Abelian integral may be expressed in terms of the function~$f$. The result is given as follows, 
\bea
\label{Gf}
G_{h+1}(x,y|\Omega) = G_h (x,y|\tau) + { 1 \over 8 \pi t} \big ( f(x)-f(y) \big )^2 + \cO(e^{-2 \pi t}) 
\eea 
One word of caution is in order here. This asymptotic evaluation is not uniform, and is valid only for points $x,y$ on $\Sigma _{h+1}$ which are fixed as $ t \to \infty$. In particular the points $x,y$ are not allowed to approach the punctures $p_a, p_b$.

\subsection{Degenerating the Arakelov metric}

The canonical K\"ahler form of (\ref{muh}) on $\Sigma _{h+1}$ will be denoted by $\kap_{h+1}$, and is given by,
\bea
\kap_{h+1} = { i \over 2(h+1)} \sum _{I,J=1}^{h+1} (Y^{-1})^{IJ} \om_I \wedge \oom_J
\eea  
Substituting the expression for $Y^{-1}$ from (\ref{invY}) and for the Abelian differentials, we obtain,
\bea
\kap_{h+1} = { i \over 2 (h+1)}  \sum _{I,J=1}^h (\tau_2 ^{-1})^{IJ} \om_I \wedge \oom_J
+ { i \over 2 (h+1) t} \left ( \om_{h+1} - u_2^t \, \om  \right ) \wedge \left ( \oom_{h+1} - u_2^t \, \oom \right ) 
\eea
Using (\ref{omt}), and the expression for the canonical K\"ahler form $\kap_h$ of (\ref{muh}) on $\Sigma _h$, the result may be arranged as follows,
\bea
\label{kapf}
\kap_{h+1} = { h \over h+1} \, \kap_h +{ i \over 2 (h+1)t} \, \om_t \wedge \oom_t
\eea
Note that the $h$-dependent normalization multiplying $\kap_h$ is due to the fact that both $\kap_{h+1}$ and $\kap_h$ are normalized to unit volume.

\subsection{Auxiliary integrals}

Before proceeding to the degeneration of the Arakelov Green function and Kawazumi-Zhang invariant, we shall derive here a number of useful results on the integration involving the form $\om_t$ and the  function $f$.
We represent the surface $\Sigma _{h+1}$ by the surface $\Sep$, depicted in figure 7, with the cycles $\mC_a$ and $\mC_b$ identified with one another. 

\subsubsection{Two simple checks}

Two integrals are well-known from the Riemann bilinear relations applied to (\ref{omt}), 
\bea
\label{RRL}
\int _\Sep \om_I \wedge \oom_t =0 \hskip 1in \int _\Sep \om _t \wedge \oom_t = - 2 i t
\eea
where $t$ was defined in (\ref{t}). It will be instructive to reproduce these results purely  in terms of our set-up. Using (\ref{tom}), the first integral may be rearranged as follows,
\bea
\int _\Sep \om_I \wedge \oom_t = - { 1 \over 2 \pi i} \int _\Sep  d \Big  ( f(z)  \om_I (z) \Big )
= - { 1 \over 2 \pi i} \oint _{\p \Sep}  f(z)  \om_I (z) 
\eea
where we have used Stokes's theorem to obtain the last equality. Next, we use the fact that the boundary of $\Sep$ is given by the union of curves $\p \Sep = \mC_a \cup \mC_b$, and taking the proper orientation conventions into account, we have, 
\bea
\label{bi}
\oint _{\p \Sigma _{ab}} \om =  - \oint _{\mC_a} \om  - \oint _{\mC_b} \om 
\eea
for an arbitrary 1-form $\omega$. We use the fact that $\mC_a$ and $\mC_b$ are level sets of the function $f$ to evaluate the function $f$ on the boundary components as follows,
\bea
\label{2pit}
2 \pi t = - f(\mC_a) = f(\mC_b) 
\eea
Substituting these results, we are left with contour integrals of $\om_I$ over $\mC_a$ and $\mC_b$, which vanish since $\om_I$ are holomorphic on $\Sigma _h$, thus recovering the first formula in (\ref{RRL}). An analogous calculation may be performed for the second integral in (\ref{RRL}), 
\bea
\int _\Sep \om _t \wedge \overline{\om _t} 
= { 1 \over 4 \pi^2 } \int _{\Sigma _{ab}} dz \p_z f (z) \wedge d f(z)
\eea
We use the differential form relation, 
\bea
dz \p_z f (z) \wedge d f(z) = - d \Big (f(z) dz \p_z f (z) \Big ) + f(z) d \Big (dz \p_z f (z) \Big )
\eea
Since the function $f$ is harmonic  in  $\Sigma _{ab}$ the second term on the right vanishes. The first term may be transformed into an   integral over the boundary $\p \Sep$ using Stokes's theorem, which in turn may be converted into a sum of two line-integrals using (\ref{bi}), and $f$ is constant on each line integral in view of (\ref{2pit}), so that we are left with, 
\bea
\label{4d16}
\int _\Sep \om _t \wedge \overline{\om _t} 
=
- { t \over 2 \pi } \oint _{\mC_a}  dz \p_z f (z)  +  { t \over 2 \pi } \oint _{\mC_b}  dz \p_z f (z) 
\eea
The differential $\p_z f (z)$ is holomorphic in $\Sep$ but on $\Sigma _h$ has simple poles at $p_a$ and $p_b$ with respective  residues $-1$ and $+1$ already given in (\ref{derf}). Evaluating the remaining integrals in (\ref{4d16}), we recover the second equation in (\ref{RRL}).

\subsubsection{Integrals of powers of $f$}

The method of calculation given above may be generalized to other integrals. The derivation is completely analogous to the one used in the preceding paragraphs, and will not be exhibited here. For arbitrary $n \geq 0$ we have,
\bea
\label{regs1}
\int _\Sep \om _I \wedge \overline{\om _t } \, \, f^n = 
\int _\Sep \om _t \wedge \overline{\om _t } \, f^{2n+1}  = 0
\eea
for all $I=1,\cdots, h$, as well as 
\bea
\label{regs2}
\int _\Sep \om _t \wedge \overline{\om _t } \, \, f^{2n} = - { 2 i \over 2n+1} (2 \pi)^{2n} t^{2 n+1}
\eea
The evaluation of all these integrals is exact. Note that the derivative with respect to $t$ of (\ref{regs2}) agrees with (\ref{fn}) obtained by using the variational method. While that method was not powerful enough to also determine the $t$-independent term, we have shown here that this constant term in fact vanishes.

\subsubsection{Integrals of functions which vary on $\mC_a$ and $\mC_b$}

Next, we consider the integral of   $\om_t \wedge \oom_t$ multiplied by a function  $\psi$ which is smooth on $\Sigma _{ab}$ but  may not be smooth at the punctures $p_a$ and $p_b$. Using the expression for $\om_t$ in  (\ref{tom}), and isolating an exact differential as in the preceding subsection, we have, 
\bea
\label{fi}
\int _\Sep \om _t  \wedge \overline{\om_t } \, \psi
& = &- { 1 \over 4 \pi^2} \int _{\Sigma_{ab}}  \Big ( d  (  f  dz \p_z f  \, \psi  ) + \half  dz \p_z f^2 \wedge d\bar z \pbz \psi \Big )
\\
& = & - { t \over 2 \pi} \oint _{\mC_a}    dz \p_z f  \, \psi  
+ { t \over 2 \pi} \oint _{\mC_b}    dz \p_z f  \, \psi  
+ {1 \over 8 \pi^2}  \int _\Sep dz \wedge d\bar z \,  f(z)^2 \p_z \pbz \psi 
\no
\eea
where we have used the fact that $f(z)$ is constant on each contour and given by (\ref{2pit}).  The contour  integrals require evaluating $\psi$  on the curves $\mC_a$ and $\mC_b$, and the function $\psi$ will be assumed to be continuous on these curves for all $t$. 

\sm

To compute these integrals, we use the results of subsection \ref{sec:3.5.1} and, in particular, the parametrization of the curves $\mC_a$ and $\mC_b$ of (\ref{CaCb}). Given that the contours $\mC_a$ and $\mC_b$ are exponentially small in the coordinate system $z$, we may use the approximation of (\ref{derf}) for $\p_z f$ and parametrize the points $z$ on the curve $\mC_a$ by $ z=p_a^{\theta_1}$ and on the curve $\mC_b$ by $ z=p_b^{\theta_2 }$ for $\theta _1, \theta _2 \in [0, 2 \pi]$, where $p_a^\theta$ and $p_b^\theta$ were defined in (\ref{papb}). The result is as follows, 
\bea
\label{fi1}
\int _\Sep \om _t  \wedge \overline{\om_t } \, \psi
& = & - { it \over 2 \pi} \int _0 ^{ 2 \pi} d \theta \Big ( \psi \left (p_b^\theta  \right ) 
+ \psi \left (p_a^\theta  \right )  \Big ) 
\no \\ &&
+ {1 \over 8 \pi^2}  \int _\Sep dz \wedge d\bar z \,  f(z)^2 \p_z \pbz \psi 
\eea
An immediate generalization  is obtained by including further factors of powers of $f$,
\bea
\label{fi2}
\int _\Sep \om _t  \wedge \overline{\om_t } \, f^n \psi
& = & -i \,  {  (2 \pi)^{n-1} t^{n+1}  \over n+1} \int _0 ^{ 2 \pi} d \theta 
\Big (  \psi \left (p_b^\theta  \right )  + (-)^n \psi \left (p_a^\theta  \right )  \Big ) 
\no \\ &&
+ {1 \over 4 \pi^2(n+1)(n+2) }  \int _\Sep dz \wedge d\bar z \,  f(z)^{n+2} \, \p_z \pbz \psi 
\eea
One readily verifies that (\ref{fi2}) reduces to (\ref{fi1}) for $n=0$ and to (\ref{regs1}) and  (\ref{regs2}) for $\psi=1$.

\sm

In the special case when  the function $\psi$ is smooth  at the punctures, the contour integrals simply pick up the value of $\psi$ at $p_a$ and $p_b$, and the above integral reduces to, 
\bea
\label{fi3}
\int _\Sep \om _t  \wedge \overline{\om_t } \, f^n \psi
& = & -i \,  {  (2 \pi)^n t^{n+1}  \over n+1} 
\Big ( \psi (p_b)  + (-)^n \psi (p_a) \Big )
\no \\ &&
+ {1 \over 4 \pi^2(n+1)(n+2) }  \int _\Sep dz \wedge d\bar z \,  f(z)^{n+2} \, \p_z \pbz \psi 
\eea
The order of approximation is governed by the Taylor expansion of $\psi$ around the punctures, and the corrections to this formula are therefore suppressed by factors of $R$.

\subsection{Degenerating the Arakelov Green function}

In this subsection, we shall obtain the non-separating degeneration limit of the Arakelov Green function $\GA_{h+1}$ on the surface $\Sigma _{h+1}$ upon degenerating the cycle $\mB_{h+1}$. 

\vskip 0.2in

\noindent
{\bf Theorem 1} ~ {\sl The non-separating degeneration of the Arakelov Green function $\GA_{h+1}(x,y)$, for fixed points $x,y$ as the cycle $\mB_{h+1}$ degenerates,  is given by}
\bea
\label{thm1}
\GA_{h+1}(x,y) & = & 
{ \pi t \over 3 (h+1)^2} + \GA_h(x,y) + { 1 \over (h+1)^2} \GA_h(p_a,p_b) 
\\ &&
-{ 1 \over 2(h+1)} \Big ( \GA_h(x,p_a) +  \GA_h(x,p_b) + \GA_h(y,p_a) +  \GA_h(y,p_b)  \Big )
\no \\ &&
+ { f(x)^2 + f(y)^2 \over 8 \pi (h+1) t} - { f(x) f(y) \over 4 \pi t} - { h \over 4 \pi (h+1)^2 t} \int _{\Sigma _h} \kap_h (z) f(z)^2 + \cO(e^{-2 \pi t}) 
\no \eea
{\sl where $\GA_h$ and $\kap_h$ are respectively the Arakelov Green function and canonical K\"ahler form on the compact surface $\Sigma _h$, and the function $f$ is given by $f(x) = \GA_h(x,p_b)-\GA_h(x,p_a)$. The integral on the last line is absolutely convergent.}

\sm

Comparing the statement of Theorem 1 with Theorem 7.2 in \cite{MR1105425}, we observe complete agreement of the terms of orders $t^1$ and $ t^0$ provided we identify our parameter $t$ and the parameter $|\tau|$ of \cite{MR1105425} by the following relation, which follows from Lemma 7.3 and 7.5 in \cite{MR1105425},
\bea
\label{ttau}
- \ln |\tau| = 2 \pi t + \GA_h(p_a,p_b)  + \cO(t^{-1})
\eea
Theorem 1 further gives the $\cO(t^{-1})$ contribution, which was not evaluated in \cite{MR1105425}, but its  real power  lies in the much stronger result that the entire $t$-dependence is given by a Laurent polynomial in $t$ of degree (1,1), up to exponentially suppressed terms. Finally, we note that the integral which enters (\ref{thm1}),
\bea
\int _{\Sigma _h} \kap_h (z) f(z)^2 = \int _{\Sigma _h} \kap_h 
\Big ( \GA_h(z,p_a)^2 + \GA_h(z,p_b)^2- 2 \GA_h(z,p_a) \GA_h(z,p_b) \Big ) 
\eea
produces modular graph functions on the surface $\Sigma _h$ respectively with one puncture  from the first two terms on the right-hand side, and two punctures from the third term.

\subsubsection{Proof of Theorem 1}

To prove Theorem 1, we use the relation between the Green functions $G$ and $\GA$ given in (\ref{GAG}), as well as the degeneration result of the string Green function $G_{h+1}$ obtained in (\ref{Gf}), and of the canonical K\"ahler form $\kap _{h+1}$ obtained  in (\ref{kapf}). Furthermore, we use the fact that $\GA(x,y)$ is unchanged by letting $G(x,y) \to G(x,y) + c(x)+c(y)$ in (\ref{GAG}) for an arbitrary function $c(x)$. Thus, the term proportional to $f(x)^2+f(y)^2$ in (\ref{Gf}) cancels out. The integration of the remaining term proportional to $f(x)f(y)$ integrates to zero against $\kap_{h+1}$ in view of the normalization of $f$ adopted in (\ref{fGreen}). As a result, we find, 
\bea
\GA_{h+1}(x,y) & = & G_h(x,y) - { 1 \over 4 \pi t} f(x) f(y) 
- \int _\Sep \kap_{h+1} (z) \Big ( G_h (x,z) + G_h(y,z) \Big )
\no \\ &&
+ \int _\Sep \kap_{h+1} (z) \int _\Sep \kap_{h+1} (w) \, G_h(z,w)
\eea
Substituting the expression for $G_h$ in terms of $\GA_h$, the function $\gamma_h$ and the constant $\gamma^1_h$, we find that all dependence on $\gamma_h$ and $\gamma^1_h$ cancels. Using next the expression for $\kap_{h+1}$ in terms of $\kap_h$ and $\om_t \wedge \oom_t$ of (\ref{kapf}), and the normalization of the Arakelov Green function $\int _{\Sigma _h} \kap_h (z) \GA_h(x,z)=0$, the expression simplifies further and we obtain, 
\bea
\GA_{h+1}(x,y) =  \GA_h(x,y) - { 1 \over 4 \pi t} f(x) f(y) - \tilde \gamma _h (x) - \tilde \gamma _h (y) + \tilde \gamma^1_h
\eea
where 
\bea
\label{tilgam}
\tilde \gamma _h(x) & = &  { i \over 2 (h+1)t} \int _\Sep \om_t \wedge \oom_t (z) \, \GA_h (x,z) 
\no \\ 
\tilde \gamma _h ^1 & = &  -{ 1 \over 4 (h+1)^2 t^2} \int _\Sep \om_t \wedge \oom_t (z)  \int _\Sep \om_t \wedge \oom_t (w) \, \GA_h(z,w)
\eea
The integral in $\tilde \gamma _h (x)$, for fixed point $x$, is evaluated using (\ref{fi3}), and we find,
\bea
\tilde \gamma _h (x) =
 { 1 \over 2(h+1)} \Big ( \GA_h(x,p_a) + \GA_h(x,p_b) \Big )
- { f(x)^2 \over 8 \pi (h+1)t}
 +{ 1 \over 8 \pi (h+1) t} \int _{\Sigma _h} \kap_h (z) f(z)^2
\eea
The integrals in $\tilde \gamma _h^1$ must be evaluated with the full expression of (\ref{fi2}), as both integration points $z,w$ may run over the funnel. Thus we use,
\bea
\int _{\Sep}\om_t \wedge \oom_t (w) \GA_h(z,w) & = & 
- { it \over 2 \pi} \int _0 ^{2 \pi} d \theta \Big ( \GA_h(z, p_a^\theta) + \GA_h(z, p_b^\theta) \Big )
\no \\ &&
+ { i \over 4 \pi} f(z)^2 -{ i \over 4 \pi } \int_{\Sigma _h}  \kap_h (w)f(w)^2 
\eea
Integration over $z$ of the last line produces, 
\bea
\int _{\Sep} \om_t \wedge \oom_t (z) \left (  { i \over 4 \pi} f(z)^2 -{ i \over 4 \pi } \int_{\Sigma _h} \kap_h (w)f(w)^2  \right ) 
= { 2 \pi \over 3} t ^3 - { t \over 2 \pi} \int _{\Sigma _h}\kap_h (w)f(w)^2
\eea
while integration over $z$ of the first line may again be carried out with the help of (\ref{fi2}), and we find after some simplifications, 
\bea
\tilde \gamma^1_h 
& = & 
-{ 2 \pi  t \over 3 (h+1)^2 } + { 1 \over 4 \pi (h+1)^2 t} \int_{\Sigma _h}  \kap_h (w)f(w)^2
 + { \GA_h(p_a,p_b) \over  2 (h+1)^2} 
\no \\ &&
 +{ 1 \over 8 \pi^2 (h+1)^2} \int _0 ^{2 \pi } d \theta_1 \int _0 ^{2 \pi } d \theta_2 \,  \GA_h(p_a^{\theta_1}  , p_a^{\theta_2}) 
\eea
To evaluate the integral on the last line, we use the fact that both arguments of $\GA_h$ are points in the funnel, so that we may use the approximation of (\ref{gnear1}). The function $\GA_h$ then involves only the difference of the angles, so that one of the integrations may be performed trivially. The remaining integral evaluates to,
\bea
\int _0 ^{2 \pi } \!\! d \theta_1 \int _0 ^{2 \pi } \!\! d \theta_2 \,  \GA_h(p_a^{ \theta_1 } , p_a^{ \theta _2}) 
& = & 
2 \pi \int _0 ^{2 \pi } \!\! d \theta_2  \,  \Big ( - \ln |1- e^{ i \theta_2} |^2  + 2 \pi t + \GA_h(p_a,p_b) \Big )
\no \\ & = &
(2 \pi )^2 (2 \pi t + \GA_h(p_a,p_b))
\eea
In going from the second to the third line, we use the fact that  the remaining integral of the $- \ln | \cdots |^2$ term gives zero. In summary we then have, 
\bea
\tilde \gamma^1 _h  =  {  \pi  t \over 3 (h+1)^2 } + { 1 \over 4 \pi (h+1)^2 t} \int_{\Sigma _h} \kap_h (w)f(w)^2
 + { 1 \over  (h+1)^2} \, \GA_h(p_a,p_b)
 \eea
Assembling the contribution of $\tilde \gamma _h(x)+ \tilde \gamma _h (y)$ and $\tilde \gamma _h^1$, we obtain the formula of Theorem 1.

\subsection{Degenerating the Kawazumi-Zhang invariant}

In this subsection, we shall use the previous results  to compute the non-separating degeneration of the Kawazumi-Zhang invariant for arbitrary genus. It will be convenient to recast the definition of the KZ-invariant of (\ref{KZ}) in terms of the following 2-form,
\bea
\nu_h (x,y) = (Y^{-1})^{IJ} \om_I(x) \wedge \oom_J(y)
\eea
Using the notation $ |\nu_h(x,y)|^2 =  \nu_h(x,y) \wedge \overline{\nu}_h (x,y)$, the KZ-invariant $\f_h$ for a surface of genus $h$ may be written as follows,
\bea
\label{KZ2}
\f_h  = - { 1 \over 4}  \int _{\Sigma _h^2}  |\nu_h(x,y)|^2  \GA_h(x,y) 
\eea

\vskip 0.05in

\noindent
{\bf Theorem 2} {\sl  The non-separating degeneration of the Kawazumi-Zhang invariant $\f_{h+1}$,  as the cycle $\mB_{h+1}$ degenerates,  is given by}
\bea
\label{thm2}
\f_{h+1} & = &{ h \pi t \over 3 (h+1)} +  \f_h + { h \over h+1} \GA_h (p_a,p_b) 
+ { h (2h+3) \over 4 \pi (h+1) t} \int _{\Sigma _h} \kap_h(x) f(x)^2 
\no \\ &&
+ { 3 \over 16 \pi t} \int _{\Sigma _h^2} |\nu _h (x,y)|^2  f(x) f(y) + \cO(e^{-2 \pi t})
\eea
{\sl where the remaining integrals are absolutely convergent on $\Sigma _h$.}

\sm

Comparing the statement of Theorem 2 with the results of the second part of Theorem A  in  \cite{zbMATH06355718}, we observe agreement  of the terms of orders $t^1$ and $ t^0$ for all genera provided we identify our parameter $t$ and the parameter $|\tau|$ as we have already done for the Arakelov Green function in (\ref{ttau}), and take into account the differences in notation and factor of $-1/2$ in the normalization for the Arakelov Green function. Furthermore, for genus two (namely for $h=1$), Theorem 2 also  agrees with  formulas (9), (16), and (21) in \cite{Pioline:2015qha}, which were obtained from the theta lift representation. Note that $\varphi_1$  vanishes in this case, as does the  integral on the second line of (\ref{thm2}).  

\sm

Just as for the case of the Arakelov Green function in Theorem 1, we observe that Theorem 2  gives the $\cO(t^{-1})$ contribution, which was not evaluated in~\cite{zbMATH06355718},  but its real power again lies in the fact that the entire $t$-dependence is given by a Laurent polynomial in~$t$ of degree (1,1), up to exponentially suppressed terms. The integrals remaining on the right-hand side of (\ref{thm2}) are absolutely convergent and produce genus-$h$ modular graph functions and their generalizations for a punctured Riemann surface of genus $h$. 

\subsubsection{Proof of Theorem 2}

The key ingredients in the proof are the results of Theorem 1 for the degeneration of the Arakelov Green function $\GA_{h+1}$ in (\ref{thm1}), and the decomposition of the volume form $ |\nu_{h+1}(x,y)|^2$, which is given by,
\bea
\label{nu}
 |\nu_{h+1}(x,y)|^2 & = &  |\nu_h(x,y)|^2  +{ 1 \over t^2} \om_t (x) \wedge \oom_t(x) \wedge \om_t(y) \wedge \oom_t(y)
\no \\ &&
 + {1 \over t} \oom_t (x) \wedge \om_t (y) \wedge \nu_h (x,y) 
 +{ 1 \over t} \om_t(x) \wedge \oom_t(y) \wedge \overline{\nu}_h (x,y)
 \eea
along with a useful result for integrating  a function which is independent of $x$ (or $y$),
\bea
- { 1 \over 4} \int _{\Sigma _{h+1}^2}   |\nu_{h+1}(x,y)|^2 \psi (y) 
= (h+1) \int _{\Sigma _{h+1}} \kap_{h+1} (y) \psi (y)
\eea 
This last result readily allows us to integrate the corresponding terms when substituting $\GA_{h+1}(x,y)$ from (\ref{thm1}) into the formula for $\f_{h+1}$,
\bea
\f_{h+1}  = - { 1 \over 4}  \int _{\Sigma _{h+1}^2} |\nu_{h+1} (x,y)|^2  \GA_{h+1}(x,y) 
\eea
and we find,
\bea
\label{fi5}
\f_{h+1} & = & - { 1 \over 4}  \int _{\Sigma_{ab} ^2} |\nu_{h+1} (x,y)|^2 \left ( \GA_h(x,y) - { f(x) f(y) \over 4 \pi t} \right )
\no \\ &&
- {  \pi t \over 3 (h+1)}
 - { \GA_h(p_a,p_b) \over h+1}
-   { 1 \over 4 \pi (h+1)t} \int_{\Sigma _h} \kap_h f^2
\eea
Here, we have used (\ref{kapf}) and (\ref{regs2}) for $n=1$ to obtain the following relation, 
\bea
\int_{\Sep} \kap_{h+1} f^2 =  { 4 \pi^2 t^2 \over 3 (h+1)} + { h \over h+1} \int _{\Sigma _h} \kap_h f^2 
\eea
and we have evaluated the integral on the second line of (\ref{fi5}), as follows,
\bea
 \int _\Sep \kap_{h+1} (x)  \GA_h(x,p_a) = { \pi t \over 2(h+1)} + { \GA_h(p_a,p_b) \over h+1} 
 + { 1 \over 8 \pi (h+1)t} \int _{\Sigma _h} \kap_h f^2
 \eea
 with an analogous formula for the integral of $\GA_h(x,p_b)$.
Both these results have been used to simplify the right-hand side of (\ref{fi5}). 

\sm

It remains to carry out the integrals in the first line of (\ref{fi5}), which we do using (\ref{nu}). The term proportional to $f(x)f(y)$ integrates to zero against the last three terms on the right-hand side  in (\ref{nu}). The integral of the first term in (\ref{nu}) against $\GA_h$ produces the KZ-invariant at genus $h$. The integral of the second term on the right-hand side of (\ref{nu}) is proportional to the quantity $\tilde \gamma^1_h$ which was already evaluated in (\ref{tilgam}), while the remaining integrals produced by the last two terms on the right-hand side of (\ref{nu}) are convergent integrals on $\Sigma _h$. Finally, these last integrals may be simplified as follows,
\bea
\int _{\Sigma_h ^2} \oom_t (x) \om_t(y)  \nu_h (x,y) \GA_h(x,y)
& = & 
{ 1 \over 4 \pi^2} \int _{\Sigma_h ^2} d \bar x  dy \pbx f(x)  \p_y f(y) \nu_h (x,y) \GA_h(x,y)
\no \\
& = & 
{ 1 \over 4 \pi^2} \int _{\Sigma_h ^2} d \bar x   dy  f(x)  f(y) \nu_h (x,y) \pbx \p_y \GA_h(x,y)
\eea
In going from the first line to the last we have integrated by parts in both $\bar x$ and $y$. This operation commutes through $\nu_h(x,y)$ since this quantity is holomorphic in $x$ and anti-holomorphic in $y$, and produces no boundary terms. We may now use the following mixed partial derivative relation for the Arakelov Green function, 
\bea
\pbx \p_y \GA_h(x,y) = \pi \delta ^{(2)}(x,y) - \pi (\overline {\nu}_h )_{\bar x y}(x,y)
\eea
where $\overline{\nu}_h (x,y) = { i \over 2} (\overline {\nu}_h )_{\bar x y}(x,y) d \bar x dy$. 
After some simplifications, we find, 
\bea
\int _{\Sigma _h^2} \oom_t (x) \om_t(y)  \nu_h (x,y) \GA_h(x,y)
=  
- { h \over \pi} \int_{\Sigma _h} \kappa _h f^2 - { 1 \over 4 \pi} \int _{\Sigma_h ^2} |\nu_h (x,y)|^2 f(x) f(y)
\eea
Collecting all contributions, we find (\ref{thm2}).


\section{Degenerating the genus-two string amplitude}
\setcounter{equation}{0}
\label{sec5}

In this section, we shall prove a general theorem on the structure of the $t$-dependence of the genus-two amplitude to arbitrary weight $w$. The starting point is the partial  amplitude, integrated over vertex  points, but unintegrated over the moduli of the underlying compact  surface, which we recall from (\ref{B2g}) along with (\ref{BY}), (\ref{Delta}) and (\ref{stu})\footnote{In this section, we set $h=1$, denote $\Sigma _{h+1}$ by $\Sigma$ and omit the corresponding subscripts on $G, \cG$ and $\kappa$.}, 
\bea
\cB^{(2)}   (s_{ij} |\Omega) = 
{ 1 \over 16} \int _{\Sigma ^4} { \cY \wedge \overline{ \cY} \over (\det Y)^2} 
\exp \left \{ \sum _{i<j} s_{ij} G(z_i,z_j|\Omega ) \right \}
\eea
The scalar Green function $G$ of (\ref{G}) may equivalently be replaced by the Arakelov Green function $\GA$ using (\ref{GAG}) and the momentum conservation relation $\sum_j s_{ij}=0$ for all $i$. 

\sm

Given $\Omega$, the integral over $\Sigma ^4$ is absolutely convergent whenever $|s_{ij} |<1$ for all $i,j=1,2,3,4$, so that $\mB^{(2)}   (s_{ij} |\Omega)$ admits a convergent Taylor series expansion in powers of $s_{ij}$.  Instead of using the expansion given earlier in (\ref{Bexp}), it will be more convenient for our purpose to expand in powers of $s_{ij}$ and arrange the terms according to the {\sl weight} $w$, 
\bea
\cB^{(2)}  (s_{ij} |\Omega) = \sum_{w=0}^\infty  \cB_w (s_{ij} |\Omega)
\eea
where  $\cB_w (s_{ij} |\Omega)$  collects all terms which are homogeneous of total degree $w$ in $G$,
\bea
\cB_w (s_{ij} |\Omega)
=
{ 1 \over 16 \, w!} \int _{\Sigma ^4} { \cY \wedge \overline{ \cY} \over (\det Y)^2} 
\left ( \sum _{i<j} s_{ij} G(z_i,z_j|\Omega) \right )^w
\eea
Manifestly, $\cB_w (s_{ij} |\Omega)$ is a homogeneous symmetric polynomial in the invariants $s_{ij}$ of total degree of homogeneity $w+2$. We shall now prove the following theorem on the structure of its expansion in powers of $t$ in the non-separating degeneration near the cusp $t \to \infty$.

\subsection{ Theorem 3}

{\sl The genus-two modular graph function $\cB_w (s_{ij} |\Omega)$ has the following asymptotic behavior under non-separating degeneration of the genus-two surface $\Sigma$,
\bea
\cB_w (s_{ij} |\Omega) = \sum _{k=-w}^w \mb ^{(k)} _w (s_{ij} | v ,\tau) \, (\pi t)^k + \cO(e^{-2 \pi t})
\eea
where the period matrix $\Omega$, its imaginary part $Y$, and the variable $t$ are parametrized by,
\bea
\Omega = \left ( \begin{matrix}\tau & v \cr v & \sigma \cr \end{matrix} \right )
\hskip 0.6in 
Y = \left ( \begin{matrix} \tau_2 & v_2 \cr v_2 & \sigma_2 \cr \end{matrix} \right )
\hskip 0.6in 
t = { \det Y \over \tau_2} = \sigma _2 - { v_2^2 \over \tau_2}
\eea
The coefficients $\mb ^{(k)} _w (s_{ij}| v ,\tau)$ are generalized modular graph functions, which  are doubly periodic in $v=p_b-p_a$, i.e. they are invariant under $v\to v+\ZZ+\tau\ZZ$ and satisfy,
\bea
\tau \to \tau' = { a \tau + b \over c \tau + d} 
\hskip 0.6in 
v \to v' = { v \over c \tau +d}
\hskip 0.6in 
\mb ^{(k)} _w (s_{ij} | v' ,\tau') = \mb ^{(k)} _w (s_{ij} | v ,\tau)
\eea
All $\sigma_1$-dependent terms  are  exponentially suppressed in $t$ so that the coefficients $\mb^{(k)}_w(s_{ij} | v,\tau)$  are independent of $\sigma_1$, as the notation indeed indicates.  }

\sm

The remainder of this section is devoted to proving this theorem.

\subsection{Parametrization by the surface $\Sigma _{ab}$ and the function $f$}

Considering (\ref{fGreen}) for $h=1$,  we define the function $f(z)= g(z-p_b|\tau) - g(z-p_a|\tau)$ on the surface $\Sep$, constructed from a genus-one surface with two punctures $p_a, p_b$.  We begin by proving the following lemma. 

\vskip 0.2in

\noindent
{\bf Lemma  1} ~
{\sl In terms of the function $f$, the volume-form $\cY \wedge \bar \cY$ on $\Sigma ^4$  is given by,
\bea
{ \cY \wedge \bar \cY \over 16 \, (\det Y)^2} =  { \tau_2^2 \over16 \pi^4 t^2} 
\left | \sum _{i<j} s_{ij} \, \p_{z_i} f (z_i) \, \p_{z_j} f(z_j) \right |^2 \prod _{i=1}^4 \kappa (z_i) 
\eea
This relation is exact, up to exponentially suppressed terms.}

\sm

To prove Lemma 1, we express $\Delta$ of (\ref{Delta}) in terms of the Abelian differential $\om_t$,
and then express $\om_t$ in terms of the function $f$ by (\ref{tom}), so that we have,
\bea
\Delta (x,y) = {  dx \wedge dy \over 2 \pi i}   \Big ( \p_x f(x) - \p_y f(y) \Big )
\eea
Substituting this expression in $\cY$, and rearranging terms, we find, 
\bea
\cY = { dz_1 \wedge dz_2 \wedge dz_3 \wedge dz_4 \over 4 \pi^2} \sum _{i<j} s_{ij} \, \p_{z_i} f(z_i) \, \p_{z_j} f(z_j)
\eea
The lemma follows from this result, using also $dz_i \wedge d \bar z_i = -2i \tau_2 \kappa (z_i)$ and $\det Y= \tau_2 t$.

\sm

Next, we make use of the uniform asymptotic expansion of the genus-two Green function,
obtained in Theorem 1 in (\ref{thm1}),
\bea
G(x,y|\Omega) = g( x-y|\tau ) + { 1 \over 8 \pi t} \Big (f(x)-f(y) \Big )^2 + \cO(e^{-2 \pi t})
\eea
and momentum conservation $\sum _i s_{ij}=0$ to write the expression for $\cB^{(2)} $ as follows,
\bea
\label{Bww}
\cB_w(s_{ij}|\Omega) = \sum _{w_f, w_g=0}^w {\delta _{w_f+w_g, w}  \over w_f! ~  w_g ! }
\left ( - { 1 \over 4 \pi t} \right )^{w_f+2} \cB_{w_f, w_g} (s_{ij} |\Omega)
+ \cO(e^{-2 \pi t})
\eea
Throughout we shall use the following notation,
\bea
\int _{\Sigma _{ab} ^4} \kappa = \prod _{i=1}^4 \int _\Sep \kappa (z_i) 
\eea
and $\cB_{w_f, w_g} (s_{ij} |\Omega)$ is given by,
\bea
\cB_{w_f, w_g} (s_{ij} |\Omega) & = & 
{ \tau_2^2 \over\pi^2} \int _{\Sigma _{ab} ^4} \kappa \, 
\left | \sum _{i<j} s_{ij} \, \p_{z_i} f (z_i) \, \p_{z_j} f(z_j) \right |^2
\left (  \sum _{i<j} s_{ij} \, f(z_i) f(z_j) \right )^{w_f}
\no \\ && \hskip 0.5in
\times  
\left ( \sum_{i<j} s_{ij} \, g(z_i-z_j) \right )^{w_g} 
\eea
To extract the $t$-dependence of $\cB_{w_f, w_g} (s_{ij} |\Omega)$, we decompose the integrand as follows, 
\bea
\label{Sk}
\Big | \sum _{i<j} s_{ij} \, \p_{z_i} f (z_i) \, \p_{z_j} f(z_j) \Big |^2
= \sum_{k=0}^2 \cS^{(k)} 
\eea
where the terms are arranged according to the number of points $z_i$ that are shared between the meromorphic factor and its complex conjugate. Explicitly, we have,
\bea
\cS^{(0)} & = &  \sum_{i<j} \, \sum _{k \not= i,j} \,  \sum _{\ell \not = i,j,k} s_{ij}^2 (\p_i f)( \p_j f) (\bar \p_k f ) (\bar \p_\ell f)
\no \\
\cS^{(1)} & = &  \half \sum_{i=1}^4 \, \sum _{j \not = i} \, \sum _{k \not= i,j} s_{ij} s_{ik} |\p_i f|^2 
\Big ( \p_j f \, \bar \p_k f + \bar \p_j f \,  \p_k f \Big )
\no \\
\cS^{(2)} & = &  \sum _{i<j} s_{ij}^2 \,  |\p_i f|^2 \, |\p_j f|^2
\eea
Here, and below,  we use the abbreviation $\p_i f = \p_{z_i} f(z_i)$ and analogously for its complex conjugate. Using this notation, we split $\cB_{w_f, w_g} (s_{ij} |\Omega)$ accordingly,
\bea
\label{Bk}
\cB_{w_f, w_g} (s_{ij} |\Omega) = \sum _{k=0}^2 \cB_{w_f, w_g} ^{(k)} (s_{ij} |\Omega) 
\eea
where,
\bea
\cB_{w_f, w_g} ^{(k)} (s_{ij} |\Omega)  =  
{ \tau_2^2 \over\pi^2} \int _{\Sigma _{ab} ^4} \kappa   \, \cS^{(k)} 
\left (  \sum _{i<j} s_{ij} \, f(z_i) f(z_j) \right )^{w_f}
\left ( \sum_{i<j} s_{ij} \, g(z_i-z_j|\tau) \right )^{w_g} 
\eea
This formula will be the starting point for the evaluation of the asymptotics as $t \to \infty$.

\subsection{The asymptotics of $\cB_{w_f, w_g} ^{(k)} (s_{ij} |\Omega)$ as $t \to \infty$}

To prove Theorem 3, we use the results of the following Lemma. 

\vskip 0.2in

\noindent
{\bf Lemma  2} ~ {\sl For $w=w_f+w_g$ with $w_f, w_g \geq 0$, and up to contributions which are exponentially suppressed as a function of $t$, the functions $\cB_{w_f, w_g} ^{(k)} (s_{ij} |\Omega) $ satisfy the following properties.
\begin{description}
\itemsep=0in
\item {\rm (a)} We have the following vanishing results for $w_g \leq 1$, 
\bea
\cB_{w, 0} ^{(0)} (s_{ij} |\Omega) = \cB_{w-1, 1} ^{(0)} (s_{ij} |\Omega) = \cB_{w, 0} ^{(1)} (s_{ij} |\Omega) =0
\eea
\item {\rm (b)} for $w_g \geq 2$ the functions $ \cB_{w_f, w_g} ^{(0)} (s_{ij} |\Omega) $ are constant in  $t$;
\item {\rm (c)}  for $w _g \geq 1$ the functions $ \cB_{w_f, w_g} ^{(1)} (s_{ij} |\Omega) $  are polynomial in $t$ of degree  $w_f+1$;
\item {\rm (d)} for $w_g \geq 0$ the functions  $ \cB_{w_f, w_g} ^{(2)} (s_{ij} |\Omega) $ are polynomial in $t$ of degree $w+w_f+2$;
\item {\rm (e)} the polynomial $ \cB_{w, 0} ^{(2)} (s_{ij} |\Omega) $ has vanishing terms of orders $t^0$ and $t^1$;
\item {\rm (f)} the polynomial $ \cB_{w-1, 1} ^{(2)} (s_{ij} |\Omega) $ has vanishing term of order $t^0$;
\item {\rm (g)} the polynomial $ \cB_{w-1, 1} ^{(1)} (s_{ij} |\Omega) $ has vanishing term of order $t^0$.
\end{description}
}

The proof of this lemma heavily relies on the variational method, which was introduced  in subsection 3.6. The proof is lengthy and technical, and will be given in  Appendix A.

\subsection{Proof of  Theorem 3}

To prove Theorem 3 we recollect the general formula for $\cB_w(s_{ij}|\Omega)$ of (\ref{Bww}), expressed in terms of the partial amplitudes $\cB_{w_f, w_g} ^{(k)} (s_{ij} |\Omega) $, 
\bea
\cB_w(s_{ij}|\Omega) = \sum _{w_f, w_g=0}^w { \delta _{w_f+w_g,w} \over w_f! ~ w_g!} 
 \left ( - { 1 \over 4 \pi t} \right )^{w_f+2} \sum _{k=0}^2 \cB_{w_f, w_g} ^{(k)} (s_{ij} |\Omega) 
+ \cO(e^{-2 \pi t})
\eea
It was proven in Lemma 2 that each function $\cB_{w_f, w_g} ^{(k)} (s_{ij} |\Omega) $ is a polynomial in $t$, whose degree is bounded from above by $2w+2$, up to contributions which are exponentially suppressed in $t$. 
To prove the theorem, it remains to make precise the degree of the resulting Laurent polynomial in $t$ for the full amplitude $\cB_w(s_{ij}|\Omega)$, which is achieved by carefully inspecting the degree of each type of contribution.  The special restrictions on degrees and on vanishing terms all occur  for $w_g=0$ and $w_g=1$ according to items (e), (f), and (g) of Lemma 2, and therefore we shall isolate these contributions explicitly below, 
\bea
\cB_w(s_{ij}|\Omega) & = &
{ 1 \over w!}  \left (  { -1 \over 4 \pi t} \right )^{w+2}  \cB_{w, 0} ^{(2)} (s_{ij} |\Omega) 
\\ &&
 + {1 \over (w-1)!}  \left (  { -1 \over 4 \pi t} \right )^{w+1} \Big (  \cB_{w-1, 1} ^{(2)} (s_{ij} |\Omega) +  \cB_{w-1, 1} ^{(1)} (s_{ij} |\Omega)  \Big ) 
\no \\ &&
+ \sum _{w_g=2}^w  { 1 \over (w-w_g)! ~ w_g!}  \left (  { -1 \over 4 \pi t} \right )^{w-w_g+2} \sum _{k=0}^2 \cB_{w-w_g, w_g} ^{(k)} (s_{ij} |\Omega) 
+ \cO(e^{-2 \pi t})
\no
\eea
Here, we have used item (a) of Lemma 2 to omit the terms with $w_g=0$ and $w_g=1$ for $\cB^{(0)}_{w-w_g, w_g}(s_{ij} |\Omega) $ and the terms with $w_g=0$ for $\cB^{(1)}_{w-w_g, w_g}(s_{ij} |\Omega) $. In the table below we list the range of exponents $t^\alpha$ to $\cB_w (s_{ij} |\Omega)$ which arises from each contribution $\cB_{w-w_g, w_g} ^{(k)} (s_{ij} |\Omega) $, labelled by $w_g$ and $k=0,1,2$,
\bea
k=0  \hskip 0.5in w_g\geq 2 & \hskip 0.5in & -w \leq \alpha \leq -2
\no \\
k=1  \hskip 0.5in w_g\geq 1 & \hskip 0.5in & -w \leq \alpha \leq -1
\no \\
k=2  \hskip 0.5in w_g\geq 0 & \hskip 0.5in & -w \leq \alpha \leq w
\eea
For $k=0$ we have used item (b) of Lemma 2; for $k=1$ we have used items (c) and (g) of Lemma 2; while for $k=2$ we have used items (d), (e), and (f) of Lemma 2. It is immediate from inspection of this table that $\cB_w(s_{ij}|\Omega)$ has a Laurent polynomial expansion in $t$ of the form stated in Theorem 1. Since the genus-two amplitude $\cB_w(s_{ij}|\Omega) $ is invariant under the genus-two modular group $Sp(4,\ZZ)$ and the parameter $t$ is invariant under the corresponding $SL(2,\ZZ)$ subgroup, it follows immediately that the expansion coefficients $\mb^{(k)} _w (s_{ij} |v, \rho)$ must be invariant under this $SL(2,\ZZ)$ as well.
This completes the proof of Theorem 3.


\section{Degenerating higher genus modular graph functions}
\setcounter{equation}{0}
\label{sec6}

In this last section, we shall obtain the structure of the non-separating degeneration of the modular graph functions belonging to the simple class put forward in section 2.5 for arbitrary genus.  The functions $\mod (s_{ij}|\Omega)$ of (\ref{h-loop}) provide, in some sense, the simplest generalization of genus-one modular graph functions. As was noted in section 2.6, this class of modular graph functions does not include superstring amplitudes at higher genus. Nonetheless, the structure of their non-separating degeneration accurately illustrates the behavior for the more general classes of modular graph functions described in section 2.8, which do include  genus-two superstring amplitudes.  

\sm

We consider the generating function of modular graph functions $\mod (s_{ij}|\Omega)$  on a surface $\Sigma _{h+1}$ of genus $h+1$ for $h \geq 1$ with at most $N$ vertices,  given by (\ref{h-loop}),
\bea
\label{h-loop1}
\mod  (s_{ij}|\Omega) = \prod _{i=1}^N \int _{\Sigma_{h+1}}  \kap_{h+1} (z_i) \, 
\exp \left \{ \sum _{1 \leq i < j \leq N} s_{ij} \, \GA_{h+1} (z_i, z_j|\Omega) \right \}
\eea
The complex parameters $s_{ij}$ are considered as  independent, unlike for the case of string amplitudes. As in the case of genus-one modular graph functions, this provides the appropriate generating function for {\sl all} modular graph functions, and not just the particular combinations that enter into string amplitudes.
The analytic properties of $\mod (s_{ij}|\Omega) $ in the variables $s_{ij}$ are identical to those of its genus-one counterpart, and its Taylor series at $s_{ij}=0$ is absolutely convergent when $|s_{ij} | <1$ for all $i,j=1,\cdots, N$. Expanding in powers of the variables $s_{ij}$ or equivalently in powers of $\GA_{h+1}$ gives, 
\bea
\mod  (s_{ij}|\Omega)  =  \sum _{w=0}^\infty { 1 \over w!} \, \mod _w (s_{ij}|\Omega) 
\eea
The weight $w$ contribution $\mod _w (s_{ij}|\Omega)$ is a homogeneous polynomial in $s_{ij}$ of degree $w$, given by the following absolutely convergent integrals on $N$ copies of $\Sigma _{h+1}$,
\bea
\mod _w (s_{ij}|\Omega)  = \prod _{i=1}^N \int _{\Sigma_{h+1}}  \kap_{h+1} (z_i) \, 
\left \{  \sum _{1 \leq i < j \leq N} s_{ij} \, \GA_{h+1} (z_i, z_j|\Omega) \right \}^w
\eea
We shall now prove the following theorem.

\vskip 0.2in

\noindent
{\bf Theorem 4} ~ {\sl The generating function $\mod _w (s_{ij}|\Omega)$ of modular graph functions  on the genus $h+1$ surface $\Sigma_{h+1}$ has the following behavior under the non-separating degeneration
of the cycle $\mB_{h+1}$ on $\Sigma _{h+1}$, 
\bea
\mod_w (s_{ij}|\Omega) = \sum _{k=-w}^w \mf _w ^{(k)} (s_{ij} |v,\tau) (\pi t)^k + \cO(e^{- 2 \pi t})
\eea
where $\Omega$ and $\tau$ are the period matrices respectively of the compact surfaces $\Sigma _{h+1}$ and $\Sigma _h$. Together with the Abelian integral $v$, these variables are related as follows,
\bea
\Omega = \left ( \begin{matrix} \tau & v \cr v^t & \sigma \cr \end{matrix} \right )
\hskip 1in 
t = { \det (\Im \Omega ) \over \det ( \Im \tau )}
\eea
The coefficients $\mf^{(k)} _w (s_{ij} | v , \tau)$ are generalized modular graph functions which depend on the punctures through $v$ and  transform under the $Sp(2h,\ZZ)\ltimes \ZZ^2$ subgroup of $Sp(2h+2,\ZZ)$ which leaves the degenerating cycle $\mB_{h+1}$ invariant, and transforms $\tau$ and $v$ as follows,
\bea
\label{mmk}
\mf_w ^{(k)} \left ( s_{ij} \Big | \left (  (c \tau +d)^{-1} \right )^t  (v+\ZZ^h+\tau\ZZ^h), \,  (a\tau + b) (c \tau +d)^{-1} \right ) 
= \mf _w ^{(k)} (s_{ij} | v ,\tau)
\eea
with the $Sp(2h,\ZZ)$ modular transformation defined in (\ref{taup}).}

\sm

The proof relies on the degeneration formulas for the canonical K\"ahler form $\kap _{h+1} $ given in (\ref{kapf}) and the Arakelov Green function given in (\ref{thm1}). Another key ingredient is the variational method with respect to the parameter $t$, which was already employed in the proof of Theorem 3 for the case of genus two. One complication here is due to the fact that the surface $\Sigma _h$ is not necessarily a torus, so that translation invariance is not available on $\Sigma _h$.  A second, combinatorial, complication is due to the fact that we may have an arbitrary number of points $z_i$ with $i=1,\cdots, N$.

\sm

Degenerating $\kap _{h+1}$ gives rise to $2^N$ terms, depending on whether  $\kap_h$ or ${ 1 \over t} \om _t \oom_t$ occurs at each point $z_i$. We collect these different contributions with the help of an integer label $n$ in the range $0 \leq n \leq N$. The contribution $n=0$ is the single term  for which each $\kap_{h+1}$ degenerates to $\kap_h$ at every point $z_i$. For $n \geq 1$, we use the freedom to permute the points $z_i$ amongst one another (as well as  the variables $s_{ij}$) to arrange the contribution so that a factor ${ 1 \over t} \om _t  \oom_t$ occurs for the points $i=1, \cdots, n$, while a factor $\kap_h$ occurs for the points $i=n+1, \cdots , N$. We shall denote these contributions as follows,
\bea
\label{modn}
\mod _{w,n}  (s_{ij}|\Omega)  = \prod _{i=1}^n \int _{\Sigma_{h+1}}  { 1 \over t} \om _t  \, \oom_t (z_i) \, 
\prod _{j=n+1}^N \int _{\Sigma_{h+1}} \kap_h (z_j)
\left \{  \sum _{1 \leq i < j \leq N} s_{ij} \, \GA_{h+1} (z_i, z_j|\Omega) \right \}^w
\eea
It will be understood that for the  case $n=0$ the first factor must be dropped entirely. The above decomposition may be viewed as the generalization of the decomposition carried out in  (\ref{Sk}) and (\ref{Bk}) for the case of genus two.

\sm

It is instructive to consider first the case $n=0$. Since all terms in the degeneration of $\GA_{h+1}$ given in (\ref{thm1}) have logarithmic behavior as $x$ and $y$ approach one another or as they approach $p_a$ or $p_b$, the integrals against $\kap_h$ are all absolutely convergent. As a result, all $t$-dependence of $\mod_{w,0}$ may be obtained directly from a multinomial expansion of the $w$-power in (\ref{modn}).
Since $\GA_{h+1}$ has power-behaved terms of orders $t^1, t^0, t^{-1}$ only, it follows that $\mod_{w,0}$ is a Laurent polynomial in $t$ of degree $(w,w)$ plus exponentially suppressed terms. The term with the highest power of $t$ arises solely from the first term on the right side of (\ref{thm1}), while the term with the lowest power of $t$ arises from the terms on the last line of (\ref{thm1}).   The coefficients in the Laurent polynomial 
are linear combinations of products of integrals of combinations of Green functions, and are modular invariants of the form (\ref{mmk}).

\sm

For the cases $n \geq 1$ we carry out a multinomial expansion of the $w$ power of the sum over Green functions $\GA_{h+1}$, using the degeneration  of (\ref{thm1}). We are interested only in 
evaluating the $t$-dependence, up to exponential corrections, and not in obtaining the complete expressions for the coefficients in the expansion. Omitting all coefficients, we schematically represent  the seven terms in the expansion (\ref{thm1}), along with their multinomial expansion power,  as follows,
\bea
\label{w}
\GA_{h+1} & = & t \, \, \, + \GA_h(x,y) + \GA_h(p,p) + \GA_h(x,p) + { 1 \over t} f(x)^2 +{1 \over t} f(x)f(y) + {1 \over t} 
\no \\
w & = & w_t + ~~~w_g ~~~ + ~~~ w_{pp} ~~~ + ~~~ w_p ~~~ + ~~~ w_{f^2} ~~ + ~~~~~~ w_{f} ~~~~ +  w_c
\eea
where $w_t, w_g, w_{pp}, w_p, w_{f^2}, w_f, w_c \geq 0$. A term with the above weight assignment has the following structure,
\bea
\label{moda}
t^{w_t-w_{f^2}-w_{f} - w_c-n}  \, \cC _{\{ w\} }
\eea
where the coefficient is given by,
\bea
\cC _{\{ w\} } & = & 
\prod _{i=1}^n \int _{\Sigma_{h+1}}   \om _t  \oom_t (z_i) \, 
\prod _{j=n+1}^N \int _{\Sigma_{h+1}} \kap_h (z_j)
\left (  \sum _{i < j } s_{ij} \, \GA_h (z_i, z_j) \right )^{w_g}
 \\ && \times
\left (  \sum _{i \not= j } s_{ij} \, \GA_h (z_i, p) \right )^{w_p}
\left (  \sum _{i \not= j } s_{ij} \, f (z_i)^2 \right )^{w_{f^2}}
\left (  \sum _{i \not= j } s_{ij} \, f (z_i)f(z_j) \right )^{w_{f}}
\no
\eea
and $\{ w \}$ stands for the assignment of the seven $w$-values in (\ref{w}). The key property of $\cC_{\{ w\}}$ is that its integrand is independent of $t$, a features which was obtained at the cost of forming the above multinomial expansion. As a result, all $t$-dependence of $\cC_{\{ w\}}$ arises from the $t$-dependence of its integrations in $z_1, \cdots, z_n$.

\sm

Using the variational arguments, developed for genus two in the proof of Theorem 3, we may now take the variations in $t$ of these integrals whose sole $t$-dependence is through its integration domain. By induction on $n$ one shows that $\cC_{\{w\}}$ is a polynomial in $t$ of degree $n+w_g+ w_p+2w_{f^2} + 2 w_f$ for $n \geq 2$, and $n+ w_p+2w_{f^2} + 2 w_f$ for $n=1$. However, $\cC_{\{w\}}$ receives vanishing contributions to the orders $t^0, \cdots, t^{n-1}$. Putting all together, we see that the contribution with the assignment $\{ w \}$ given above has the following $t$-dependence, 
\bea
t^{w_t-w_{f^2}-w_{f} - w_c-n}   \sum_{k=n}^{n+w_g + w_p+2w_{f^2} + 2 w_f} \cC_k t^k
= 
  \sum_{k=k_- }^{ k_+ } \tilde \cC _k \,  t^k
\eea
for some $t$-independent coefficients $\cC_k$ and $\tilde \cC_k$ and with the following expressions for $k_\pm$,
\bea
k_+ & = & w -w_{pp} - 2w_c
\no \\
k_- & = & -w +2w_t +w_g+w_{pp}+w_p
\eea
Clearly, since all partial weights are positive or zero, the largest value which can be attained for $k_+$ is $w$ requiring $w_{pp} =w_c=0$, while the smallest value that can be attained by $k_-$ is $-w$ for $w_t=w_g=w_{pp}=w_p=0$. These conditions are compatible with one another, so that as all possible values of the partial weights are considered, we find that $k_+ \leq w$ and $k_- \geq -w$ for all cases. This result concludes the proof of Theorem 4, as we already know that all coefficients are modular graph functions and their non-holomorphic Jacobi form generalizations.


\appendix

\section{Appendix: Proof of Lemma 2}
\setcounter{equation}{0}
\label{appA}

In this appendix, we present the proof of each one of the statements in Lemma 2. 

\subsection{Proof of {\rm (a)}}

The proof of (a) is based on the following identity,
 \bea
\label{id5}
 \int _\Sep \kappa (z) \, \p_z f(z) \, f(z)^n =0
 \eea
 valid for any $n \geq 0$, up to contributions which are exponentially suppressed in $t$. For $w_g=0$, the identity (\ref{id5}) may be applied to the integration over any one of the points $z_i$ in the function $\cB_{w, 0} ^{(0)} (s_{ij} |\Omega)$ and gives zero. For $w_g=1$, the integrand of  $\cB_{w-1, 1} ^{(0)} (s_{ij} |\Omega)$ contains  one factor of the Green function  $g(z_i-z_j)$, and the identity (\ref{id5}) may be applied to either one of the remaining points $z_k$ with $k\not=i,j$, and gives zero. Finally, for $w_g=0$, the integrand of $\cB_{w, 0} ^{(1)} (s_{ij} |\Omega)$ contains a factor of $|\p_i f|^2  ( \p_j f \, \bar \p_k f + \bar \p_j f \,  \p_k f ) $ with $j\not= k$ and $j,k \not= i$. The identity (\ref{id5}) may be applied to the integration over the points $z_j$ and $z_k$ and gives zero.
 
\subsection{Proof of {\rm (b)}}
 
The proof of (b) is straightforward, as the integrations over all vertex insertion points $z_i$ are convergent as $t \to \infty$ on the compact torus. Since the integrals are convergent, the difference between the contributions from the surface $\Sep$ and  the genus-one surface $\Sigma $ is exponentially suppressed (see the results of section 3.5), and thus only a contribution of order $t^0$ is generated.

\subsection{Proof of {\rm (c)}}

The proof of (c) will be obtained by considering the variation $\delta \cB_{w_f, w_g} ^{(1)} (s_{ij} |\Omega) $ under a variation of $t$.  We concentrate on the contribution from the term $s_{12} s_{13}|\p_1 f|^2 ( \p_2 f \, \bar \p_3 f + \bar \p_2 f \,  \p_3 f ) $ in $\cS^{(1)}$, the other contributions being obtained by permutations of the integration points and permutations of $s_{ij}$. The variation in $t$ is obtained entirely,  up to exponential corrections,  from the variation of the integration domain for $z_1$, and is given by,
\bea
\delta \mB_{w_f, w_g} ^{(1)} (s_{ij} |\Omega)  & = &  
s_{12} s_{13} \, { \tau_2^2 \over\pi^2} \int _{\delta D_a\cup \, \delta D_b} \kappa (z_1)  |\p_1 f|^2 \, \prod _{i=2}^4 \int_\Sigma  \kappa (z_i) \, 
  \Big ( \p_2 f \, \bar \p_3 f + \bar \p_2 f \,  \p_3 f \Big ) 
\no \\ && \times 
\left (  \sum _{i<j} s_{ij} \, f(z_i) f(z_j) \right )^{w_f}
\left ( \sum_{i<j} s_{ij} \, g(z_i-z_j) \right )^{w_g} + \hbox{11 perms} 
\eea
The integrations over $z_2, z_3, z_4$ are all convergent as $t \to \infty$, and the integration domains for these variables may be extended from $\Sep$ to the full torus $\Sigma$, up to exponentially suppressed terms (see subsection 3.5). We compute the integrals over $\delta D_a$ and $\delta D_b$ following the methods used earlier in subsection 3.6. In particular, for $z_1 \in \delta D_a$, we have $f(z_1)=-2\pi t$, while for $z_1 \in \delta D_b$ we have $f(z)=2\pi t$.   We can write out these contributions explicitly,
\bea
{ \p \over \p t}  \cB_{w_f, w_g} ^{(1)} (s_{ij} |\Omega)  =  
\cL_+^{(1)}  (s_{ij}|\Omega) + \cL_-^{(1)}  (s_{ij}|\Omega) + \hbox{ 11 permutations} 
\eea
where 
\bea
\cL _\pm^{(1)}  (s_{ij}|\Omega)  & = & 
s_{12} s_{13} \, { \tau_2  \over \pi} \int _0^{2 \pi} d \theta  \, \prod _{i=2}^4 \int_\Sigma  \kappa (z_i) \, 
   \Big ( \p_2 f \, \bar \p_3 f + \bar \p_2 f \,  \p_3 f \Big ) \left ( \sum_{i<j} s_{ij} \, g(z_i-z_j) \right )^{w_g} 
\no \\ && \hskip 0.5in \times 
\left (  \sum _{1<i<j} s_{ij} \, f(z_i) f(z_j) \pm 2 \pi t \sum _{j=2}^4 s_{1j} f(z_j)  \right )^{w_f}
\eea
In $\cL_-^{(1)} $, the point $z_1$ is parametrized  by $z_1 = p_a^\theta $ while in $\cL_+^{(1)} $ the point $z_1$ is parametrized by $z_1 = p_b^\theta $. All remaining integrals have  finite limits as $t \to \infty$. The integrals over $\theta$ localize their integrand at $p_a$ in $\cL_-^{(1)} $ and at $p_b$ in $\cL_+^{(1)}$ and evaluate to $2 \pi$.  The highest power of $f(z_1)= \pm 2 \pi t$ that can occur is $w_f$, so that the $t$-derivative is a polynomial in $t$ of degree $w_f$, and $ \cB_{w_f, w_g} ^{(1)} (s_{ij} |\Omega)$ is a polynomial in $t$ of degree $w_f+1$ as long as $w_g >0$, and up  to terms which are exponentially suppressed  in $t$.

\subsection{Proof of {\rm (d)}}

The proof of (d) will  be obtained by considering the  variation in $t$ of the first variation $\delta \cB_{w_f, w_g} ^{(2)} (s_{ij} |\Omega) $. We concentrate on the contribution from the term  $s_{12}^2 |\p_1f|^2 |\p_2 f|^2$ in $\cS^{(2)}$, the other contributions being obtained by two permutations. The variation in $t$ receives contributions from the variations of the integration domains of both $z_1$ and $z_2$, but again these two contributions are related by permutations. Thus, in total we vary only the integration domain of $z_1$ and include 5 permutations. Putting all together, we have, 
\bea
\delta \mB_{w_f, w_g} ^{(2)} (s_{ij} |\Omega)  & = &  
s_{12}^2 \, { \tau_2^2 \over\pi^2} \int _{\delta D_a\cup \, \delta D_b} \kappa (z_1)  |\p_1 f|^2 \prod _{i=2}^4 \int_\Sep  \kappa (z_i) \,    |\p_2 f|^2 \left ( \sum_{i<j} s_{ij} \, g(z_i-z_j) \right )^{w_g} 
\no \\ && \times 
\left (  \sum _{i<j} s_{ij} \, f(z_i) f(z_j) \right )^{w_f}
+ \hbox{5 permutations} 
\eea
Separating out the contributions from $\delta D_a$ and $\delta D_b$, we have,
\bea
{ \p \over \p t}  \cB_{w_f, w_g} ^{(2)} (s_{ij} |\Omega)  =  
\cL_+^{(2)}  (s_{ij}|\Omega) + \cL_-^{(2)}  (s_{ij}|\Omega) + \hbox{ 5 permutations} 
\eea
where
\bea
\cL_\pm ^{(2)} & = & 
s_{12}^2 \, { \tau_2  \over \pi} \int _0^{2 \pi} d \theta _1 \, \prod _{i=2}^4 \int_\Sep  \kappa (z_i) \,    |\p_2 f|^2 
\left ( \sum_{i<j} s_{ij} \, g(z_i-z_j) \right )^{w_g} 
\no \\ && \times 
\left (  \sum _{1<i<j} s_{ij} \, f(z_i) f(z_j) \pm 2 \pi t \sum_{j=2}^4 s_{1j} f(z_j) \right )^{w_f}
\eea
The point $z_1$ is parametrized in $\cL_-^{(2)}$ and $\cL_+^{(2)}$ respectively by $z_1 = p_a^{\theta_1}$ and $z_1 = p_b^{ \theta_1}$. Next, we expand the integrands in order to pick up all its polynomial dependence on $t$, 
\bea
\cL_-^{(2)}   & = &  
s_{12}^2 \sum _{k=0}^ {w_f} \left ( \begin{matrix}w_f \cr k \cr \end{matrix} \right ) ( -2 \pi t)^k \cC_{w_f,w_g} ^{(k)} (p_a) 
\no \\ 
\cL_+^{(2)}   & = &  
s_{12}^2  \sum _{k=0}^ {w_f} \left ( \begin{matrix}w_f \cr k \cr \end{matrix} \right ) ( +2 \pi t)^k \cC_{w_f,w_g} ^{(k)} (p_b) 
\eea
and 
\bea
\cC_{w_f,w_g} ^{(k)} (p_a)  
& = & 
{\tau_2 \over \pi} \int _0^{2 \pi} d \theta _1 \, \prod _{i=2}^4 \int_\Sep  \kappa (z_i) \,    |\p_2 f|^2 
\left ( \sum_{i<j} s_{ij} \, g(z_i-z_j) \right )^{w_g} 
\no \\ && \times 
\left (  \sum _{1<i<j} s_{ij} \, f(z_i) f(z_j) \right )^{w_f-k} \left ( \sum_{j=2}^4 s_{1j} f(z_j) \right )^k
\eea
where the point $z_1$ is parametrized by $z_1=p_a^{ \theta _1}$, and enters only as an argument of the Green function $g(z_1-z_j)$. We have a similar formula for $\cC_{w_f,w_g} ^{(k)} (p_b) $, where the point $z_1$ is parametrized by $z_1=p_b^{ \theta _1}$.

\sm

Having extracted all the explicit $t$-dependence, we now consider the variation in $t$ of the coefficients
$\cC_{w_f,w_g} ^{(k)} (p_a)  $ and $\cC_{w_f,w_g} ^{(k)} (p_b)  $. Their dependence on $t$ is through two different sources. The first is the dependence of the point $z_1$ on $R$ which itself depends on~$t$. This dependence, however, is exponentially suppressed in $t$, and may be neglected. The other source is the dependence of the integration domain for $z_2$ on $t$, and is given by,
\bea
{ \p \over \p t} \cC_{w_f,w_g} ^{(k)} (p_a)  & = & \cL^{(3)}_+ (p_a) + \cL^{(3)}_- (p_a)
\no \\
{ \p \over \p t} \cC_{w_f,w_g} ^{(k)} (p_b)  & = & \cL^{(3)}_+ (p_b) + \cL^{(3)}_- (p_b)
\eea
The combinations $\cL^{(3)}_\pm (p_a)$ are given by,
\bea
\cL^{(3)}_\pm (p_a) & = & 
 \int _0^{2 \pi} d \theta _1 \, \int _0^{2 \pi} d \theta _2 \, \prod _{i=3}^4 \int_\Sigma  \kappa (z_i) \,  
\left ( \sum_{i<j} s_{ij} \, g(z_i-z_j) \right )^{w_g} 
\\ && \times 
\left (  s_{34} \, f(z_3) f(z_4) \pm 2 \pi t \sum _{j=3,4} s_{2j} f(z_j) \right )^{w_f-k} 
\left ( \sum_{j=3}^4 s_{1j} f(z_j) \pm 2 \pi t s_{12} \right )^k
\no
\eea
where we set $z_1=p_a^{\theta _1}$ in both $\cL^{(3)}_\pm$,  while in $\cL^{(3)}_- $ we set $z_2 = p_a^{ \theta _2}$ and in $\cL^{(3)}_+ $ we set $z_2 = p_b^{ \theta _2}$. The combinations $\cL^{(3)}_\pm (p_b)$ are given by, 
\bea
\cL^{(3)}_\pm (p_b) & = & 
 \int _0^{2 \pi} d \theta _1 \, \int _0^{2 \pi} d \theta _2 \, \prod _{i=3}^4 \int_\Sigma  \kappa (z_i) \,  
\left ( \sum_{i<j} s_{ij} \, g(z_i-z_j) \right )^{w_g} 
\\ && \times 
\left (  s_{34} \, f(z_3) f(z_4) \pm 2 \pi t \sum _{j=3,4} s_{2j} f(z_j) \right )^{w_f-k} 
\left ( \sum_{j=3}^4 s_{1j} f(z_j) \pm 2 \pi t s_{12} \right )^k
\no
\eea
where we set $z_1=p_b^{ \theta _1}$ in both $\cL^{(3)}_\pm$,  while in $\cL^{(3)}_- $ we set $z_2 = p_a^{ \theta _2}$ and in $\cL^{(3)}_+ $ we set $z_2 = p_b^{ \theta _2}$. 

\sm

In the functions $\cL_+^{(3)}(p_a)$ and $\cL_- ^{(3)}(p_b)$, we may simply set $(z_1, z_2) =(p_a,p_b)$ and $(z_1, z_2) =(p_b,p_a)$ respectively, within the approximation where exponentials in $t$ are omitted.   The  integrations over $\theta _1$ and $\theta_2$ are then trivial to carry out and each give a factor of $2 \pi$, while the remaining integrations over $z_3$ and $z_4$ are convergent in the $t \to \infty $ limit. Their dependence on $t$ is then through a polynomial of degree $w_f$.

\sm

In the functions $\cL_-^{(3)}(p_a)$ and $\cL_+ ^{(3)}(p_b)$ on the other hand, setting $z_1$ and $z_2$ simply to $p_a$ or to $p_b$ will produce a singularity in the Green function $g(z_1-z_2)$. Therefore, the full dependence on the angles $\theta_1, \theta _2$ must be retained and integrated over. To do so, we first prove the following Lemma.

\subsubsection{Lemma  3 ~ {\sl Integrals inside the funnels, defined by,}}
\bea
V_N =  \int _0^{2 \pi} d \theta _1 \, \int _0^{2 \pi} d \theta _2 \, 
g \left (R e^{i \theta _1}- R e^{i \theta _2} \right ) ^N
\eea
{\sl evaluate as follows, }
\bea
V_N = 
 4 \pi^2 \sum _{n=0}^N \left ( \begin{matrix}N \cr n \cr \end{matrix} \right ) 
 \Big (-2 \pi t - g(p_a-p_b) \Big )^{N-n} \sum_{k_1, \cdots, k_n \not=0}  
 { \delta _{k_1 + \cdots + k_n,0} \over  |k_1| \cdots |k_n| } 
\eea

\sm

To prove this lemma, we use the results of subsection 3.5.1 and in particular the fact that the argument of the Green function under the integral is exponentially suppressed so that the following approximate expression for the Green function may be used reliably within the exponential approximation, 
\bea
g(x-y)= - \ln |x-y|^2-  2\ln R - 2 \pi t - g(p_a-p_b) + \cO(x-y)
\eea
Clearly, the $R$-dependence cancels, and we are left with the simplified expression, 
\bea
V_N =  2 \pi \int _0^{2 \pi} d \theta  
 \left ( - \ln \left | 1- e^{i \theta } \right |^2  - 2 \pi t - g(p_a-p_b) \right ) ^N
\eea
Expanding the $N$-th power, we find, 
\bea
V_N =  2 \pi \sum _{n=0}^N \left ( \begin{matrix}N \cr n \cr \end{matrix} \right ) (- 2 \pi t - g(p_a-p_b))^{N-n}  
\int _0^{2 \pi} d \theta  
 \left ( - \ln \left | 1- e^{i \theta } \right |^2   \right ) ^n
\eea
To compute the $\theta$-integral we use the power expansion of each logarithmic factor, 
which upon minor rearrangements gives the statement of the Lemma.

\subsubsection{Completing the proof of {\rm (d)}}

Using the result of Lemma 3, we may now extract the $t$-dependence of the remaining functions. We shall concentrate on $\cL^{(3)}_-(p_a)$, the case of $\cL^{(3)}_+(p_b)$ being analogous. We begin by making the integrals explicit, upon neglecting contributions which are exponentially suppressed, and we have, 
\bea
\cL^{(3)}_-(p_a) & = & 
 \int _0^{2 \pi} d \theta _1 \, \int _0^{2 \pi} d \theta _2 \, \prod _{i=3}^4 \int_\Sigma  \kappa (z_i) 
\\ && \times 
\left ( s_{12} \, g(R \, e^{i \theta _1}- R\, e^{i \theta _2}) +  \sum _{j=3,4} (s_{1j}+s_{2j}) g(p_a-z_j) + s_{34} g(z_3-z_4)  \right )^{w_g} 
\no \\ && \times 
\left (  s_{34} \, f(z_3) f(z_4) - 2 \pi t \sum _{j=3,4} s_{2j} f(z_j) \right )^{w_f-k} \left ( \sum_{j=3}^4 s_{1j} f(z_j) -2 \pi t s_{12} \right )^k
\no
\eea
The integrals over $z_3$ and $z_4$ are convergent. The highest degree dependence on $t$ is obtained when the  Green function $g(R \, e^{\theta _1}- R \, e^{i \theta _2})$ is raised to the maximal power $w_g$ on the second line, while the maximal power of $t$ available from the remaining two factors in $w_f$. 
We are now in  a position to complete the counting for the degree in $t$. Putting together the results we have obtained for the integrals inside the funnel, we see that $\p_t \cC_{w_f,w_g} ^{(k)} (p_a) $ and $\p_t \cC_{w_f,w_g} ^{(k)} (p_b) $ are polynomials in $t$ of degree $w$.  Therefore, $\cC_{w_f,w_g} ^{(k)} (p_a) $ and $\cC_{w_f,w_g} ^{(k)} (p_b) $ are polynomials of degree $w+1$, while $\p_t \cB _{w_f, w_g} ^{(2)} (s_{ij}|\Omega)$ is of degree $w+w_f+1$ and $\cB _{w_f, w_g} ^{(2)} (s_{ij}|\Omega)$ is of degree $w+w_f+2$, which proves item (d).

\subsection{Proof of {\rm (e)}}

The proof of (e) relies on the exact results, up to exponentially suppressed terms,  which are available 
for $w_g=0$, and we  have,
\bea
\cB_{w, 0}^{(2)} (s_{ij} |\Omega)  =  
{ \tau_2^2 \over\pi^2}  \sum _{i<j} s_{ij}^2 \int _{\Sigma _{ab} ^4} \kappa \,  |\p_i f|^2 \, |\p_j f|^2
\left (  \sum _{k < \ell } s_{k \ell} \, f(z_k) f(z_\ell) \right )^w
\eea
By expanding the $w$-th power of the sum in the last factor, we see that $\cB_{w, 0}^{(2)} (s_{ij} |\Omega)$ is a linear combination of terms with $i \not= j$ of the following form, 
\bea
\cI^{(2)} (a_i, a_j, a_k, a_\ell) = \tau_2^2 \int _{\Sigma _{ab} ^4} \kappa \,  |\p_i f|^2 \, |\p_j f|^2 f(z_i)^{a_i} f(z_j)^{a_j} f(z_k)^{a_k} f(z_\ell)^{a_\ell}
\eea
where $k \not= \ell$ as well as $k, \ell \not= i,j$, and we have $a_i,a_j,a_k,a_\ell \geq 0$ with 
\bea
a_i+a_j+a_k+a_\ell = 2w
\eea
Using the  formulas of (\ref{regs1}) and (\ref{regs2}), recast in the notation used here, we find that,
\bea
\tau_2 \int _\Sep \kappa (z) |\p_z f(z)|^2 f(z)^n = { (2 \pi)^{n+2} \over (n+1)} t^{n+1} 
\eea
when $n$ is even and vanishes when $n$ is odd. Using this result, the  integral $\cI^{(2)} $ is found to vanish whenever $a_i$ or $a_j$ are odd, while when $a_i$ and $a_j$ are both even, it is given as follows,
\bea
\cI^{(2)}(a_i, a_j, a_k, a_\ell) = { (2\pi)^{a_i+a_j+2} \over (a_i+1)(a_j+1)} \, t^{a_i+a_j+2} 
\prod _{n=k,\ell} \int _\Sigma \kappa (z_n) f(z_n)^{a_n}
\eea
Since the possible values of $a_i+a_j$ range between $0$ and $2w$, we see that $\cB_{w, 0}^{(2)} (s_{ij} |\Omega)$ is a polynomial in $t$ of degree $2w+2$ with vanishing monomials of order $t^0$ and $t^1$.

\subsection{Proof of {\rm (f)}}

The proof of (f) proceeds analogously to the proof of (e),
\bea
\cB_{w-1, 1}^{(2)} (s_{ij} |\Omega)  =  
{ \tau_2^2 \over\pi^2}  \sum_{\alpha < \beta} s_{\alpha \beta} 
\sum _{i<j} s_{ij}^2 \int _{\Sigma _{ab} ^4} \kappa \,  |\p_i f|^2 \, |\p_j f|^2 g(z_\alpha - z_\beta) 
\left (  \sum _{k < \ell } s_{k \ell} \, f(z_k) f(z_\ell) \right )^{w-1}
\eea
By expanding the $(w-1)$-th power of the sum in the last factor, we see that $\cB_{w-1, 1}^{(2)} (s_{ij} |\Omega)$ is a linear combination of terms with $i \not= j$ of the following form, 
\bea
 \int _{\Sigma _{ab} ^4} \kappa \,  |\p_i f|^2 \, |\p_j f|^2 g(z_\alpha - z _\beta) f(z_i)^{a_i} f(z_j)^{a_j} f(z_k)^{a_k} f(z_\ell)^{a_\ell}
\eea
where $k \not= \ell$ as well as $k, \ell \not= i,j$, and we have $a_i,a_j,a_k,a_\ell \geq 0$ with 
\bea
a_i+a_j+a_k+a_\ell = 2w-2
\eea
When $\{ \alpha, \beta \} \cap \{ i, j \} = \emptyset$, the integrations in $i$ and $j$ each provide at least one factor of $t$, so that terms of orders $t^0$ and $t^1$ vanish. When $\{ \alpha, \beta \} \cap \{ i, j \} =\{ \alpha =i\}$ (or any permutation thereof), the integration over $z_j$ will produce at least one factor of $t$, and the term of order $t^0$ will vanish. Finally, when $\{ \alpha, \beta \} = \{ i, j \} $, we focus on the integrals with $z=z_i$ and $w=z_j$,
\bea
\tau_2^2  \int _\Sep  \kappa (z) \int _\Sep \kappa (w)   |\p_z f|^2 \, |\p_w f|^2 g(z - w) f(z)^{a_i} f(w)^{a_j}
\eea
which multiply integrals which converge to finite factors as $t \to \infty$. Fortunately, these integrals can be computed exactly.  Carrying out the integral in $w$, we have,
\bea
&&
\tau_2 \int _\Sep \kappa (w)   |\p_w f|^2 g(z - w)  f(w)^{a_j}
\no \\ && \qquad = 
{ (2 \pi t)^{a_j+1} \over 2 (a_j+1)} \int _0 ^{2 \pi} d \theta 
\Big ( g(p_a^\theta -w) +(-)^{a_j} g(p_b^\theta  -w) \Big )
\no \\ && \hskip 0.6in 
+ { \pi \over (a_j+1)(a_j+2)} \Big ( -f(z)^{a_j+2} + \int _\Sigma \kappa (w) f(w)^{a_j+2} \Big )
 \eea
 Clearly the first term has an overall factor of $t$, so its contribution has vanishing order $t^0$ term. 
 For the remaining terms, it is straightforward to see that they also factor out at least one power of $t$ upon further integration over $z$, thereby proving part (f).

\subsection{Proof of {\rm (g)}}

The proof of {\rm (g)} similarly relies on exact results, up to exponentially suppressed terms,  which are available 
for $w_g=1$, and we  have,
\bea
\cB_{w-1, 1} ^{(1)} (s_{ij} |\Omega)  & = &  
{ \tau_2^2 \over 2 \pi^2} \sum _{\alpha < \beta } 
\sum_{i=1}^4 \sum _{j \not = i} \sum _{k \not= i,j} s_{\alpha \beta}  s_{ij} s_{ik}
\int _{\Sigma _{ab} ^4} d\mu  \,   g(z_\alpha -z_\beta) 
 |\p_i f|^2 
\no \\ && \times 
\Big ( \p_j f \, \bar \p_k f + \bar \p_j f \,  \p_k f \Big )
\left (  \sum _{m<n} s_{mn} \, f(z_m) f(z_n) \right )^{w-1}
\eea
By analogy with {\sl (e)}, upon expanding the $(w-1)$-th power in the last factor, we see that $\cB_{w-1, 1} ^{(1)} (s_{ij} |\Omega) $ is a linear combination of terms with $j \not=k$ and $j,k \not= i$ of the following form,
\bea 
\cI^{(1)} (\alpha , \beta ; b_i, b_j, b_k, b_\ell) & = &
 \rho_2^2\int _{\Sigma _{ab} ^4} \kappa \,   g(z_\alpha -z_\beta) 
 |\p_i f|^2  \Big ( \p_j f \, \bar \p_k f + \bar \p_j f \,  \p_k f \Big )
 \no \\ && \hskip 1in  \times 
f(z_i)^{b_i} f(z_j)^{b_j} f(z_k)^{b_k} f(z_\ell)^{b_\ell} 
\eea
with $b_i, b_j, b_k, b_\ell \geq 0$ and 
\bea
b_i+b_j+b_k+b_\ell = 2w-2
\eea
Whenever the set $\{ \alpha , \beta \}$ is different from the set $\{ j,k\}$, the integral vanishes identically with the help of the identity (\ref{id5}), applied to the integration over the point $z_j$ or $z_k$ whichever does not belong to the set $\{ \alpha , \beta \}$. The only remaining non-vanishing integral is therefore for  $\alpha =j$ and $\beta =k$ (or the swapped version thereof) without loss of generality, 
\bea
\cI^{(1)} (j , k ; b_i, b_j, b_k, b_\ell) & = &
 \tau_2^2\int _{\Sigma _{ab} ^4} \kappa \,   g(z_j -z_k) 
 |\p_i f|^2  \Big ( \p_j f \, \bar \p_k f + \bar \p_j f \,  \p_k f \Big )
 \no \\ && \hskip 1in  \times 
f(z_i)^{b_i} f(z_j)^{b_j} f(z_k)^{b_k} f(z_\ell)^{b_\ell} 
\eea
The integrals over $z_i$ and $z_\ell$ factor out of the combined integral over $z_j$ and $z_k$.  When $b_i$ is odd, the integrals vanishes. When $b_i$ is even, it evaluates to,
\bea
\mI^{(1)} (j , k ; b_i, b_j, b_k, b_\ell) = 
{ (2 \pi)^{b_i+2} \over b_i+1} \, t^{b_i+1}\,  \cL(b_j,b_k) \, \int _\Sigma \kappa (z_\ell) f(z_\ell )^{b_\ell} 
\eea
where we have defined the integral over $z_j$ and $z_k$ by, 
\bea
\cL(b_j,b_k) = \tau_2 \int _\Sep \kappa (z_j) \int _\Sep \kappa (z_k) \, g(z_j -z_k)  
 \Big ( \p_j f \, \bar \p_k f + \bar \p_j f \,  \p_k f \Big ) f(z_j)^{b_j} f(z_k)^{b_k} 
\eea
Let us now show that this integral evaluates to a finite limit as $t \to \infty$. To do so, we combine the factors of $f$ with the derivatives,  
\bea
\cL(b_j,b_k) = \tau_2  \int _\Sep \kappa (z_j) \int _\Sep \kappa (z_k) \, g(z_j -z_k)  
{ \p_j f^{b_j+1}  \, \bar \p_k f^{b_k+1}  + \bar \p_j f^{b_j+1}  \,  \p_k f^{b_k+1}  \over (b_j+1)(b_k+1)}
\eea
Next, we integrate by parts in both $z_j$ and $z_k$. Surface terms are exponentially suppressed in $t$, and may be neglected, and we find,
\bea
\cL(b_j,b_k) = { 2 \tau_2 \over (b_j+1)(b_k+1)}  \int _\Sep \kappa (z_j) \int _\Sep \kappa (z_k) \, \p_j \bar \p_k g(z_j -z_k)  
 f^{b_j+1} (z_j)  f^{b_k+1} (z_k) 
\eea
Using
\bea
\tau_2 \p_j \bar \p_k g(z_j-z_k) = \pi \delta (z_j-z_k) - \pi 
\eea
it is clear that this integral is convergent as $t \to \infty$.



\providecommand{\href}[2]{#2}\begingroup\raggedright\endgroup

\end{document}